\documentclass[aps,prb,twocolumn,amsmath,amssymb,nofootinbib,eqsecnum,superscriptaddress,floatfix]{revtex4}

\usepackage{amsmath}
\usepackage{amssymb}
\usepackage[dvips]{color}
\usepackage{graphicx}
\usepackage{dcolumn} 
\usepackage{bm} 
\usepackage[hypertex]{hyperref}
\usepackage{longtable}
\newcommand{\Slash}[1]{\ooalign{\hfil/\hfil\crcr$#1$}}

\begin{document}

\title{Masses in graphene-like two-dimensional electronic systems: topological
defects in order parameters and their fractional exchange statistics}

\author{Shinsei Ryu} 
\affiliation{Department of Physics, University of California,
Berkeley, CA 94720, USA}

\author{Christopher Mudry} 
\affiliation{
Condensed Matter Theory Group,
Paul Scherrer Institut, CH-5232 Villigen PSI, Switzerland
        }

\author{Chang-Yu Hou}
\author{Claudio Chamon}
\affiliation{
Physics Department, Boston University, Boston, MA 02215, USA
        }

\date{\today}

\begin{abstract}
We classify all possible 36 gap-opening instabilities in graphene-like
structures in two dimensions, i.e., masses of Dirac
Hamiltonian when the spin, valley, and superconducting channels are
included. These 36 order parameters break up into 56 possible
quintuplets of masses that add in quadrature, and hence do not compete
and thus can coexist. There is additionally a 6th competing mass, the
one added by Haldane to obtain the quantum Hall effect in graphene
without magnetic fields, that breaks time-reversal symmetry and
competes with all other masses in any of the quintuplets. Topological
defects in these 5-dimensional order parameters can generically bind
excitations with fractionalized quantum numbers. The problem
simplifies greatly if we consider spin-rotation invariant systems
without superconductivity. In such simplified systems, the possible
masses are only 4 and correspond to the Kekul\'e dimerization pattern,
the staggered chemical potential, and the Haldane mass. Vortices in
the Kekul\'e pattern are topological defects that have Abelian
fractional statistics in the presence of the Haldane term. We
calculate the statistical angle by integrating out the massive
fermions and constructing the effective field theory for the
system. Finally, we discuss how one can have generically 
non-Landau-Ginzburg-type transitions, with direct
transitions between phases characterized by distinct order parameters.
\end{abstract}

\maketitle

\section{
Introduction
        }
\label{sec: Introduction}

Many of the physical properties of graphene are captured by
a one-band tight-binding electronic Hamiltonian
with uniform, 
real-valued, 
and nearest-neighbor hopping amplitude 
whereby:
(i) electron-electron interactions are ignored;
(ii) spin-orbit interactions are ignored;
(iii) the electronic band structure is replaced
by two conical dispersions centered about 
two non-equivalent points,
the Dirac points, in the first Brillouin zone;
and 
(iv) the coupling to electro-magnetic external fields
is governed by the minimal substitution.
For instance, graphene displays an integer quantum Hall effect (IQHE)
as a function of the applied bias voltage,%
~\cite{Novoselov05,Zhang05}
and it shows an universal optical conductivity.%
~\cite{Nair08}
Both these properties can be understood within the 
non-interacting electron picture.

Although most experiments observe the massless Dirac
spectrum assumed in (iii), electronic instabilities
in the form of single-particle spectral gaps (mass gaps in short) 
can be triggered by external perturbations such as some commensurate
substrates,~\cite{Lanzara07} or large enough magnetic fields that can
change the balance between the kinetic and the potential energy.%
~\cite{Checkelsky09,Nomura09,Chamon09} In this paper we study a number of
issues pertaining to Dirac fermions in two-dimensions when a mass gap is
opened in the fermionic spectrum by different non-vanishing order
parameters. In particular, we shall study in great detail the simpler
case when there is no superconducting instabilities and spin-rotation
invariance is maintained, in which case there are only 4 possible
masses. We derive in this simpler case the effective action when the
massive fermions are integrated out, and read from this action the
fractional statistics of topological defects in the mass order
parameters. We also present a complete classification of all possible
masses (36 in total) in the general case where any spin, valley, and
superconducting instabilities are permitted.

In the simpler spinless problem (or, more realistically, the problem
when spin-rotation invariance is never broken), the 4 different
masses that can be added to the two-dimensional Dirac equation
representing graphene are the following. One perturbation is a
staggered chemical potential, taking values $+\mu^{\ }_{\mathrm{s}}$
and $-\mu^{\ }_{\mathrm{s}}$ in the two sublattices of the honeycomb
lattice of graphene say. It opens a gap $2|\mu^{\ }_{\mathrm{s}}|$ 
at the two Dirac points.%
~\cite{Semenoff84} 
A second mass gap $2|\eta|$ arises by adding directed next-nearest-neighbor
hopping amplitudes in the presence of fluxes, but such that no net
magnetic flux threads a hexagonal Wigner-Seitz unit cell of graphene say.  
This perturbation breaks time-reversal symmetry (TRS).%
~\cite{Haldane88} 
Finally, a real-valued modulation of the nearest-neighbor hopping 
amplitude with a wave vector connecting the two Dirac points 
(i.e., a Kekul\'e dimerization pattern for graphene) also opens a gap
$2|\Delta|$.~\cite{Hou07} This real-valued modulation
of the nearest-neighbor hoppings is parametrized by
the complex order parameter
$\Delta=\mathrm{Re}\,\Delta+{i}\mathrm{Im}\,\Delta$ whose phase
controls the angles of the dimerization pattern.  This mass
corresponds to two real masses $\mathrm{Re}\,\Delta$ and
$\mathrm{Im}\,\Delta$, bringing the total number of
real-valued masses that conserve the electron number and spin-rotation
symmetry (SRS) to four.

If the order parameters $\mu^{\ }_{\mathrm{s}}$, $\eta$, and $\Delta$
are not uniform, but vary in space and contain topological
textures, then midgap states in the massive Dirac spectrum can
appear. Examples are static line defects at which $\mu^{\
}_{\mathrm{s}}$ and $\eta$ change signs,~\cite{Callan85} and static
point defects represented by vortices in the phase of
$\Delta$.~\cite{Hou07} As occurs at a static domain wall in
one-dimensional polyacetylene,~\cite{Jackiw76,Su79,Jackiw81npb} a
fractional electronic charge is exponentially localized in the
vicinity of a static charge $\pm1$ vortex in the phase of
$\Delta$.~\cite{Hou07} 

The value of the fractional charge that is bound to a vortex in the
phase of $\Delta$ also depends on whether the vortex is dressed with 
a half flux of the axial vector potential $\boldsymbol{a}^{\ }_{5}$ 
or not.~\cite{Chamon08a,Chamon08b} 
When the axial gauge flux is absent
(logarithmically confined case), 
the value of the charge can be tuned
continuously as a function of the ratio
$\mu^{\ }_{\mathrm{s}}/m$
where $m:=\sqrt{|\Delta|^{2}+\mu^{2}_{\mathrm{s}}}$.%
~\cite{Chamon08a,Chamon08b} 
It is independent of the 
ratio $\mu^{\ }_{\mathrm{s}}/m$
when the axial gauge half flux is present (deconfined case), 
for the charge is then pinned to the rational values 
$Q=\pm 1/2$.%
~\cite{Chamon08a,Chamon08b}
These values of the fractional charges persist as long as the
magnitude of the TRS-breaking mass $|\eta|$ is smaller than the mass
scale $m$.~\cite{Chamon08a,Chamon08b}  
There is a phase transition at $|\eta|=m$. 
For $|\eta|>m$ the fractional charge bound to the vortices vanishes.
~\cite{Chamon08a,Chamon08b}

\begin{figure}
\includegraphics[angle=0,scale=0.6]{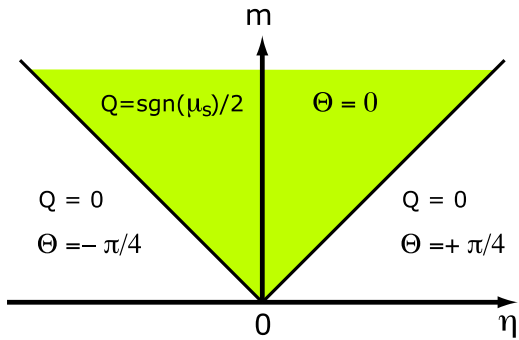}
\caption{ 
Phase diagram parametrized by the TRS mass $m$ and the
TRS-breaking mass $\eta$.  There are three regions delimited by the
boundaries $|\eta|=m$, in each of which the spectral gap does not
close.  The boundaries $|\eta|=m$ are lines of critical points at
which the spectral gap closes.  When the vortices are screened by half
of an axial gauge flux, they carry a fractional fermionic charge of
$|Q|=1/2$ with the vanishing statistical phase $\Theta=0$ under pairwise
exchange in regions for which TRS is weakly broken, i.e.,
the painted region $m>|\eta|$. 
Unit vortices are charge neutral but acquire the
non-vanishing statistical phase $|\Theta|=\pi/4$ under pairwise
exchange in regions for which TRS is strongly broken, i.e.,
$|\eta|>m$. [See Eqs.~(\ref{eq: main result}).] When the vortices are
not screened by the axial gauge flux, the charge $Q$ acquires a dependence
on the ratio of the chemical potential $\mu^{\ }_\mathrm{s}$ and $m$, for
$|\eta|<m$, and $Q$ vanishes for $|\eta|>m$. The statistics also
depend on which phase one sits, but it is non-zero for any $\eta\ne
0$, and it is related to the value of the charge, as shown in
Secs.~\ref{sec: Fractional statistical angle} and
\ref{eq: Numerical calculation of the charge and Berry phase}.}
\label{fig:phase-diagram}
\end{figure}

Just like the charge, the statistical phase $\Theta$ acquired upon the
exchange of two vortices depends on whether the vortex in the phase of
$\Delta$ is screened or not by the axial gauge flux. In this paper,
we derive the statistical angle from the effective action
obtained upon integrating out the massive fermions. 
(We thereby resolve conflicting claims about $\Theta$ in the
literature.~\cite{Chamon08a,Seradjeh08,Milovanovic08}) The statistical
angle depends on the interplay between the magnitude of the
TRS-breaking mass $\eta$ and the magnitude $m$ of the TRS masses.  
There are phase transitions at the lines $|\eta|=m$
depicted in Fig.~\ref{fig:phase-diagram} that separates regions
dominated by the TRS-breaking masses and those
dominated by the TRS-preserving mass $\eta$. The
statistics $\Theta$ jumps for both the screened and unscreened
vortices at the phase boundaries.

When unit vortices in $\Delta$ are screened by an axial gauge flux,
they are deconfined.~\cite{Jackiw07} Their statistics is well-defined
in a dynamical sense and it takes universal values independent of the
ratio $\mu^{\ }_{\mathrm{s}}$ on both sides of the transition. We show
that 
\begin{subequations}
\label{eq: main result}
\begin{equation}
\Theta=0
\hbox{
when
$m>|\eta|$
     }
\end{equation}
and that 
\begin{equation}
\frac{\Theta}{\pi}=
\mathrm{sgn}(\eta)\,
Q^{2}=
\frac{\mathrm{sgn}\,\eta}{4}
\hbox{
when
$|\eta|>m$.
     }
\end{equation}
\end{subequations}
Along the lines $|\eta|=m$ in the zero-temperature phase diagram of
Fig.~\ref{fig:phase-diagram}, the gap in the Dirac spectrum vanishes.
At criticality, the notion of point-particles is moot and so is the
question of their quantum numbers.

A remarkable complementarity has emerged. 
Defects carry either a fractional
charge $Q=\pm 1/2$ but no fractional statistical phase when the breaking
of TRS is not too strong ($|\eta|<m$), or no fractional charge but a
fractional statistics $\Theta/\pi=\pm 1/4$ when the breaking of TRS is
dominant ($|\eta|>m$).

When unit vortices in the order parameter $\Delta$ are not accompanied
by an axial gauge flux, they are logarithmically
confined.~\cite{Hou07} Although their statistics is not well-defined
dynamically, it is nevertheless possible to create them and exchange
them by external means. If so, both their charges and statistics
acquire a dependence on all masses $\eta$, $\mu^{\ }_{\mathrm{s}}$,
and $\Delta$, that we compute analytically and test numerically here
in this paper.

We then go beyond the simpler spinless case with only 4 masses, and we
classify all 36 masses in the general case where any spin, valley, and
superconducting instabilities are allowed. These 36 order parameters
break up into 56 possible quintuplets of masses that add in quadrature
(to a value $m^2$), and thus do not compete with one another. The
Haldane mass, the generalization of the $\eta$ mass above, competes
with all the other 35 masses, and thus one has generically a quantum
phase transition when $|\eta|=m$. We argue that these 5-tuplets provide
a rich playground for Landau-forbidden continuous phase transitions.
We discuss in the paper how any U(1) order parameter in a 5-tuplet can
be assigned a conserved charge and supports topological defects in the
form of vortices. A pair of U(1) order parameters in a 5-tuplet is
said to be dual if the vortices of one order parameter binds the
charge of the other order parameter and vice versa.  A continuous
phase transition can then connect directly the two dual U(1) ordered
phases through a confining-deconfining transition of their vortices.

This paper is organized as follows.  We define the relevant continuum
Dirac Hamiltonian and review its symmetries for the simpler problem
with only 4 masses, that encodes the competition between a
charge-density, a bond-density, and an integer-quantum-Hall
instability at the Dirac (charge neutral) point of any graphene-like
two-dimensional electronic system in Sec.~\ref{sec: Hamiltonian and
symmetries}. We reveal a hidden non-Abelian structure of the field
theory in Sec.~\ref{sec: Path integral formulation of the model} that
plays an important role when deriving the charge and statistics of
quasiparticles. The fermions are integrated in the background of
these 4 order parameters and of the U(1)$\times$U(1) gauge fields to
leading order in a gradient expansion in Sec.%
~\ref{sec: Derivative expansion}. 
The effective low-energy and long-wave length interacting
field theory thereby obtained is a Anderson-Higgs-Chern-Simons field
theory for bosonic fields: two U(1) gauge fields and one phase
field. The induced fractional fermion number and the induced
fractional Abelian statistical phase in the
Anderson-Higgs-Chern-Simons field theory of Sec.%
~\ref{sec: Derivative expansion} are computed in 
Sec.~\ref{sec: Fractional charge quantum number} and 
Sec.~\ref{sec: Fractional statistical angle}, respectively. The
numerical calculation of the fractional charges and statistical phases
within a single-particle (mean-field) approximation that violates the
U(1)$\times$U(1) gauge symmetry is presented in 
Sec.~\ref{eq: Numerical calculation of the charge and Berry phase}. A
microscopic (lattice) model sharing the same U(1)$\times$U(1) gauge
symmetry and low-energy long-wave-length particle content as the
Anderson-Higgs-Chern-Simons field theory is constructed in
Sec.~\ref{sec: Microscopic models}. Either by enlarging the particle
content of the lattice model from Sec.~\ref{sec: Microscopic models} or
by allowing additional magnetic, spin-orbit, or superconductivity
instabilities to compete with the charge-density, bond-density, and
integer-quantum-Hall instabilities in graphene-like two-dimensional
systems, we are lead to a classification presented in 
Sec.~\ref{sec: Reinstating the electron spin} of all 36 competing
orders of a Dirac Hamiltonian represented by 16-dimensional Dirac
matrices that encodes the quantum dynamics of electrons constrained to
a two-dimensional space, as occurs in graphene at the charge neutral
point say. We conclude in Sec.~\ref{sec: Summary} and relegate some
intermediary steps to the Appendix.


\section{
Hamiltonian and symmetries: spinless case with 4 masses
        }
\label{sec: Hamiltonian and symmetries}

The continuum model under consideration in this 
paper is defined by the second-quantized 
planar Hamiltonian
$
\hat{H}:=
\int d^{2}\boldsymbol{r}\,
\hat{\mathcal{H}}
$
where\cite{footnote: covariant covention for gradient}
\begin{subequations}
\label{eq: def quantum Hamiltonian}
\begin{equation}
\begin{split}
&
\hat{\mathcal{H}}:=
\hat{\mathcal{H}}^{\ }_{0}
+
\hat{\mathcal{H}}^{\ }_{\mathrm{gauge}}
+
\hat{\mathcal{H}}^{\ }_{\mathrm{scalar}},
\\
&
\hat{\mathcal{H}}^{\ }_{0}:=
\hat{\psi}^{\dag}
\boldsymbol{\alpha}
\cdot
\left(
-
{i}
\boldsymbol{\partial}
\right)
\hat{\psi},
\\
&
\hat{\mathcal{H}}^{\ }_{\mathrm{gauge}}:=
\hat{\psi}^{\dag}
\boldsymbol{\alpha}
\cdot
\left(
\boldsymbol{a}
+
\boldsymbol{a}^{\ }_{5}
\gamma^{\ }_{5}
\right)
\hat{\psi},
\\
&
\hat{\mathcal{H}}^{\ }_{\mathrm{scalar}}:=
\hat{\psi}^{\dag}
\left(
|\Delta|
\beta
e^{{i}\theta\gamma^{\ }_{5}}
+
\mu^{\ }_{\mathrm{s}}
R
+
{i}
\eta
\alpha^{\ }_{1}
\alpha^{\ }_{2}
\right)
\hat{\psi}.
\end{split}
\label{eq: def quantum Hamiltonian a}
\end{equation}
The 4 components of the spinor-valued operator
\begin{equation}
\hat{\psi}(\boldsymbol{r})=
\begin{pmatrix}
\hat{\psi}^{\ }_{\mathrm{+A}}(\boldsymbol{r})
\\
\hat{\psi}^{\ }_{\mathrm{+B}}(\boldsymbol{r})
\\
\hat{\psi}^{\ }_{\mathrm{-B}}(\boldsymbol{r})
\\
\hat{\psi}^{\ }_{\mathrm{-A}}(\boldsymbol{r})
\end{pmatrix}
\equiv
\begin{pmatrix}
\hat{\psi}^{\ }_{\ell}(\boldsymbol{r})
\end{pmatrix}
\label{eq: def quantum Hamiltonian b}
\end{equation}
obey the equal-time fermion algebra
\begin{equation}
\begin{split}
&
\{
\hat{\psi}^{\   }_{\ell }(\boldsymbol{r} ),
\hat{\psi}^{\dag}_{\ell'}(\boldsymbol{r}')
\}=
\delta^{\ }_{\ell,\ell'}
\delta(\boldsymbol{r}-\boldsymbol{r}'),
\\
&
\{
\hat{\psi}^{\dag}_{\ell }(\boldsymbol{r} ),
\hat{\psi}^{\dag}_{\ell'}(\boldsymbol{r}')
\}=
\{
\hat{\psi}^{\   }_{\ell }(\boldsymbol{r}),
\hat{\psi}^{\   }_{\ell'}(\boldsymbol{r}')
\}=0.
\end{split}
\label{eq: def quantum Hamiltonian c}
\end{equation}
The representation (\ref{eq: def quantum Hamiltonian b})
is here fixed by the indices
$\mathrm{A}$ and $\mathrm{B}$ that distinguish
the two triangular sublattices of the
honeycomb lattice and the indices
$+$ and $-$ that distinguish
the two inequivalent Dirac points
(valleys) of graphene.
With this choice, the 4 Dirac matrices 
$\alpha^{x}\equiv\alpha^{1}$,
$\alpha^{y}\equiv\alpha^{2}$,
$\alpha^{z}\equiv\alpha^{3}\equiv R$,
and
$\beta$
are defined by their 4-dimensional chiral 
representation\cite{Itzykson80}
\begin{equation}
\begin{split}
&
\boldsymbol{\alpha}:=
\begin{pmatrix}
\boldsymbol{\tau}
&
0
\\
0
&
-
\boldsymbol{\tau}
\end{pmatrix}\equiv
\sigma^{\ }_{3}
\otimes
\boldsymbol{\tau}\equiv
\begin{pmatrix}
\alpha^{1},
&
\alpha^{2}
\end{pmatrix},
\\
&
\alpha^{3}:=
\begin{pmatrix}
\tau^{\ }_{3}
&
0
\\
0
&
-
\tau^{\ }_{3}
\end{pmatrix}\equiv
\sigma^{\ }_{3}
\otimes
\tau^{\ }_{3}\equiv
R,
\\
&
\beta:=
\begin{pmatrix}
0
&
\tau^{\ }_{0}
\\
\tau^{\ }_{0}
&
0
\end{pmatrix}\equiv
\sigma^{\ }_{1}
\otimes
\tau^{\ }_{0},
\end{split}
\label{eq: def quantum Hamiltonian d}
\end{equation}
where the $2\times2$ unit matrix $\tau^{\ }_{0}$
and the three Pauli matrices 
$\tau^{\ }_{1}$,
$\tau^{\ }_{2}$,
and
$\tau^{\ }_{3}$
act on the sublattices indices ($\mathrm{A},\mathrm{B}$)
while
the $2\times2$ unit matrix $\sigma^{\ }_{0}$
and the three Pauli matrices 
$\sigma^{\ }_{1}$,
$\sigma^{\ }_{2}$,
and
$\sigma^{\ }_{3}$
act on the valley indices ($+,-$).
The matrix
\begin{equation}
\gamma^{\ }_{5}\equiv
\gamma^{5}:=
-{i}
\alpha^{1}
\alpha^{2}
\alpha^{3}=
\begin{pmatrix}
\tau^{\ }_{0}
&
0
\\
0
&
-
\tau^{\ }_{0}
\end{pmatrix}\equiv
\sigma^{\ }_{3}
\otimes
\tau^{\ }_{0}
\end{equation}
\label{eq: def quantum Hamiltonian e}
\end{subequations}

\noindent
acts trivially on the sublattices indices
while it acts non-trivially on the valley indices,
i.e., $(1\pm \gamma^{\ }_{5})/2$ is a projector on the
$+$ and $-$ valley indices, respectively.
In (3+1)-dimensional space and time quantum electrodynamics, 
the eigenspaces of $(1\pm \gamma^{\ }_{5})/2$
define the chiral indices, a terminology 
that we shall also use in this paper. 
The external (background) real-valued fields
$\boldsymbol{a}         =(a_{1},a_{2})$,
$\boldsymbol{a}^{\ }_{5}=(a_{51},a_{52})$,
$|\Delta|$,
$\theta\equiv
-\mathrm{arg}\,\Delta$,
$\mu^{\ }_{\mathrm{s}}$,
and 
$\eta$
are space- and time-dependent fields.
Their microscopic interpretation is the following. 

A strong uniform magnetic field (rotational of
$\boldsymbol{a}$) is responsible
for the IQHE in graphene.%
\cite{McClure56}
A vector field
$\boldsymbol{a}^{\ }_{5}$
encodes changes in the curvature (ripples) of graphene,
~\cite{Morozov06,Morpurgo06} and it can also encode 
defective coordination numbers at apical defects.
~\cite{Gonzalez92,Lammert00,Pachos08}
A constant $\mu^{\ }_{\mathrm{s}}$ realizes
in graphene a staggered chemical potential 
and opens an electronic spectral gap.%
\cite{Semenoff84}
A constant $\eta$ realizes in graphene
a directed next-nearest-neighbor hopping amplitude
without net magnetic flux through the Wigner-Seitz cell
of the honeycomb lattice and it also 
opens an electronic spectral gap.%
\cite{Haldane88}
A constant $\Delta$ realizes
in graphene a Kekul\'e distortion
of the nearest-neighbor hopping amplitude
and, again, opens an electronic spectral gap.%
\cite{Hou07}
The 4 space- and time-independent 
$\mathrm{Re}\,\Delta$,
$\mathrm{Im}\,\Delta$,
$\mu^{\ }_{\mathrm{s}}$,
and
$\eta$
exhaust all possible ways for the opening of
a spectral gap in the single-particle spectrum of
the kinetic Dirac kernel
$\boldsymbol{\alpha}\cdot(-{i}\boldsymbol{\partial})$,
as
$\beta$, $\beta\gamma^{\ }_{5}$, $R$, and
${i}\alpha^{1}\alpha^{2}$
generate the largest set of traceless and Hermitian
$4\times4$ matrices that anticommutes with
$\boldsymbol{\alpha}\cdot(-{i}\boldsymbol{\partial})$.
The 3 masses 
$\mathrm{Re}\,\Delta$,
$\mathrm{Im}\,\Delta$,
and
$\mu^{\ }_{\mathrm{s}}$
are compatible, i.e., they open
the gap $2m$, where
\begin{equation}
m:=
\sqrt{|\Delta|^{2}+\mu^{2}_{\mathrm{s}}},
\label{eq: def m}
\end{equation}
for $\beta$,
$\beta\gamma^{\ }_{5}$,
and $R$ anticommute pairwise.
On the other hand,
the mass $\eta$ competes with the mass $m$,
as ${i}\alpha^{\ }_{1}\alpha^{\ }_{2}$
commutes with 
$\beta$,
$\beta\gamma^{\ }_{5}$,
and $R$ (the competition between $\eta$ and $m$ leads to a phase
transition when $|\eta|=m$, which shall be important
in the discussion of fractional statistics in this paper). 
The fields $\boldsymbol{a}$,
$\boldsymbol{a}^{\ }_{5}$,
$\Delta$,
$\mu^{\ }_{\mathrm{s}}$,
and $\eta$
have also appeared in the context of 
(a) slave-boson 
treatments of antiferromagnetic spin-1/2 Heisenberg model
on the square lattice in the $\pi$-flux phase,%
\cite{Affleck88,Wen89,Fradkin91,Mudry94} 
and (b) Anderson localization for electrons hopping
on a square lattice with a flux of half a magnetic
flux quantum per plaquette,
i.e, the square lattice with $\pi$-flux phase.%
\cite{Ludwig94,Hatsugai97,Guruswamy00}

\subsection{
Symmetries
           }
\label{subsec: Symmetries}

The model defined in 
Eq.~(\ref{eq: def quantum Hamiltonian})
possesses a number of symmetry operations that we list below and utilize
in the paper.

\subsubsection{
Time-reversal symmetry
              }

In the Heisenberg representation,
\begin{equation}
\hat{H}(t)\to
\hat{H}(-t)
\end{equation}
under the anti-unitary transformation
\begin{equation}
\begin{split}
&
\hat{\psi}(\boldsymbol{r},t)\to
\left(
TK
\hat{\psi}
\right)
(\boldsymbol{r},-t),
\\
&
\boldsymbol{a}(\boldsymbol{r},t)\to
-
\boldsymbol{a}(\boldsymbol{r},-t),
\qquad
\eta(\boldsymbol{r},t)\to
-\eta(\boldsymbol{r},-t),
\\
&
\boldsymbol{a}^{\ }_{5}(\boldsymbol{r},t)\to
\boldsymbol{a}^{\ }_{5}(\boldsymbol{r},-t),
\qquad
\theta(\boldsymbol{r},t)\to
\theta(\boldsymbol{r},-t),
\\
&
|\Delta|(\boldsymbol{r},t)\to
|\Delta|(\boldsymbol{r},-t),
\qquad
\mu^{\ }_{\mathrm{s}}(\boldsymbol{r},t)\to
\mu^{\ }_{\mathrm{s}}(\boldsymbol{r},-t),
\end{split}
\label{eq: def time reversal}
\end{equation}
where complex conjugation is represented by $K$ and 
\begin{equation}
T:=
\beta\alpha^{1}\gamma^{\ }_{5}=
\begin{pmatrix}
0
&
\tau^{\ }_{1}
\\
\tau^{\ }_{1}
&
0
\end{pmatrix}\equiv
\sigma^{\ }_{1}
\otimes
\tau^{\ }_{1}=
T^{t}
\label{eq: def mathcal T}
\end{equation}
is a unitary, Hermitian 
(and thus symmetric) matrix.
Transformation (\ref{eq: def time reversal})
realizes reversal of time in graphene, for
$T$ exchanges 
the two valleys while acting trivially 
on the sublattice indices. Moreover,
transformation (\ref{eq: def time reversal})
realizes reversal of time
for an effectively spinless single particle,
for $T$ is symmetric.
Hamiltonian
$\hat{H}$ is time-reversal symmetric and
can be represented by real-valued matrix elements,%
\cite{footnote H is real valued}
if all background fields are static while 
\begin{equation}
a^{\ }_{1}=
a^{\ }_{2}=
\eta=0.
\end{equation}

\subsubsection{
Sublattice symmetry
              }

Always in the Heisenberg representation,
\begin{equation}
\hat{H}(t)\to
-
\hat{H}(t)
\end{equation}
under the unitary transformation
\begin{equation}
\begin{split}
&
\hat{\psi}(\boldsymbol{r},t)\to
\left(
R
\hat{\psi}
\right)
(\boldsymbol{r},t),
\\
&
\boldsymbol{a}(\boldsymbol{r},t)\to
\boldsymbol{a}(\boldsymbol{r},t),
\qquad
\eta(\boldsymbol{r},t)\to
-\eta(\boldsymbol{r},t),
\\
&
\boldsymbol{a}^{\ }_{5}(\boldsymbol{r},t)\to
\boldsymbol{a}^{\ }_{5}(\boldsymbol{r},t),
\qquad
\theta(\boldsymbol{r},t)\to
\theta(\boldsymbol{r},t),
\\
&
|\Delta|(\boldsymbol{r},t)\to
|\Delta|(\boldsymbol{r},t),
\qquad
\mu^{\ }_{\mathrm{s}}(\boldsymbol{r},t)\to
-\mu^{\ }_{\mathrm{s}}(\boldsymbol{r},t),
\end{split}
\label{eq: def SL}
\end{equation}
where 
\begin{equation}
R:=
\alpha^{\ }_{3}=
\begin{pmatrix}
\tau^{\ }_{3}
&
0
\\
0
&
-
\tau^{\ }_{3}
\end{pmatrix}\equiv
\sigma^{\ }_{3}
\otimes
\tau^{\ }_{3}=
R^{t}
\label{eq: def m,athcal R}
\end{equation}
is a diagonal, unitary, and Hermitian matrix.
Transformation (\ref{eq: def SL})
realizes in graphene the change of sign
of the single-particle wave functions
on every sites of the honeycomb lattice
belonging to one and only one triangular sublattice.
The single-particle eigenstates of the conserved Hamiltonian
$\hat{H}$ obey the spectral symmetry (SLS)
by which
any single-particle eigenstate $|\Psi\rangle$ 
with a non-vanishing energy eigenvalue $\varepsilon$ 
has the mirror eigenstate $R|\Psi\rangle$  with 
the non-vanishing energy eigenvalue $-\varepsilon$,
if all background fields are static while
\begin{equation}
\mu^{\ }_{\mathrm{s}}=
\eta=0.
\end{equation}

\subsubsection{
Continuous gauge symmetries
              }

We now turn to the continuous symmetries
obeyed by the Dirac Hamiltonian%
~(\ref{eq: def quantum Hamiltonian})
in the Heisenberg representation.
To this end, we make use of
\begin{equation}
0=
[\gamma^{\ }_{5},\boldsymbol{\alpha}]=
\{\gamma^{\ }_{5},\beta\}=
[\gamma^{\ }_{5},R].
\label{eq: gamma5 either commutes or anticommutes}
\end{equation}
The commutators and anticommutator%
~(\ref{eq: gamma5 either commutes or anticommutes})
imply that
\begin{equation}
\hat{H}(t)\to
\hat{H}(t)
\end{equation}
under the U(1)$\otimes$U(1) 
local gauge transformation
\begin{equation}
\begin{split}
&
\hat{\psi}\to
e^{{i}\left(\phi+\phi^{\ }_{5}\gamma^{\ }_{5}\right)}
\hat{\psi},
\qquad
\boldsymbol{a}\to
\boldsymbol{a}
-
\boldsymbol{\partial}\phi,
\\
&
\boldsymbol{a}^{\ }_{5}\to
\boldsymbol{a}^{\ }_{5}
-
\boldsymbol{\partial}\phi^{\ }_{5},
\qquad
\theta\to
\theta
-
2
\phi^{\ }_{5},
\\
&
|\Delta|\to
|\Delta|,
\qquad
\mu^{\ }_{\mathrm{s}}\to
\mu^{\ }_{\mathrm{s}},
\qquad
\eta\to
\eta,
\end{split}
\label{eq: local gauge symmetries}
\end{equation}
generated by the two space- and time-dependent
real-valued smooth 
functions $\phi$ and $\phi^{\ }_{5}$.
The microscopic origin of the global U(1)
gauge symmetry generated by $\phi$
is conservation of the electron number in graphene. 
For planar graphene,
the continuous global axial U(1) 
gauge symmetry generated by $\phi^{\ }_{5}$
is broken as soon as the curvature of the tight-binding
dispersion is accounted for
so that the Dirac points
are not anymore decoupled.
We shall nevertheless impose the
local axial U(1) gauge symmetry
at the level of the approximation captured
by the Dirac Hamiltonian%
~(\ref{eq: def quantum Hamiltonian})
and see through its consequences in this paper. 
(We do provide a microscopic example of a lattice model that realizes
the local axial U(1) gauge symmetry in Sec.~\ref{sec: Microscopic models}.)

\section{
Path integral formulation of the model with 4 masses
        }
\label{sec: Path integral formulation of the model}

For our purposes, it will be more convenient to
trade the operator formalism for an
effective partition function defined by
integrating over the Dirac fermions in the background 
of the gauge fields 
$\boldsymbol{a}$
and
$\boldsymbol{a}^{\ }_{5}$
and of the scalar fields
$\Delta$, $\mu^{\ }_{\mathrm{s}}$, and $\eta$.
We will demand that this effective theory 
captures the U(1)$\otimes$U(1) 
local gauge symmetry%
~(\ref{eq: local gauge symmetries}).
This is possible in odd-dimensional space and time,%
\cite{Fujikawa04}
for the Grassmann measure can be 
regularized without breaking the
U(1)$\otimes$U(1)
local gauge symmetry of the Lagrangian. 
Of course, maintaining the
U(1)$\otimes$U(1)
local gauge symmetry
can only be achieved if the phase 
$\theta=-\mathrm{arg}\,\Delta$ 
of the Kekul\'e background field
$\Delta$ is also included as a dynamical field.
For simplicity but without loss of generality
as far as the computation of the charge quantum number
and statistical phase are concerned, 
the masses $m$ and $\eta$ will be
taken to be space- and time-independent 
parameters, while $\Delta$ and $\mu^{\ }_{\mathrm{s}}$
vary in space and time (with
$m=\sqrt{|\Delta|^2+\mu^{2}_{\mathrm{s}}}$ constant)
through $\theta\equiv -\mathrm{arg}\,\Delta$ 
and $\cos\alpha\equiv\mu^{\ }_{\mathrm{s}}/m$. (For simplicity, 
we shall also focus on the case where $\mu^{\ }_{\mathrm{s}}$ is also 
constant in space and time, with the exception of near the vortex core, 
where $\Delta\to 0$, so $\mu^{\ }_{\mathrm{s}}$ has to adjust
as to keep $m$ constant.)

Thus, we seek the effective field theory defined
by the Grassmann path integral
\begin{subequations}
\label{eq: def Z}
\begin{equation}
\begin{split}
&
Z^{\ }_{m,\eta}[a^{\ }_{\mu},a^{\ }_{5\mu},\theta,\alpha]
:=
\int\mathcal{D}[\bar\psi,\psi]
\exp
\left(
{i}
\int d^3x\,
\mathcal{L}^{\ }_{m,\eta}
\right),
\\
&
\mathcal{L}^{\ }_{m,\eta}:=
\bar{\psi}
\left(
\gamma^{\mu}{i}\partial^{\ }_{\mu}
-
\gamma^{\mu} 
a^{\ }_{\mu}
-
\gamma^{\mu} 
\gamma^{\  }_{5} 
a^{\ }_{5\mu}
-
M^{\ }_{m,\eta}
\right)
\psi,
\end{split}
\end{equation}
where we have also included the time-components 
$a^{\ }_{0}$ (TRS but SLS breaking) and ${a_{5}}_{0}$
(SLS but TRS breaking) of the 
U(1)$\otimes$U(1)
gauge fields to 
maintain space and time covariance.
The independent Grassmann-valued fields
over which the path integral is performed are
the 4-components spinors
$\bar\psi$
and
$\psi$.
They depend on the contravariant 3-vectors
$x^{\mu}=(t,\boldsymbol{r})$ 
[covariant 3-vectors
$x^{\ }_{\mu}=(t,-\boldsymbol{r})$]
and we will use the repeated summation convention
$x^{\mu}y_{\mu}=
 x^{0}y^{0}
-x^{1}y^{1}
-x^{2}y^{2}$.
We have defined the four gamma matrices
\begin{equation}
\gamma^{0}:=
\beta,
\quad
\gamma^{1}:=
\beta
\alpha^{1},
\quad
\gamma^{2}:=
\beta
\alpha^{2}
\quad
\gamma^{3}:=
\beta
\alpha^{3},
\end{equation}
for which lowering and raising of the 
greek indices $\mu,\nu=0,1,2$
is achieved with the Lorentz metric 
$g^{\ }_{\mu\nu}=\mathrm{diag}(1,-1,-1)$.
The 4 matrices 
$\gamma^{0}$,
$\gamma^{1}$,
$\gamma^{2}$,
and
$\gamma^{3}$
obey the usual Clifford algebra
in Minkowsky space 
in the chiral representation,
i.e.,
$\gamma^{\ }_{5}={i}
\gamma^{0}\gamma^{1}
\gamma^{2}\gamma^{3}$
is diagonal.
We have also defined the matrix
\begin{equation}
\begin{split}
&
M^{\ }_{m,\eta}:=
m
\left(
n^{\ }_{1}M^{\ }_{1}
+
n^{\ }_{2}M^{\ }_{2}
+
n^{\ }_{3}M^{\ }_{3}
\right)
+
\eta\gamma^{\ }_{5}\gamma^{3},
\\
&
M^{\ }_{1}:=
1,
\qquad
M^{\ }_{2}:=
-{i}\gamma^{\ }_{5},
\qquad
M^{\ }_{3}:=
\beta
\alpha^{3}\equiv
\gamma^{3},
\end{split}
\end{equation}
for which we do not distinguish upper and lower
latin indices $\mathrm{a},\mathrm{b}=1,2,3$ 
as they are contracted
with the Euclidean metric
$\delta^{\ }_{\mathrm{a}\mathrm{b}}=
\mathrm{diag}(1,1,1)$.
(Notice that because space and time is (2+1) dimensional,
we can use the gamma matrix
$\gamma^{3}$
to open a spectral gap by taking $M^{\ }_{3}=\gamma^{3}$.)
The space and time dependencies 
in $M^{\ }_{m,\eta}$ follow 
entirely from those of the phase 
$\mathrm{arg}\,\Delta$. Indeed,
while the masses $\eta$ and $m$
are constant in space and time,
the direction of the unit vector 
$\boldsymbol{n}$ 
with the 3 components
\begin{equation}
n^{\ }_{1}:=
\frac{|\Delta|\cos\theta}{m},
\quad
n^{\ }_{2}:=
-\frac{|\Delta|\sin\theta}{m},
\quad
n^{\ }_{3}:=
\frac{\mu^{\ }_{\mathrm{s}}}{m}
\label{eq: para unit vector}
\end{equation}
\end{subequations}
can vary in space and time.

The U(1)$\otimes$U(1)
local gauge symmetry~(\ref{eq: local gauge symmetries})
has become the invariance of the 
Lagrangian in Eq.~(\ref{eq: def Z})
under the 
U(1)$\otimes$U(1)
local gauge transformation
\begin{equation}
\begin{split}
&
\bar\psi\to
\bar\psi\,
e^{-{i}(\phi-\phi^{\ }_{5}\gamma^{\ }_{5})},
\qquad
\psi\to
e^{{i}(\phi+\phi^{\ }_{5}\gamma^{\ }_{5})}\,
\psi,
\\
&
a^{\ }_{ \mu}\to
a^{\ }_{ \mu}
-
\partial^{\ }_{\mu}\phi,
\quad
a^{\ }_{5\mu}\to
a^{\ }_{5\mu}
-
\partial^{\ }_{\mu}\phi^{\ }_{5},
\\&
\theta\to
\theta
-
2\phi^{\ }_{5}.
\end{split}
\label{eq: local gauge symmetries bis}
\end{equation}
In spite of appearances
[$\bar\psi\psi\to
\bar\psi\exp(2{i}\phi^{\ }_{5}\gamma^{\ }_{5})\,\psi$],
the Grassmann Jacobian induced
by the U(1)$\otimes$U(1)
local gauge transformation~(\ref{eq: local gauge symmetries})
is unity and does not produce a quantum anomaly  
in (2+1) dimensions (odd space-time dimension).%
~\cite{Fujikawa04}

We take advantage of the fact that
$\bar\psi$ and $\psi$ are independent Grassmann 
integration variables to bring the algebra obeyed 
by the 6 matrices
$\gamma^{\mu}$ $\mu=0,1,2$
and
$M^{\ }_{\mathrm{a}}$ $\mathrm{a}=1,2,3$
to a form that will simplify greatly
the evaluation of the partition function%
~(\ref{eq: def Z}). Under the non-unitary
change of integration variable
\begin{equation}
\bar\psi=:
\bar\chi\gamma^{\ }_{5}\gamma^{3},
\qquad
\psi=:\chi,
\end{equation}
the partition function~(\ref{eq: def Z})
becomes
\begin{subequations}
\label{eq: def Z bis}
\begin{equation}
\begin{split}
&
Z^{\ }_{m,\eta}
\left[
B^{\ }_{\mu},
n^{\ }_{\mathrm{a}}
\right]=
\int\mathcal{D}[\bar\chi,\chi]
\exp
\left(
{i}
\int d^3x\,
\mathcal{L}^{\ }_{m,\eta}
\right),
\\
&
\mathcal{L}^{\ }_{m,\eta}=
\bar{\chi}
\left(
\Gamma^{\mu}{i}\partial^{\ }_{\mu}
+
\Gamma^{\mu} 
B^{\ }_{\mu}
-
m\,
n^{\ }_{\mathrm{a}}
\Sigma^{\ }_{\mathrm{a}}
-
\eta
\right)
\chi,
\end{split}
\end{equation}
where the matrices
\begin{equation}
\Gamma^{\mu}:=
\gamma^{\ }_{5}
\gamma^{3}
\gamma^{\mu},
\qquad
\Sigma^{\ }_{\mathrm{a}}:=
\gamma^{\ }_{5}
\gamma^{3}
M^{\ }_{\mathrm{a}},
\end{equation}
obey 
\begin{equation}
\{\Gamma^{\mu},\Gamma^{\nu}\}=
2g^{\mu\nu},
\quad
[\Sigma^{\ }_{\mathrm{a}},\Sigma^{\ }_{\mathrm{b}}]=
{i}\epsilon^{\ }_{\mathrm{abc}}\Sigma^{\ }_{\mathrm{c}},
\quad
[\Gamma^{\mu},\Sigma^{\ }_{\mathrm{a}}]=0,
\end{equation}
for $\mu,\nu=0,1,2$ 
and $\mathrm{a},\mathrm{b},\mathrm{c}=1,2,3$ 
and we have regrouped the gauge fields into
\begin{equation}
B^{\ }_{\mu}\equiv
b^{0 }_{\mu}
+
b^{\mathrm{a}}_{\mu}\Sigma^{\mathrm{a}}
\label{eq: from B's to a's and a5's d}
\end{equation}
following the prescription
\begin{equation}
\begin{split}
&
b^{0 }_{\mu}:=
-
a^{\ }_{\mu},
\qquad
b^{1}_{\mu}:=b^{2}_{\mu}:=0,
\qquad
b^{3 }_{\mu}:=
+
a^{\ }_{5\mu}.
\label{eq: from B's to a's and a5's e}
\end{split}
\end{equation}
\end{subequations}
Notice that
\begin{equation}
\Sigma^{\ }_{3}=
-\gamma^{\ }_{5}
\end{equation}
so that the symmetry under the 
U(1)$\otimes$U(1)
local gauge transformation~(\ref{eq: local gauge symmetries bis})
has become the invariance of the
Lagrangian in Eq.~(\ref{eq: def Z bis})
under
\begin{equation}
\begin{split}
&
\bar\chi\to
\bar\chi\,
e^{-{i}(\phi-\phi^{\ }_{5}\Sigma^{\ }_{3})},
\qquad
\chi\to
e^{+{i}(\phi-\phi^{\ }_{5}\Sigma^{\ }_{3})}
\chi,
\\
&
b^{0}_{\mu}\to
b^{0}_{\mu}
+
\partial^{\ }_{\mu}\phi,
\qquad
b^{3}_{\mu}\to
b^{3}_{\mu}
+
\partial^{\ }_{\mu}\phi^{\ }_{5},
\\
&
b^{1}_{\mu}\to
b^{1}_{\mu},
\qquad
b^{2}_{\mu}\to
b^{2}_{\mu},
\qquad
\theta\to
\theta
-2\phi^{\ }_{5}.
\end{split}
\label{eq: local gauge symmetries bis bis}
\end{equation}

\subsection{
Hidden U(2) non-Abelian structure
           }
\label{subsec: Hidden U(2) non-Abelian structure}

To make the U(2) non-Abelian structure explicit,
observe first that the mass
$
mn^{\ }_{\mathrm{a}}\Sigma^{\ }_{\mathrm{a}}
$
is an element of an su(2) Lie algebra.
Indeed, there exists a $4\times 4$ matrix
$U$ representing an element of SU(2)
generated by $\Sigma^{\ }_{\mathrm{a}}$
$\mathrm{a}=1,2,3$ such that
\begin{equation}
m\,n^{\ }_{\mathrm{a}}\Sigma^{\ }_{\mathrm{a}}=
m\,U\Sigma^{\ }_{3} U^{\dag}.
\end{equation}
We then infer that the partition functions
(\ref{eq: def Z})
or, equivalently,
(\ref{eq: def Z bis})
are special cases of the more general partition
function
\begin{subequations}
\label{eq: def Z bis bis}
\begin{equation}
\begin{split}
&
Z:=
\int\mathcal{D}[\bar\chi,\chi]
\exp
\left(
{i}
\int d^3x\,
\mathcal{L}^{\ }_{m,\eta}
\right),
\\
&
\mathcal{L}^{\ }_{m,\eta}:=
\bar{\chi}
\left(
\Gamma^{\mu}{i}\partial^{\ }_{\mu}
+
\Gamma^{\mu} 
B^{\ }_{\mu}
-
m\,
U
\Sigma^{\ }_{3}
U^{\dag}
-
\eta
\right)
\chi,
\end{split}
\end{equation}
where 
\begin{equation}
B^{\ }_{\mu}(x)=
b^{0}_{\mu}(x)
+
b^{\mathrm{a}}_{\mu}(x)
\Sigma^{\mathrm{a}},
\qquad
\mu=0,1,2,
\end{equation}
are arbitrary elements of the Lie algebra 
$\mathrm{u}(2)=\mathrm{u}(1)\oplus\mathrm{su}(2)$
and
\begin{equation}
U(x)=
e^{
{i}
u^{\ }_{\mathrm{0}}(x)
  }
e^{
{i}
u^{\ }_{\mathrm{a}}(x)
\Sigma^{\ }_{\mathrm{a}}
  },
\qquad
u^{\ }_{\mathrm{0}}(x),
u^{\ }_{\mathrm{a}}(x)\in\mathbb{R},
\end{equation} 
\end{subequations}
is an arbitrary element of U(2).
As the mapping between the unit vector
$\boldsymbol{n}(x)$ and $U(x)$
is one to many, the Lagrangian 
and the Grassmann measure in
Eq.~(\ref{eq: def Z bis bis})
are both invariant under the local U(2) gauge transformation
\begin{subequations}
\label{eq: symmetries bis bis bis}
\begin{equation}
\begin{split}
&
\bar\chi\to
\bar\chi\,V^{\dag},
\qquad
\chi\to
V\chi,
\\
&
B^{\ }_{\mu}\to
V
B^{\ }_{\mu}
V^{\dag}
-
{i}
V^{\dag}
\partial^{\ }_{\mu}
V,
\\
&
U\to VU,
\end{split}
\end{equation}
parametrized by the smooth space- and time-dependent
\begin{equation}
V(x):=
e^{
{i}
\left[
v^{\ }_{0}(x)
+
v^{\ }_{\mathrm{a}}(x)
\Sigma^{\ }_{\mathrm{a}}
\right]
  }
\in\mathrm{U}(2),
\end{equation}
and under the global 
U(1)$\times$U(1) transformation
\begin{equation}
U\to
U\, W,
\qquad
W:=
e^{{i}\phi^{\ }_{0}}
e^{{i}\phi^{\ }_{3}\Sigma^{\ }_{3}},
\end{equation}
\end{subequations}
parametrized by the real-valued \textit{numbers}
$\phi^{\ }_{0}$ and $\phi^{\ }_{3}$.

The transformation~(\ref{eq: local gauge symmetries bis})
or, equivalently,
(\ref{eq: local gauge symmetries bis bis}) 
is represented by the transformation%
~(\ref{eq: symmetries bis bis bis})
with $B^{\ }_{\mu}$
given in Eqs.~(\ref{eq: from B's to a's and a5's d})
and (\ref{eq: from B's to a's and a5's e})
and $U$ given by
\begin{equation}
U=
e^{+{i}\theta\Sigma^{\ }_{3}/2}
e^{-{i}\alpha\Sigma^{\ }_{2}/2}
e^{-{i}\theta\Sigma^{\ }_{3}/2}
\label{eq: parametrization U}
\end{equation}
whereby the unit vector~(\ref{eq: para unit vector})
is parametrized by
\begin{equation}
\boldsymbol{n}=
\left(
 \sin\alpha\cos\theta,
-\sin\alpha\sin\theta,
 \cos\alpha
\right)^{t}.
\end{equation}
(Recall that 
$\cos\alpha:=\mu^{\ }_{\mathrm{s}}/m$, 
$\sin\alpha:=|\Delta|/m$, 
and that the phase 
$\theta=-\mathrm{arg}\,\Delta$ 
is space and time dependent.)

A gradient expansion for the partition function%
~(\ref{eq: def Z bis bis})
with an arbitrary space and time dependent 
$U\in\mathrm{SU}(2)$ but with 
$B^{\ }_{\mu}=0$
and 
$\eta=0$
was performed by Jaroszewicz
and shown to produce the effective action
for the O(3) non-linear-sigma model (NLSM)
modified by a Hopf term.%
\cite{Jaroszewicz84,Chen89,Hlousek90,Yakovenko90,Abanov00} 
This Hopf term was shown by Chen and Wilczek to vanish
as soon as the TRS-breaking mass $\eta$ is larger 
in magnitude than the TRS mass $m$. 
Chen and Wilczek also
showed that an Abelian Chern-Simons term for a non-vanishing
$b^{0}_{\mu}\equiv a^{\ }_{\mu}$ is present if and only if
$|\eta|>m$.

Hopf or Chern-Simons terms can cause the fractionalization
of quantum numbers. Although charge fractionalization can here
also be deduced from the presence of midgap
single-particle states 
of the Dirac Hamiltonian%
~(\ref{eq: def quantum Hamiltonian})
in static backgrounds,%
~\cite{Hou07,Jackiw07,Chamon08a,Chamon08b,Jaroszewicz84} 
it is natural to explore
the emergence of fractional statistics 
under the exchange of point-like quasiparticles by exploring
the fully dynamical theory encoded by the partition function%
~(\ref{eq: def Z bis}).
To this end,
it is essential to preserve all symmetries as we did
up to now. The point-like quasiparticle 
whose braiding statistics
we shall derive are vortices~\cite{Hou07} in the dynamical phase
$\theta=-\mathrm{arg}\,\Delta$, including the case when
they are accompanied by axial gauge half fluxes in 
$a^{\ }_{5\mu}$ that screen the interactions between vortices.%
~\cite{Jackiw07}

\section{
Derivative expansion and the effective action
        }
\label{sec: Derivative expansion}

It is known
that the Dirac Hamiltonian%
~(\ref{eq: def quantum Hamiltonian})
with static backgrounds can support zero modes.%
~\cite{Hou07,Jackiw07,Chamon08a,Chamon08b,Jaroszewicz84}
This can be of a nuisance when computing
a fermion determinant. However, it is possible
to elegantly dispose of this difficulty with
the help of the observation made by Jaroszewicz
that a non-singular U(2) gauge transformation
on the Dirac Kernel in the partition function%
~(\ref{eq: def Z bis bis})
can turn a single-particle midgap state into 
a single-particle threshold state 
without changing the spectral asymmetry.%
~\cite{Jaroszewicz84,Jaroszewicz86}
This is achieved by redefining the 
Grassmann integration variables
in the partition function%
~(\ref{eq: def Z bis bis})
according to
\begin{equation}
\bar\chi=:
\bar\chi'U^{\dag},
\qquad
\chi=:
U\chi'.
\label{eq: U(2) trsf on chi's}
\end{equation}
The partition function%
~(\ref{eq: def Z bis bis})
becomes
\begin{subequations}
\label{eq: def Z bis bis bis}
\begin{equation}
\begin{split}
&
Z^{\prime}_{m,\eta}[B^{\prime}_{\mu}]:=
\int\mathcal{D}[\bar\chi',\chi']
\exp
\left(
{i}
\int d^3x\,
\mathcal{L}^{\prime}_{m,\eta}
\right),
\\
&
\mathcal{L}^{\prime}_{m,\eta}:=
\bar{\chi}'
\left(
\Gamma^{\mu}{i}\partial^{\ }_{\mu}
+
\Gamma^{\mu} 
B^{\prime}_{\mu}
-
m\,
\Sigma^{\ }_{3}
-
\eta
\right)
\chi',
\end{split}
\end{equation}
where 
\begin{equation}
B^{\prime}_{\mu}=
U^{\dag}
B^{\     }_{\mu}
U
+
U^{\dag}
{i}\partial^{\ }_{\mu}
U
\label{eq: def B'mu}
\end{equation}
\end{subequations}
need not be a pure gauge because of the term
$U^{\dag}B^{\     }_{\mu}U$.

The symmetries~(\ref{eq: symmetries bis bis bis})
of the Lagrangian 
and the Grassmann measure in
Eq.~(\ref{eq: def Z bis bis})
become the invariance of the Lagrangian 
and the Grassmann measure in
Eq.~(\ref{eq: def Z bis bis bis})
under the local U(2) gauge symmetry
\begin{subequations}
\label{eq: symmetries bis bis bis bis}
\begin{equation}
\begin{split}
&
\bar\chi'\to
\bar\chi',
\qquad
\chi'\to
\chi',
\\
&
B^{\ }_{\mu}\to
V
B^{\ }_{\mu}
V^{\dag}
-
{i}
V^{\dag}
\partial^{\ }_{\mu}
V,
\\
&
U\to VU,
\end{split}
\label{eq: symmetries bis bis bis bis a}
\end{equation}
parametrized by the space- and time-dependent
$
V(x)
\in\mathrm{U}(2)
$
and under the global 
U(1)$\times$U(1) gauge symmetry
\begin{equation}
\begin{split}
&
\bar\chi'\to
\bar\chi'\,W,
\qquad
\chi'\to
W^{\dag}\chi',
\qquad
U\to
U\, 
W,
\end{split}
\label{eq: U(1) U(1) b}
\end{equation}
\label{eq: symmetries bis bis bis bis b}
\end{subequations}

\noindent
parametrized by 
the space and time independent
$W:=
\exp({i}\phi^{\ }_{0})
\exp({i}\phi^{\ }_{3}\Sigma^{\ }_{3})$.
Notice that
\begin{equation}
B^{\prime}_{\mu}\to
W^{\dag}\,B^{\prime}_{\mu}\,W
\end{equation}
under the transformation%
~(\ref{eq: symmetries bis bis bis bis}).

Evidently, the transformed Dirac fermions are local
U(2) gauge singlets. Thus, by dressing the original
Dirac fermions into local U(2) gauge singlets,
any midgap single-particle states from the original
static Dirac Hamiltonian has migrated to
the threshold of the continuum part of the
transformed single-particle spectrum, provided
the single-particle spectral gap has not closed,
i.e., $m\neq|\eta|$
in the parameter space $(m,\eta)\in\mathbb{R}^{2}$ of 
Fig.~\ref{fig:phase-diagram}. 
This dressing is achieved
without changing the spectral asymmetry
in any region of Fig.~\ref{fig:phase-diagram}
in which the single-particle gap remains open, 
for the U(2) gauge transformation is not singular.

The parametrization
\begin{equation}
\begin{split}
&
b^{\prime 0}_{\mu}=
-
a^{\ }_{\mu},
\\
&
b^{\prime 1}_{\mu}=
-
\sin \alpha
\cos \theta
\left(
a^{\ }_{5 \mu}
-
\frac{1}{2}
\partial^{\ }_{\mu}\theta 
\right),
\\
&
b^{\prime 2}_{\mu}=  
+
\sin \alpha 
\sin \theta
\left(
a^{\ }_{5 \mu}
-
\frac{1}{2}
\partial^{\ }_{\mu}\theta 
\right),
\\
&
b^{\prime 3}_{\mu}=
+
\left(
\cos \alpha\, 
a^{\ }_{5 \mu}
+
\frac{
1
-
\cos\alpha
     }
     {
2
     }
\partial^{\ }_{\mu} \theta
\right),
\end{split}
\label{eq: paramertization B'mu}
\end{equation}
of 
$B^{\prime}_{\mu}=
b^{\prime\mathrm{0}}_{\mu}
+
b^{\prime\mathrm{a}}_{\mu}\Sigma^{\mathrm{a}}$
where $\mu=0,1,2$
follows from inserting 
Eqs.~(\ref{eq: from B's to a's and a5's d}),
(\ref{eq: from B's to a's and a5's e}),
and
(\ref{eq: parametrization U})
into Eq.~(\ref{eq: def B'mu}).
The transformation law of
Eq.~(\ref{eq: paramertization B'mu})
under the local U(1)$\otimes$U(1)
gauge transformation~(\ref{eq: local gauge symmetries bis})
is
\begin{equation}
\begin{split}
&
b^{\prime 0}_{\mu}\to
b^{\prime 0}_{\mu}
+
\partial^{\ }_{\mu}\phi,
\\
&
b^{\prime 1}_{\mu}\to
\cos(2\phi^{\ }_{5})\,
b^{\prime 1}_{\mu}
-
\sin(2\phi^{\ }_{5})\,
b^{\prime 2}_{\mu},
\\
&
b^{\prime 2}_{\mu}\to
\sin(2\phi^{\ }_{5})\,
b^{\prime 1}_{\mu}
+
\cos(2\phi^{\ }_{5})\,
b^{\prime 2}_{\mu},
\\
&
b^{\prime 3}_{\mu}\to
b^{\prime 3}_{\mu}
-
\partial^{\ }_{\mu}\phi^{\ }_{5}.
\end{split}
\label{eq: gauge trs law parametrization b'mu}
\end{equation}

At this stage, it is convenient to define the effective 
action (Lagrangian) 
\begin{equation}
S^{\mathrm{eff}}_{m,\eta}[B^{\prime}_{\mu}]\equiv
\int d^{3}x\,
\mathcal{L}^{\mathrm{eff}}_{m,\eta}:=
-{i}\ln Z^{\prime}_{m,\eta}[B^{\prime}_{\mu}]
\end{equation}
in the background field $B^{\prime}_{\mu}$
given by 
Eq.~(\ref{eq: gauge trs law parametrization b'mu}).
This effective action is constrained by the 
gauge symmetries in the following way.

Any transformation of the Grassmann integration variables 
$\bar\chi'$ and $\chi'$ with unity for the Jacobian
leaves the numerical value of the partition function%
~(\ref{eq: def Z bis bis bis})
unchanged. As the Grassmann measure
in the partition function%
~(\ref{eq: def Z bis bis bis}) 
is invariant under the local 
U(1)$\otimes$U(1) 
transformation
\begin{equation}
\begin{split}
&
\bar\chi'\to
\bar\chi'\,
V^{\dag},
\quad
\chi'\to
V\,
\chi',
\quad
V:=
e^{+{i}(\phi-\phi^{\ }_{5}\Sigma^{\ }_{3})},
\end{split}
\label{eq: dressing chi'}
\end{equation}
it follows that
\begin{equation}
Z^{\prime}_{m,\eta}[B^{\prime}_{\mu}]=
Z^{\prime}_{m,\eta}[
V^{\dag}B^{\prime}_{\mu}V-V^{\dag}{i}\partial^{\ }_{\mu}V].
\end{equation}
The partition function%
~(\ref{eq: def Z bis bis bis})
thus takes the form
\begin{subequations}
\label{eq: L eff if m neq 0}
\begin{equation}
Z^{\prime}_{m,\eta}[B^{\prime}_{\mu}]=
\exp
\left(
{i}
\int d^{3}x\,
\mathcal{L}^{\mathrm{eff}}_{m,\eta}
\right)
\end{equation}
where
\begin{equation}
\begin{split}
\mathcal{L}^{\mathrm{eff}}_{m,\eta}=&\,
\hphantom{+}
C^{(0)}_{11}
\left(
b^{\prime 1 \rho} 
b^{\prime 1}_{\rho}
+
b^{\prime 2 \rho} 
b^{\prime 2}_{\rho}
\right)
\\
&\,
+
C^{(1)}_{00}
\epsilon^{\nu\rho\kappa} 
b^{\prime 0}_{\nu} 
\partial^{\ }_{\rho} 
b^{\prime 0}_{\kappa}
+
C^{(1)}_{33}
\epsilon^{\nu\rho\kappa} 
b^{\prime 3}_{\nu}  
\partial^{\ }_{\rho} 
b^{\prime 3}_{\kappa}
\\
&\,
+
C^{(1)}_{11}
\epsilon^{\nu\rho\kappa}
\left(
b^{\prime 1}_{\nu} 
\partial^{\ }_{\rho} 
b^{\prime 1}_{\kappa}
+
b^{\prime 2}_{\nu} 
\partial^{\ }_{\rho} 
b^{\prime 2}_{\kappa}
-
2 \epsilon^{\mathrm{a}\mathrm{b}3}
b^{\prime \mathrm{a}}_{\nu} 
b^{\prime \mathrm{b}}_{\rho} 
b^{\prime 3}_{\kappa}
\right)
\\
&\,
+
C^{(1)}_{03}
\epsilon^{\nu\rho\kappa} 
b^{\prime 0}_{\nu}
\partial^{\ }_{\rho}
b^{\prime 3}_{\kappa}
+\ldots,
\end{split}
\label{eq: L eff m not 0}
\end{equation}
\end{subequations}
up to first order in a derivative expansion.
This Lagrangian changes by the usual Abelian Chern-Simons 
boundary terms under the gauge transformation%
~(\ref{eq: gauge trs law parametrization b'mu}).
The real-valued coefficients
$C^{(0)}_{11}$,
$C^{(1)}_{00}$,
$C^{(1)}_{33}$,
$C^{(1)}_{11}$,
and
$C^{(1)}_{03}$
are functions of the parameters $m\in\mathbb{R}$
and $\eta\in\mathbb{R}$ with $m\neq|\eta|$.
A tedious calculation, summarized in Appendix%
~\ref{appsec: coefficients},
yields the values shown in Table~\ref{tab:Cs}.


\begin{table*}
\caption{
\label{tab:Cs} 
Coefficients for the effective action in
Eq.~(\ref{eq: L eff if m neq 0}). The calculation leading to these
values is presented in Appendix~\ref{appsec: coefficients}.
      \\}
\begin{ruledtabular}
\begin{tabular}{ccccc}
&$C^{(0)}_{11}$&
$C^{(1)}_{00} =
C^{(1)}_{33}$&
$C^{(1)}_{11}$&
$C^{(1)}_{03}$\\ 
&&&\\
$|\eta|<m$&
$\frac{3m^2-\eta^2}{6\pi m}$
&
$0$
&
$\frac{\eta}{6\pi m}$
&
$\frac{1}{2\pi}\;\mathrm{sgn}\,\mu^{\ }_{\mathrm{s}}$
\\
\\
$m<|\eta|$&
$\frac{m^2}{3\pi |\eta|}$
&
$\frac{1}{4\pi}\;{\rm sgn}\,\eta$
&
$\frac{3\eta^2-m^2}{12\pi \eta^2}\;{\rm sgn}\,\eta$
&
$0$
\\
\\
\end{tabular}
\end{ruledtabular}
\end{table*}

Observe that the coefficients $C^{(1)}_{00}$, $C^{(1)}_{33}$, and
$C^{(1)}_{03}$ that multiply the terms fixed by the local
U(1)$\otimes$U(1) gauge invariance in the effective
Lagrangian~(\ref{eq: L eff m not 0}) can only take a discrete set of
values, while the coefficients $C^{(0)}_{11}$ and $C^{(1)}_{11}$ that
multiply the terms fixed by the global U(1) gauge
invariance can vary continuously with $m$ and $\eta$.

The case $m=0$ when TRS is maximally broken
is special as the symmetry-breaking term
$m\Sigma^{\ }_{3}$ drops out from the Lagrangian in
Eq.~(\ref{eq: def Z bis bis bis}). The matrix $V$
in the change of Grassmann variables%
~(\ref{eq: dressing chi'})
is then not restricted to the Abelian subgroup
U(1)$\otimes$ U(1) of 
U(1)$\otimes$SU(2)
but can be arbitrarily chosen in U(2).
Consequently, 
$C^{(1)}_{33}=C^{(1)}_{11}$ in this limit, which is consistent with the 
values in Table~\ref{tab:Cs}. These (equal) coefficients then 
multiply an SU(2) non-Abelian Chern-Simons term when $m=0$, and hence
must be quantized~\cite{Deser82}, i.e.,
\begin{equation}
\begin{split}
\mathcal{L}^{\mathrm{eff}}_{m=0,\eta}=&\,
\hphantom{+}
\frac{\mathrm{sgn}\,\eta}{4\pi}
\epsilon^{\nu\rho\kappa}
\left(
\delta^{\mathrm{a}\mathrm{b}}
b^{\prime \mathrm{a}}_{\nu} 
\partial^{\ }_{\rho} 
b^{\prime \mathrm{b}}_{\kappa}
-
\frac{2}{3} 
\epsilon^{\mathrm{a}\mathrm{b}\mathrm{c}}
b^{\prime \mathrm{a}}_{\nu} 
b^{\prime \mathrm{b}}_{\rho} 
b^{\prime \mathrm{c}}_{\kappa}
\right)
\\
&\,
+
\frac{\mathrm{sgn}\,\eta}{4\pi}
\epsilon^{\nu\rho\kappa} 
b^{\prime 0}_{\nu} 
\partial^{\ }_{\rho} 
b^{\prime 0}_{\kappa}
+\ldots
\end{split}
\label{eq: L eff if m = 0 final}
\end{equation}
where the second line on the right-hand side is nothing
but the level 1 SU(2) Chern-Simons term.

In the case $\eta=0$ when TRS holds 
Eq.~(\ref{eq: L eff m not 0}) simplifies to
\begin{equation}
\begin{split}
\mathcal{L}^{\mathrm{eff}}_{m,\eta=0}=&\,
\hphantom{+}
\frac{m}{2\pi}
\left(
b^{\prime 1 \rho} 
b^{\prime 1}_{\rho}
+
b^{\prime 2 \rho} 
b^{\prime 2}_{\rho}
\right)
\\
&\,
+
\frac{\mathrm{sgn}\,\mu^{\ }_{\mathrm{s}}}{2\pi}\;
\epsilon^{\nu\rho\kappa} 
b^{\prime 0}_{\nu}
\partial^{\ }_{\rho}
b^{\prime 3}_{\kappa}
+\ldots.
\end{split}
\label{eq: L eff if eta = 0 final}
\end{equation}
Notice that the second line is a double Chern-Simons term on the
fields $b^{\prime 0}$ and $b^{\prime 3}$ which is also called a
BF Chern-Simons theory.%
~\cite{Blau91,Hansson04}

We close this section with
the main intermediary step of this paper
from which the fractionalization of the 
fermion charge and statistical phase follows.
Insertion of
Eq.~(\ref{eq: paramertization B'mu})
into
Eq.~(\ref{eq: L eff if m neq 0})
gives the effective action
\begin{widetext}
\begin{equation}
\begin{split}
\mathcal{L}^{\mathrm{eff}}_{m,\eta}=&\,
\hphantom{+}
C^{(0)}_{11}
\sin^2 \alpha
\left(
a^{\rho}_{5}
-
\frac{1}{2}
\partial^{\rho}\theta 
\right)
\left(
a^{\ }_{5\rho}
-
\frac{1}{2}
\partial^{\ }_{\rho}\theta 
\right)
\\
&\,
+
C^{(1)}_{00}
\epsilon^{\nu\rho\kappa} 
a^{\ }_{\nu} 
\partial^{\ }_{\rho} 
a^{\ }_{\kappa}
+
C^{(1)}_{33}   
\epsilon^{\nu\rho\kappa} 
\left(
\cos\alpha\,
a^{\ }_{5\nu}
+
\frac{1-\cos\alpha}{2}
\partial^{\ }_{\nu} 
\theta
\right)
\partial^{\ }_{\rho} 
\left(
\cos\alpha\,
a^{\ }_{5\kappa}
+
\frac{1-\cos\alpha}{2}
\partial^{\ }_{\kappa}\theta
\right)
\\
&\,
+ 
C^{(1)}_{11}   
\sin^2\alpha\,
\epsilon^{\nu\rho\kappa}
\left(
a^{\ }_{5\nu}
-
\frac{1}{2}
\partial^{\ }_{\nu} 
\theta
\right)
\partial^{\ }_{\rho}
\left(
a^{\ }_{5\kappa}
-
\frac{1}{2}
\partial^{\ }_{\kappa} 
\theta
\right)
\\
&\,
-
C^{(1)}_{03}
\epsilon^{\nu\rho\kappa} 
a^{\ }_{\nu}
\partial^{\ }_{\rho}
\left(
\cos\alpha\,
a^{\ }_{5\kappa}
+
\frac{1-\cos\alpha}{2}
\partial^{\ }_{\kappa} 
\theta
\right)
+
\ldots
\end{split}
\label{eq: final effective action}
\end{equation}
\end{widetext}
with the local U(1)$\otimes$U(1) gauge invariance%
\begin{equation}
\begin{split}
&
a^{\ }_{ \mu}\to
a^{\ }_{ \mu}
-
\partial^{\ }_{\mu}\phi,
\\
&
a^{\ }_{5\mu}\to
a^{\ }_{5\mu}
-
\partial^{\ }_{\mu}\phi^{\ }_{5},
\qquad
\theta\to
\theta-2\phi^{\ }_{5},
\end{split}
\label{eq: effective local gauge symmetries final}
\end{equation}
for any compact and boundary-less manifold in (2+1)-dimensional space
and time. 

Some comments are of order here. First,
the coefficient $C^{(0)}_{11}$ controls the
axial phase stiffness of the Anderson-Higgs contribution to the
effective action. 
Second, each of the coefficients
$C^{(1)}_{00}$, $C^{(1)}_{33}$, and $C^{(1)}_{11}$
multiplies a Chern-Simons term that is diagonal
with respect to the gauge fields. The coefficient
$C^{(1)}_{03}$ is different in that regard since
it couples the gauge field $a^{\ }_{\mu}$
responsible for the conservation of the fermion number
to the axial gauge field $a^{\ }_{5\mu}$ on the one hand,
and the axial singlet linear combination
$\tilde{a}^{\ }_{5\mu}\equiv a^{\ }_{5\mu}-\partial^{\ }_{\mu}\theta/2$
on the other hand. Such an off-diagonal coupling is
reminiscent of so-called BF Chern-Simons theories.%
~\cite{Blau91,Hansson04}
It is the coefficient $C^{(1)}_{03}$ that controls
the charge assignments in the field theory%
~(\ref{eq: final effective action})
and, for later convenience, we break its contribution
to the induced fermionic charge into two pieces,
\begin{equation}
\begin{split}
&
\mathcal{L}^{\ }_{\mathrm{BF}}:=
\mathcal{L}^{(1)}_{\mathrm{BF}}
+
\mathcal{L}^{(2)}_{\mathrm{BF}},
\\
&
\mathcal{L}^{(1)}_{\mathrm{BF}}:=
C^{(1)}_{03}
\left(1-\cos\alpha\right)
a
d 
\tilde{a}^{\ }_{5},
\\
&
\mathcal{L}^{(2)}_{\mathrm{BF}}:=
-
C^{(1)}_{03}
a
d
a^{\ }_{5}.
\end{split}
\label{eq: def LBF}
\end{equation}
Here, we have introduced the short-hand notation
$adb\equiv 
\epsilon^{\mu\nu\rho}a^{\ }_{\mu}\partial^{\ }_{\nu}b^{\ }_{\rho}$.

\section{
Fractional fermion charge
        }
\label{sec: Fractional charge quantum number}

Equipped with Eq.~(\ref{eq: final effective action}) and Table~\ref{tab:Cs}
we compute in this section
the leading contributions
in the gradient expansion to the 
expectation value of the conserved charge current
\begin{equation}
\left\langle
j^{\mu}
\right\rangle^{\ }_{m,\eta}:=
-{i}
\left.
\frac{
\delta
\ln Z^{\prime}_{m,\eta}[B']
     }
     {
\delta a^{\ }_{\mu}
     }
\right|^{ }_{a^{\ }_{\mu}=0}.
\end{equation}

The induced fermion charge current is 
\begin{equation}
j^{\mu}= 
-
C^{(1)}_{03}
\epsilon^{\mu\rho\kappa} 
\partial^{\ }_{\rho}
\left(
\cos\alpha\,
a^{\ }_{5\kappa}
+
\frac{1-\cos\alpha}{2}
\partial^{\ }_{\kappa} 
\theta
\right)
+\ldots.
\label{eq: final induced fermion current}
\end{equation}
It obeys the continuity equation
\begin{equation}
\partial^{\ }_{\mu}j^{\mu}=0.
\end{equation}
The total induced fermionic charge 
\begin{equation}
Q:=
\int d^{2}\boldsymbol{r}\,
j^{0}(\boldsymbol{r},t)
\end{equation}
is thus time-independent
and given by
\begin{equation}
Q=
-
C^{(1)}_{03}
\oint d\boldsymbol{l}
\cdot
\left(
\cos\alpha\,
\boldsymbol{a}^{\ }_{5}
+
\frac{1-\cos\alpha}{2}
\boldsymbol{\partial}
\theta
\right)
\end{equation}
with the help of Stokes' theorem.
The induced fermionic charge is
\begin{subequations}
\begin{equation}
Q= 
-
2\pi\;C^{(1)}_{03}
\left(
\frac{n^{\ }_{\theta}}{2}
+
\frac{1}{2}
\left(
n^{\ }_{5}
-
n^{\ }_{\theta}
\right)
\cos\alpha
\right)
\end{equation}
for the special case when the vector fields
$\boldsymbol{a}^{\ }_{5}$
and 
$\boldsymbol{\partial}\theta$
support, on a circular boundary at infinity,
the net vorticity
\begin{equation}
\begin{split}
&
a^{i}_{5}\to 
-
\frac{n^{\ }_{5}}{2}
\epsilon^{ij}
\frac{r^{j}}{\boldsymbol{r}^{2}},
\qquad
n^{\ }_{5}\in\mathbb{Z},
\\
&
\partial^{i}\theta\to 
-
n^{\ }_{\theta}
\epsilon^{ij}
\frac{r^{j}}{\boldsymbol{r}^{2}},
\qquad
n^{\ }_{\theta}\in\mathbb{Z},
\end{split}
\label{eq: net vorticity}
\end{equation}
\end{subequations}
respectively. In the absence of the axial gauge flux $n^{\ }_{5}=0$,
while the condition for the axial vorticity to screen 
the (Kekul\'e) vorticity
is $n^{\ }_{5} = n^{\ }_{\theta}$. Notice that because $C^{(1)}_{03}$
vanishes for $|\eta|>m$, there is no charge bound to the vortices in
that regime. In contrast, when $|\eta|<m$, the charge bound to the
topological defect is
\begin{equation}
\begin{split}
Q=-\mathrm{sgn}\,\mu^{\ }_\mathrm{s}\times
\begin{cases}
\sin^2\frac{\alpha}{2} \;n^{\ }_{\theta} 
\;,
&
\hbox{unscreened $(n^{\ }_{5} = 0$)},
\\
\\
\frac{1}{2} \;n^{\ }_{\theta} 
\;,
&\hbox{screened $(n^{\ }_{5} = n^{\ }_{\theta}$).}
\end{cases}
\end{split}
\label{eq: derivation of Q}
\end{equation}
These results are consistent with those in
Refs.~\onlinecite{Hou07},
\onlinecite{Chamon08a},
and \onlinecite{Chamon08b}. 

\section{
Fractional statistical angle
        }
\label{sec: Fractional statistical angle}

We start from the effective partition function
\begin{equation}
Z^{\mathrm{eff}}_{m,\eta}:=
\int\mathcal{D}[a^{\mu},a^{\mu}_{5},\theta]
\exp
\left(
{i}
\int d^{3}x\,
\mathcal{L}^{\mathrm{eff}}_{m,\eta}
\right)
\label{eq: def Z eff m eta}
\end{equation}
with the Lagrangian given by
Eq.~(\ref{eq: final effective action})
and the coefficients in Table~\ref{tab:Cs}.
In a static approximation,
i.e., if we ignore
dynamics as we did when computing the
fractional charge%
~(\ref{eq: derivation of Q}), 
vortices are independently supported by
the axial gauge field $a^{\mu}_{5}$
or by the phase $\theta$. 

We will analyze the exchange statistics in two separate cases. The
first is when the $\theta$ vortices are dynamically screened by the
half fluxes in the axial gauge field $a^{\mu}_{5}$. The second case is
when the axial gauge field is suppressed, and the $\theta$ vortex is
unscreened; this situation does not arise from the effective Lagrangian%
~(\ref{eq: final effective action}) itself, but it can occur when
one goes beyond the linearized Dirac approximation or
includes other lattice effects.

\subsection{
Screened vortices
           }
\label{subsec: Screened vortices}

The exchange statistics 
of vortices and axial gauge fluxes
follows from the effective Lagrangian for the so-called vortex currents. 
One way to obtain this effective Lagrangian in the
screened case is to notice that the 
local axial gauge invariance together with the first line in 
Eq.~(\ref{eq: final effective action}) 
provides the screening condition, for
the axial gauge potential must then track the $\theta$ field
and, in particular,
vortices in $\theta$ must be screened by half fluxes in 
$a^{\mu}_{5}$. 

One way to impose this screening is to replace
\begin{equation}
\partial^{\ }_{\mu}
\theta
- 
2
a^{\ }_{5\mu}
\to 0
\label{eq:screening-cond}
\end{equation}
in Eq.~(\ref{eq: final effective action}). This can be justified
more precisely by using the (vortex) dual description of the $XY$ model,
as presented in Appendix~\ref{sec:duality}. In effect, the
fluctuations away from the condition~(\ref{eq:screening-cond}),
which are penalized by the finite stiffness coefficient
$C^{(0)}_{11}\sin^{2}\alpha$, 
can be accounted through a Maxwell term in the dual description. However, 
the Maxwell term does not enter the exchange statistics. Thus, we can
simply use the infinite stiffness limit or, equivalently, the
condition~(\ref{eq:screening-cond}).

The Lagrangian given by
Eq.~(\ref{eq: final effective action})
in the screening limit%
~(\ref{eq:screening-cond}) 
is
\begin{equation}
\begin{split}
\mathcal{L}^{\mathrm{eff}}_{m,\eta}=&\,
\hphantom{+}
C^{(1)}_{00}\,
\epsilon^{\nu\rho\kappa} \,
a^{\ }_{\nu} \,
\partial^{\ }_{\rho} 
a^{\ }_{\kappa}
\\
&\,
+
\frac{1}{4}\;C^{(1)}_{33}\;
\epsilon^{\nu\rho\kappa} \,
\left(\partial^{\ }_{\nu} 
\theta\right)\,
\partial^{\ }_{\rho} 
\left(\partial^{\ }_{\kappa}\theta
\right)
\\
&\,
-
\frac{1}{2}\;
C^{(1)}_{03}\;
\epsilon^{\nu\rho\kappa} \,
a^{\ }_{\nu}\;
\partial^{\ }_{\rho}
\left(
\partial^{\ }_{\kappa} 
\theta
\right)
+\ldots.
\end{split}
\label{eq:effective-screened-action}
\end{equation}
The Lagrangian can be written in terms of the vortex current
\begin{equation}
\bar{j}^{\mu}_{\mathrm{vrt}}:=
\frac{1}{2\pi}
\epsilon^{\mu\nu\lambda}
\partial^{\ }_{\nu}
\partial^{\ }_{\lambda}
\theta,
\label{eq:def-j-theta-screened}
\end{equation}
that obeys the conservation law
\begin{equation}
\partial^{\ }_{\mu}\bar{j}^{\mu}_{\mathrm{vrt}}=0,
\end{equation}
using the duality representation of
the $XY$ model supplemented by a Chern-Simons term
in (2+1) space and time as done in Appendix~\ref{sec:duality}. 
This leads to the Chern-Simons Lagrangian
\begin{equation}
\begin{split}
\mathcal{L}^{\mathrm{eff}}_{m,\eta}
=&\,
\hphantom{+}
C^{(1)}_{00}
\epsilon^{\nu\rho\kappa}
a^{\ }_{\nu}
\partial^{\ }_{\rho} 
a^{\ }_{\kappa}
-
\pi
C^{(1)}_{03}  
a^{\ }_{\nu}
\bar{j}^{\nu}_{\mathrm{vrt}}
\\
&\,
+
\frac{1}{4}
C^{(1)}_{33}
\left(
\epsilon^{\nu\rho\kappa}
d^{\ }_{\nu}
\partial^{\ }_{\rho} 
d^{\ }_{\kappa}
+
4\pi
d^{\ }_{\nu}
\bar{j}^{\nu}_{\mathrm{vrt}}
\right)
\\
&\,
+\ldots,
\end{split}
\label{eq: screened dual lagrangian}
\end{equation}
from which the statistics carried by 
screened quasiparticles with
the current $\bar{j}^{\nu}_{\mathrm{vrt}}$
follows. This statistics depends 
on the coefficients in Table~\ref{tab:Cs}. We treat
separately the two phases of Fig.~\ref{fig:phase-diagram}.

\subsubsection{
Weak time-reversal symmetry breaking: $|\eta|<m$
              }

In this limit,
$
C^{(1)}_{00}=
C^{(1)}_{33}=
0
$
and
the effective Lagrangian~(\ref{eq: screened dual lagrangian})
reduces to
\begin{equation}
\begin{split}
\mathcal{L}^{\mathrm{eff}}_{m,\eta}=&\, 
-
\frac{1}{2}
\mathrm{sgn}\,\mu^{\ }_{\mathrm{s}}
a^{\ }_{\nu}
\bar{j}^{\nu}_{\mathrm{vrt}}
+
\ldots.
\end{split}
\end{equation}
Thus, because of the absence of the Chern-Simons terms, 
the statistical angle $\Theta$
under exchange of any two screened quasiparticles is bosonic,
\begin{equation}
\frac{\Theta}{\pi}=0.
\end{equation}
Notice that it also follows that the induced fermionic U(1) current
\begin{equation}
j^{\nu}= 
-
\frac{1}{2}
\mathrm{sgn}\,
\mu^{\ }_{\mathrm{s}}\
\bar{j}^{\nu}_{\mathrm{vrt}}
\end{equation}
that couples  linearly to $a^{\ }_\mu$, 
is tied to the vortex current. In other words,
screened quasiparticles
with unit vorticity are charged objects with charge 
$Q=\pm1/2$ as found in Refs.%
~\onlinecite{Hou07},
\onlinecite{Chamon08a},
\onlinecite{Chamon08b}, 
and in Sec.~\ref{sec: Fractional charge quantum number}.

\subsubsection{
Strong time-reversal symmetry breaking: $|\eta|>m$
              }

In this limit,
$
C^{(1)}_{00}=
C^{(1)}_{33}=
(4\pi)^{-1}\mathrm{sgn}\,\eta
$
and the effective Lagrangian~(\ref{eq: screened dual lagrangian})
reduces to
\begin{equation}
\begin{split}
\mathcal{L}^{\mathrm{eff}}_{m,\eta}=&\,
\hphantom{+}
\frac{1}{4\pi} 
\mathrm{sgn}\,\eta\,
\epsilon^{\nu\rho\kappa}
a^{\ }_{\nu}
\partial^{\ }_{\rho} 
a^{\ }_{\kappa}
\\
&\,
+
\frac{1}{16\pi}
\mathrm{sgn}\,\eta
\left(
\epsilon^{\nu\rho\kappa}
d^{\ }_{\nu}
\partial^{\ }_{\rho} d^{\ }_{\kappa}
+
4\pi
d^{\ }_{\nu}
\bar{j}^{\nu}_{\mathrm{vrt}}
\right)
\\
&\,
+\ldots.
\end{split}
\end{equation}
Using the coefficient of the Chern-Simons
Lagrangian for the gauge field $d^{\ }_{\nu}$ 
and its coupling to the vortex current
$\bar{j}^{\nu}_{\mathrm{vrt}}$ 
(see appendix~\ref{sec:duality} for the
relation between the statistical angle and the coefficient in front of
the Chern-Simons term), 
the statistical angle $\Theta$
under exchange of two screened quasiparticles
with unit vorticity is
\begin{equation}
\frac{\Theta}{\pi}=
\frac{1}{4}
\mathrm{sgn}\,\eta.
\end{equation}
Notice that the U(1) current now vanishes, i.e., screened
quasiparticles carrying fractional statistics are now charge neutral.

\subsection{
Unscreened vortices
           }
\label{subsec; Unscreened vortices}

We turn to the situation when the axial gauge half fluxes
are suppressed,
while $\theta$ vortices are still present. 
We call these vortices unscreened quasiparticles.
This situation arises if,
in addition to the effective Lagrangian~(\ref{eq: final effective action}) 
which followed from integrating out the Dirac fermions, there are
terms in the effective Lagrangian due to lattice degrees of freedom that
break the axial gauge symmetry. For instance, acoustic phonons and
ripples in graphene can bring about the axial vector potential 
$a^{\mu}_{5}$; however, in these cases there is an energy penalty of the form
$a^{\ }_{5\mu}a^{\mu}_{5}$ 
that breaks the axial gauge invariance
due to contributions to the elastic energy.

The case when the axial gauge potential is absent,
i.e., the quasiparticles are unscreened, is implemented by
the replacement
\begin{equation}
a^{\ }_{5\mu}
\to 
0
\label{eq:unscreening-cond}
\end{equation}
in Eq.~(\ref{eq: final effective action}).
There follows
\begin{widetext}
\begin{equation}
\begin{split}
\mathcal{L}^{\mathrm{eff}}_{m,\eta}=&\,
\hphantom{+}
\frac{1}{4}\;C^{(0)}_{11}\;
\sin^2 \alpha
\left(\partial^{\rho}\theta\right)
\left(\partial^{\ }_{\rho}\theta\right)
\\
&\,
+
\left(
C^{(1)}_{33}\sin^2\frac{\alpha}{2}
+
C^{(1)}_{11}
\cos^2\frac{\alpha}{2}
\right)
\;\sin^2\frac{\alpha}{2}
\;\;
\epsilon^{\nu\rho\kappa} 
\left(
\partial^{\ }_{\nu} 
\theta
\right)
\partial^{\ }_{\rho} 
\left(
\partial^{\ }_{\kappa}
\theta
\right)
\\
&\,
+
C^{(1)}_{00}\;
\epsilon^{\nu\rho\kappa} \;
a^{\ }_{\nu} \;
\partial^{\ }_{\rho} 
a^{\ }_{\kappa}
\\
&\,
-
C^{(1)}_{03}\;\sin^2\frac{\alpha}{2}\;\;
\epsilon^{\nu\rho\kappa} 
a^{\ }_{\nu}
\partial^{\ }_{\rho}
\left(
\partial^{\ }_{\kappa} 
\theta
\right)
+\ldots.
\end{split}
\label{eq: final effective action unscreened}
\end{equation}
This Lagrangian can be dualized with the help of the vortex current%
~(\ref{eq:def-j-theta-screened})
(see appendix~\ref{sec:duality})
\begin{equation}
\begin{split}
\mathcal{L}^{\mathrm{eff}}_{m,\eta}
=&\,
-({8\pi^2 \;C^{(0)}_{11}\;\sin^2 \alpha})^{-1}\;
f^{\mu\nu}\,f_{\mu\nu}
+c_\mu\;\bar{j}^{\mu}_{\mathrm{vrt}}
\\
&+
C^{(1)}_{00}\,
\epsilon^{\nu\rho\kappa} \,
a^{\ }_{\nu} \,
\partial^{\ }_{\rho} 
a^{\ }_{\kappa}
-
2\pi
C^{(1)}_{03}\sin^2\frac{\alpha}{2}\;
a^{\ }_{\nu}\;\bar{j}^{\nu}_{\mathrm{vrt}}
\\
&\,
+
\left(
C^{(1)}_{33}\sin^2\frac{\alpha}{2}
+
C^{(1)}_{11}
\cos^2\frac{\alpha}{2}
\right)
\;\sin^2\frac{\alpha}{2}
\;\;
\left(
\epsilon^{\nu\rho\kappa} \,
d_\nu\,
\partial^{\ }_{\rho} d_\kappa
+
4\pi\;
d^{\ }_{\nu}\;\bar{j}^{\nu}_{\mathrm{vrt}}
\right)
+\ldots.
\end{split}
\label{eq:unscreened-eff-L}
\end{equation}
\end{widetext}
(The Maxwell term $f^{\ }_{\mu\nu}f^{\mu\nu}$ is associated to the gauge
potential $c^{\ }_{\mu}$, see appendix~\ref{sec:duality}). 
We shall denote with
$\mathcal{L}^{\ }_{c}$ 
the first line of Eq.~(\ref{eq:unscreened-eff-L}).
The statistics carried by unscreened quasiparticles 
with the current
$j^{\mu}_{\mathrm{vrt}}$ follows.
This statistics depends on the
coefficients in Table~\ref{tab:Cs}. 
We treat separately the two phases of Fig.~\ref{fig:phase-diagram}.

\subsubsection{
Weak time-reversal symmetry breaking: $|\eta|<m$
              }
\label{subsubsec: Weak time-reversal symmetry breaking}

In this limit, 
$C^{(1)}_{00}=C^{(1)}_{33}=0$,
$C^{(1)}_{11}=\eta/(6\pi m)$,
$C^{(1)}_{03}=(2\pi)^{-1}\mathrm{sgn}\,\mu^{\ }_{s}$,
and the effective Lagrangian~(\ref{eq:unscreened-eff-L}) reduces to
\begin{equation}
\begin{split}
\mathcal{L}^{\mathrm{eff}}_{m,\eta}=&\,
\mathcal{L}^{\ }_{c}
-
\mathrm{sgn}\,\mu^{\ }_{\mathrm{s}}\,
\sin^2\frac{\alpha}{2}\,
a^{\ }_{\nu}\,
\bar{j}^{\nu}_{\mathrm{vrt}}
\\
&\,
+
\frac{\eta}{24\pi m}
\sin^2\alpha
\left(
\epsilon^{\nu\rho\kappa}
d^{\ }_{\nu}\,
\partial^{\ }_{\rho} d^{\ }_{\kappa}
+
4\pi\;
d^{\ }_{\nu}
\bar{j}^{\nu}_{\mathrm{vrt}}
\right).
\end{split}
\label{eq:eff-un-2}
\end{equation}
Using the coefficient of the Chern-Simons
Lagrangian for the gauge field $d^{\ }_{\nu}$ 
and its coupling to the vortex current
$\bar{j}^{\nu}_{\mathrm{vrt}}$ 
(see appendix~\ref{sec:duality} for the
relation between the statistical angle and the coefficient in front of
the Chern-Simons term), 
the statistical angle $\Theta$
under exchange of two unscreened quasiparticles 
with unit vorticity is
\begin{equation}
\frac{\Theta}{\pi}=
\frac{\eta}{6m}\;\sin^2\alpha
\\
=
\frac{2\eta}{3m}\;|Q|(1-|Q|)
\label{eq:statistical-angle-eta-smaller-m}
\end{equation}
by Eq.~(\ref{eq: derivation of Q}).
Notice that it also follows that the induced fermionic U(1) current, 
\begin{equation}
j^{\nu}=
-
\mathrm{sgn}\,\mu^{\ }_{\mathrm{s}}\,
\sin^2\frac{\alpha}{2}\,
\bar{j}^{\nu}_{\mathrm{vrt}}
\end{equation}
that couples  linearly to $a^{\ }_\mu$, 
is tied up to the vortex current. In other words,
unscreened quasiparticles
with unit vorticity are charged objects with charge 
$Q=\pm\sin^2(\alpha/2)$ that varies continuously as a
function of the ratio
$\mu^{\ }_\mathrm{s}/m$ [see Eq.~(\ref{eq: derivation of Q})] as
found in Refs.%
~\onlinecite{Hou07},
\onlinecite{Chamon08a},
and
\onlinecite{Chamon08b}.

\subsubsection{
Strong time-reversal symmetry breaking: $|\eta|>m$
              }
\label{subsubsec: Strong time-reversal symmetry breaking}

In this limit, 
$C^{(1)}_{00}=C^{(1)}_{33}=(4\pi)^{-1}\mathrm{sgn}\,\eta$,
$C^{(1)}_{11}=(3\eta^{2}-m^{2})/(12\pi\eta^{2})\,\mathrm{sgn}\,\eta$,
$C^{(1)}_{03}=0$,
and the effective Lagrangian~(\ref{eq:unscreened-eff-L}) reduces to
\begin{widetext}
\begin{equation}
\begin{split}
\mathcal{L}^{\mathrm{eff}}_{m,\eta}
=&\,
\hphantom{+}
\mathcal{L}^{\ }_{c}
+
\frac{1}{4\pi}\; \mathrm{sgn}\,\eta
\;
\epsilon^{\nu\rho\kappa} \,
a^{\ }_{\nu} \,
\partial^{\ }_{\rho} 
a^{\ }_{\kappa}
\\
&\,
+\frac{1}{4\pi}\; \mathrm{sgn}\,\eta
\;
\left[
\left(
1-\frac{m^2}{3\eta^2}
\right)
+
\frac{m^2}{3\eta^2}
\sin^2\frac{\alpha}{2}
\right]
\;\sin^2\frac{\alpha}{2}\;\;
\left(
\epsilon^{\nu\rho\kappa} \,
d^{\ }_{\nu}\,
\partial^{\ }_{\rho} d^{\ }_{\kappa}
+
4\pi
d^{\ }_{\nu}\;\bar{j}^{\nu}_{\mathrm{vrt}}
\right)
+\ldots.
\end{split}
\end{equation}
\end{widetext}
Using the coefficient of the Chern-Simons
Lagrangian for the gauge field $d^{\ }_{\nu}$ 
and its coupling to the vortex current
$\bar{j}^{\nu}_{\mathrm{vrt}}$ 
(see appendix~\ref{sec:duality} for the
relation between the statistical angle and the coefficient in front of
the Chern-Simons term), 
the statistical angle under exchange of two unscreened 
quasiparticles with unit vorticity is
\begin{equation}
\begin{split}
\frac{\Theta}{\pi}&=\,
\mathrm{sgn}\,\eta
\;
\left[
\left(
1
-
\frac{m^2}{3\eta^2}
\right)
+
\frac{m^2}{3\eta^2}
\sin^2\frac{\alpha}{2}
\right]
\;\sin^2\frac{\alpha}{2}
\\
&=\,
\mathrm{sgn}\,\eta
\;
\left[
\left(
1
-
\frac{m^2}{3\eta^2}
\right)
+
\frac{m^2}{3\eta^2}
|Q|
\right]
\;|Q|.
\end{split}
\label{eq: fractional Theta unscreened phase}
\end{equation}
Here, we have used the value of the charge 
$|Q|=\sin^2(\alpha/2)$
for the complementary phase $|\eta|<m$.
Notice that the induced fermionic charge current $j^{\ }_{\mu}$
now vanishes. The unscreened quasiparticles
carrying the fractional statistics%
~(\ref{eq: fractional Theta unscreened phase}) 
are thus charge neutral, i.e., $|Q|$ in
Eq.~(\ref{eq: fractional Theta unscreened phase})
should not be confused with the (now vanishing)
electronic charge of unscreened quasiparticles.

We stress that the quenching of the dynamics
in the axial gauge field $a^{\mu}_{5}$
implies the breaking of the axial gauge symmetry.
It can be thought of as a mean-field approximation
needed to interpret the numerical simulations 
of the Berry phase acquired by the Slater determinant 
of lattice fermions when one vortex is moved 
in a quasi-static way along a closed curved around 
another vortex.
The quench approximation can
also be justified if terms that
break explicitly the axial gauge symmetry such as mass
term for $a^{\ }_{5}$ were added
to the Lagrangian~(\ref{eq: final effective action}). 
After all, from a microscopic point of view, 
axial gauge symmetry is by no means generic.
The axial gauge fields can be viewed as
phonon-induced fluctuations in the average
separations between ions that an elastic
theory generically induces. A mass term for these 
phonons cannot be ruled out by symmetry.

\subsection{
Adding one more fermion to the midgap states
           }
\label{subsec:one-more-electron}

All calculations for the fractional charge and exchange statistics
done so far apply at zero chemical potential
$\mu=0$,
and at some finite staggered chemical potential 
$\mu^{\ }_{\mathrm{s}}\neq0$,
assuming global vortex neutrality.
Global vortex neutrality is imposed to bound the energy from above
in the thermodynamic limit or if periodic boundary conditions are
imposed. A staggered chemical potential is needed to lift
the near degeneracy between the two single-particle midgap states that 
are exponentially localized about a vortex and anti-vortex
in the bond-density-wave (Kekul\'e for graphene)  
order parameter $\Delta$, respectively, 
whose separation $r$ is much larger than $1/m$. On the one hand, when
$\mu^{\ }_{\mathrm{s}}=0$,
the two single-particle midgap levels are,
up to exponentially small corrections in $mr$, 
pinned to the band center $E=0$. In the thermodynamic limit,
their occupancy when $\mu=0$ is then ambiguous.
On the other  hand, when
$\mu^{\ }_{\mathrm{s}}\neq0$,
the two single-particle midgap levels
get pushed in opposite directions, one to $E>0$ 
and the other to $E<0$ 
(which one goes which way depends on the sign of $\mu^{\ }_{\mathrm{s}}$).
The single-particle midgap level with $E<0$ is then occupied, 
the other empty, when $\mu=0$ and the results of Secs.%
~\ref{sec: Fractional charge quantum number}
and%
~\ref{subsec; Unscreened vortices}
for the fractional and exchange statistics, respectively, apply.
We are going to prove that when 
$m>|\mu|>|\mu^{\ }_{\mathrm{s}}|$,
so that the two single-particle midgap levels
are either both empty or both occupied,
the exchange statistics is that of semions.

Suppose one adds one more electron to the Dirac sea
(here defined to be the Fermi sea at $\mu=0$), 
filling the single-particle midgap state at $E>0$. 
What happens to the exchange statistics?

The easiest way to answer this question is by realizing that the Berry
phase accumulated by a many-body wave function that can be written as a
single Slater determinant (the case in hand) is just the sum of the
Berry phases for single-particle states. If we fill one more level, 
we only need to add the Berry phase due to that single-particle state 
to that of the filled Dirac sea that we already computed. 
The contribution from the extra level can be obtained as follows. 
(Here we focus on the case $\eta=0$.
A generalization to $\eta\ne 0$ can be similarly formulated.)

A single-particle midgap wave function is localized near a vortex, 
i.e., its spatial extent is of order $1/m$. 
Details on $\Delta$ for distances much larger than $1/m$ away 
do not matter. Hence, when winding another far-away vortex around
the first one, the local order parameter $\Delta$ in the vicinity of
the first vortex just sees its phase change by $2\pi$. This allows us
to focus solely on the problem of determining what happens to the
single-particle midgap wave function 
as the phase of the order parameter near a vortex is rotated by $2\pi$.

The solution for the single-particle midgap wave function when 
$\mu^{\ }_{\mathrm{s}}=0$ and in the Dirac approximation
was obtained 
in Ref.~\onlinecite{Hou07} 
for the unscreened vortex and in Ref.~\onlinecite{Jackiw07} 
for the screened vortex. In both cases, 
the wave function picks up a phase of $\pi$ when the phase of
$\Delta$ changes by $2\pi$.
If $\mu^{\ }_{\mathrm{s}}\ne0$, 
the result remains the same, because while the midgap level moves 
with $\mu^{\ }_{\mathrm{s}}$
the wave function is independent of $\mu^{\ }_{\mathrm{s}}$
(the wave function has support in only one of the sublattices 
$\Lambda^{\ }_{\mathrm{A}}$
or 
$\Lambda^{\ }_{\mathrm{B}}$ of the underlying lattice model, 
so the finite value of the
staggered chemical potential does not perturb
the single-particle midgap wave function).

In conclusion, occupying one additional single-particle
fermion level adds a phase of
$\pi$ to the many-body Berry phase when $\eta=0$. 
This means that the statistical angle shifts by $\Delta\Theta=\pm\pi/2$,
the statistical angle for a semion, when one fermion is added (removed)
to (from) the Dirac sea.

\begin{figure}
\includegraphics[angle=0,scale=0.7]{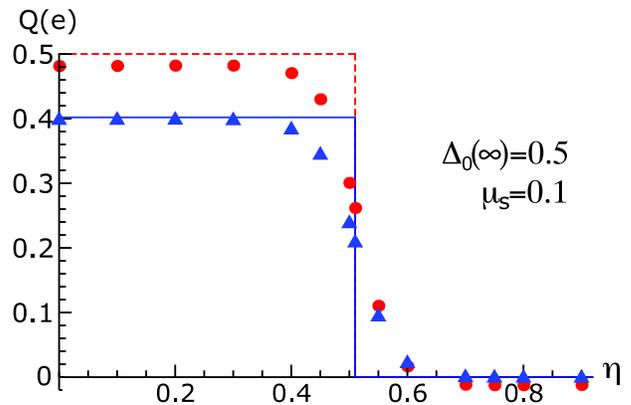}
\caption{
(Color online)
The induced fermionic charge of a quasiparticle,
a unit vortex in $\Delta(\boldsymbol{r})$ 
with or without attachment of 
an axial gauge half flux in $\boldsymbol{a}^{\ }_{5}(\boldsymbol{r})$,
as a function of $\eta$.
This charge is computed from the spectral asymmetry
of spinless fermions hopping on the square lattice
with lattice spacing $\mathfrak{a}$ and with a magnetic flux
of $\pi$ in units of the flux quantum $\phi^{\ }_{0}=hc/e$
threading each elementary plaquette
in the \textit{static background} of a unit vortex in 
$\Delta(\boldsymbol{r})$
with or without attachment of 
an axial gauge half flux in $\boldsymbol{a}^{\ }_{5}(\boldsymbol{r})$
and for a uniform value of $\eta$.
The square lattice is $100\times100$ 
in units of the lattice spacing
and the area used for integrating the 
local density of states is a square of size $50\times 50$ 
centered around a unit vortex. 
The following parameters were chosen: the hopping $t=1$,
the magnitude of $\Delta(\boldsymbol{r})$ on the boundary is  
$\Delta^{\ }_{0}(\infty)=0.5$
while the magnitude of the staggered chemical potential
is $\mu^{\ }_{\mathrm{s}}=0.1$ ($m\approx0.51$). 
Each (red) filled circle is the induced fermionic charge of a unit vortex
in $\Delta(\boldsymbol{r})$ to which an axial gauge half flux 
in $\boldsymbol{a}^{\ }_{5}(\boldsymbol{r})$ is attached.
The (red) dashed line for $\eta\leq m$ represents the $Q=1/2$ line. 
Each (blue) filled triangle is the induced fermionic charge of an unscreened 
unit vortex in $\Delta(\boldsymbol{r})$, 
i.e., the vector axial flux in $\boldsymbol{a}^{\ }_{5}(\boldsymbol{r})$
vanishes everywhere.
The (blue) solid line for $\eta\leq m$ represents $Q=0.402$,
i.e., the predicted value from the field theory with the input parameters.
When $m\gg\eta$, the induced fractional charge vanishes.
A quantum phase transition at $m=\eta$, as measured by the jumped
in the induced fermionic charge, is smeared by finite size effects.
        }
\label{fig:charge-eta-pi}
\end{figure}

\section{
Numerical calculation of the charge and Berry phase
        }
\label{eq: Numerical calculation of the charge and Berry phase}

We are going to present numerical results on the charge and
statistics of unscreened vortices supported by the 
bond-density-wave (Kekul\'e for graphene) 
order parameter $\Delta$
in the presence of the compatible and competing order parameters 
(masses when space and time independent)
$\mu^{\ }_\mathrm{s}$ and $\eta$, respectively. 
The dependence of the induced fermionic charge of vortices 
in $\Delta$ as a function of the staggered chemical potential 
$\mu^{\ }_\mathrm{s}$ was studied in Refs.~\onlinecite{Chamon08a} 
and \onlinecite{Chamon08b} 
(see also Ref.~\onlinecite{Weeks08} 
when $\mu^{\ }_{\mathrm{s}}=0$).
The following numerical results with the competing mass $\eta$ are new.

Our studies have been carried out for the honeycomb lattice, 
which is of direct relevance to graphene, 
and the square lattice with $\pi$-flux phase. 
Both lattice models yield consistent numerical results. 
In this paper, only the results for $\pi$-flux phase are presented. 
The relevant technical details for our numerical calculations 
are summarized in Appendix%
~\ref{appsec: Numerical Berry phase in the single-particle approximation}. 
To compare the numerical with our analytical results, 
derived from the Dirac Hamiltonian~(\ref{eq: def quantum Hamiltonian}), 
which is the continuum limit of the linearized lattice Hamiltonian, 
two important issues arise.

The first one is that \textit{all} band curvature effects,
present in \textit{any} microscopic lattice model,
are absent in the continuum model. Here we expect that, 
as long as the characteristic sizes over which the order parameters 
vary are large compared to the size of the unscreened vortex core, 
\textit{static} results obtained within
the continuum approximation should capture some static long-wave length 
properties of the lattice model.
This first expectation
can be concretely addressed by the numerical studies 
of the induced fermionic charge of unscreened vortices below.

The second issue that arises when one starts from a lattice model is
the assumed axial gauge invariance of Hamiltonian%
~(\ref{eq: def quantum Hamiltonian}). 
This issue is subtle and substantial. 
The Dirac Hamiltonian%
~(\ref{eq: def quantum Hamiltonian}) 
has a local U(1)$\times$U(1) gauge symmetry, 
while this symmetry is absent in graphene, say. 
Although the vector axial gauge field 
$\boldsymbol{a}^{\ }_{5}$ 
is realized in graphene, 
say through acoustic 
phonons generating ripples,
and thus couples in an axial-gauge-invariant way
to the fermions in the linear approximation,
its kinetic energy is by no means required to be gauge invariant.
For example, the kinetic energy of 
$\boldsymbol{a}^{\ }_{5}$ 
is expected to contain the axial-gauge-symmetry-breaking mass term
$|\boldsymbol{a}_{5}|^2$.
It is thus difficult to justify the axial gauge invariance
of Dirac Hamiltonian%
~(\ref{eq: def quantum Hamiltonian}) 
in a lattice model as simple as graphene.

We do not expect predictions based on Hamiltonian%
~(\ref{eq: def quantum Hamiltonian}) 
that rely crucially on the 
dynamics of the axial vector gauge field
$\boldsymbol{a}^{\ }_{5}$ to capture the 
corresponding low-energy and
long wave-length dynamical properties of graphene.
We will verify this expectation with lattice computations
that require the dynamics of the axial gauge field, for example the
induced Berry phase as one moves a composite particle made of
a vortex and an axial gauge flux around another composite particle.

In Sec.~\ref{sec: Microscopic models}, we will
present a lattice model that, by construction,
has the desired local U(1)$\times$U(1) gauge symmetry. 
This model can be used to compute numerically 
the statistical phases of unit 
bond-density-wave (Kekul\'e for graphene)
vortices screened by axial gauge half fluxes
and to verify that non-linearities in the many-body excitation 
spectrum do not affect the exchange statistics of vortices
separated by distances much larger than their vortex core,
i.e., this is one model that regularizes Hamiltonian%
~(\ref{eq: def quantum Hamiltonian})
on the lattice. 

While the system presented in Sec.~\ref{sec: Microscopic models}
serves by itself as a proof of principle 
that one can realize the local axial gauge invariance on the lattice, 
the computation of the exchange statistics of vortices in
this lattice model is a computational challenge in lattice gauge theory, 
as opposed to the much simpler exercise in exact diagonalization 
for any non-interacting lattice model.

For this reason, we now limit the numerical studies of the statistical phases
to the simpler case when $\boldsymbol{a}^{\ }_{5}\to0$, 
i.e., the case of unscreened vortices.
In effect, we are ignoring all many-body effects imposed by the
local axial gauge invariance 
and thus treating the problem at the mean-field level. 
By comparing the charge obtained from the Aharonov-Bohm
effect with that obtained directly from the local density of states, 
we will show that this approximation is qualitatively 
(but not quantitatively) justified for dynamical properties
of bond-density-wave (Kekul\'e for graphene) vortices,
whereas it fails dramatically for dynamical properties
of the axial gauge half fluxes.

\subsection{Static calculation of the charge}
\label{subsec: Static calculation of the charge}

We begin with the study of static properties, 
when the vortices or axial gauge half flux tubes are not moved, 
so that the dynamics of the axial gauge potential is not relevant. 
One physical quantity that can be studied in the static limit 
is the induced fermionic fractional charge. 
It is obtained by summing up the local fermionic density of states 
in a region of space that encloses the core of the vortex.

In our numerical studies, a vortex is placed at the center of 
the square lattice system of size $100 \times 100$
in units of the lattice spacing $\mathfrak{a}$
while a flux of $\pi$ in units of the flux quantum $\phi^{\ }_{0}=hc/e$
threads each elementary plaquette.
An area of integration, $50\times 50$, 
centered around the vortex is used for summing 
the local fermionic density of states. 
We fixed the strength of the 
bond-density-wave (Kekul\'e for graphene)
order parameter $\Delta=0.5$ 
and staggered chemical potential $\mu^{\ }_{\mathrm{s}}=0.1$. 

Figure~\ref{fig:charge-eta-pi} 
shows the value of the induced fermionic charge as a function of $\eta$ 
with and without the  axial gauge half flux. 
A clear normalization of the fractional charge to 1/2 
follows from adding an axial gauge half flux. 
Notice that there is a (smoothed) step as the
mass $\eta$ becomes comparable to $m$. This is the finite-size
signature of a quantum phase transition at $|\eta|=m$.
The results in Sec.%
~\ref{sec: Fractional charge quantum number} 
are displayed in 
Fig.~\ref{fig:charge-eta-pi}.
They correspond to sharp step functions at the transition
point $|\eta|=m$. The numerical results displayed in
Fig.~\ref{fig:charge-eta-pi} are consistent with the analytical
results~(\ref{eq: derivation of Q}), 
keeping in mind that the lattices studied are finite and thus
quantum transitions are smeared. 
For that matter, notice that the agreement between the field-theory 
prediction and numerics is best away from the critical point $|\eta|=m$.

\begin{figure}
\includegraphics[angle=0,scale=0.6]{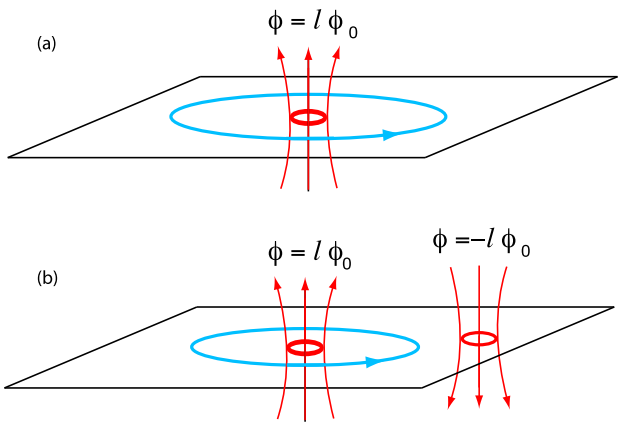}
\caption{
(Color online)
Schematics of the \textit{static} magnetic flux tubes 
inserted to probe the induced fermionic charge of a quasiparticle, 
a mass vortex with or without the attachment of an axial gauge half flux,
using the Aharonov-Bohm effect in the second set-up described in the text. 
(a) We insert one \textit{static} magnetic flux tube
(colored in red) 
with the flux $\phi=l\phi^{\ }_{0}$ 
(the flux quantum is $\phi^{\ }_{0}=hc/e$)
while a quasiparticle encircles dynamically this magnetic flux 
with the trajectory indicated by the directed loop
(colored in blue). 
(b) We insert two \textit{static} magnetic flux tubes 
(colored in red) 
with the fluxes $\phi= \pm l\phi^{\ }_{0}$ 
while a quasiparticle encircles dynamically 
one and only one magnetic flux tube with the trajectory 
indicated by the directed loop (colored in blue).
        }
\label{fig:flux-tube}
\end{figure}

\subsection{Dynamic calculation of the charge}
\label{subsec: Dynamic calculation of the charge}

A dynamical alternative to computing the induced fermionic charge through the integrated local density of states is the following. If we take a unit vortex in $\Delta$ 
(with or without an accompanying axial gauge half flux) 
around a circle of radius $r$ that encircles a magnetic flux, 
then an Aharonov-Bohm phase accumulates. 
The value of the charge induced near the vortex 
follows after matching the Berry phase
computed numerically to the analytical value of the Aharonov phase.

We carry out this approach in two different set-ups. In the first, we
apply a uniform magnetic field to the system, i.e.,
we fix a given electromagnetic flux 
\begin{equation}
l \phi^{\ }_{0},
\qquad
\hbox{$l\in\mathbb{R}$, $\phi^{\ }_{0}$ the quantum of flux},  
\end{equation}
per elementary unit cell on the lattice.
The Aharonov-Bohm phase $\gamma^{\ }_{\mathrm{AB}}$ 
that is picked up depends on the radius of the path
since the encircled magnetic flux scales with the
area. The Aharonov-Bohm phase in this case is thus given by
\begin{equation}
\label{eq:AB-phase}
\gamma^{\ }_{\mathrm{AB}}= 
2 \pi \times Q \times (\pi r^2) \times l.
\end{equation} 
Here, $Q$ is the charge bound to the unit vortex in $\Delta$. 

A second set-up is shown in Fig.~\ref{fig:flux-tube}a. 
We insert an electromagnetic flux tube with flux 
\begin{equation}
l \phi^{\ }_{0},
\qquad
\hbox{$l\in\mathbb{R}$, $\phi^{\ }_{0}$ the quantum of flux},  
\end{equation}
through the elementary unit cell on the lattice
at the center of the system. All other elementary unit cells
are free of any magnetic flux.
We then move the unit vortex in the 
bond-density-wave (Kekul\'e for graphene)
order parameter
around a path enclosing this flux. Notice that the Aharonov-Bohm phase
$\gamma^{\ }_{\mathrm{AB}}$ 
is independent of the path as long as it strictly contains the 
magnetic flux tube, i.e., the elementary unit cell at the center
of the lattice. It is expected to have the value
\begin{equation}
\label{eq:AB-phase-tube}
\gamma^{\ }_{\mathrm{AB}}= 
2 \pi \times Q \times l.
\end{equation} 

We also study the case displayed in Fig.~\ref{fig:flux-tube}b. The
reason for it is that we want to ensure that compensating 
fermionic charges on the edges of the sample do not contribute 
a phase as well. In the set up of Fig.~\ref{fig:flux-tube}b, 
whatever happens with the fermionic edge charges 
does not lead to an Aharonov phase 
because their path would encircle (even if they move) 
the vanishing total flux 
\begin{equation}
\phi= 
l\phi^{\ }_{0}
-
l\phi^{\ }_{0}=0.
\end{equation}

\begin{figure}
\includegraphics[angle=0,scale=0.75]{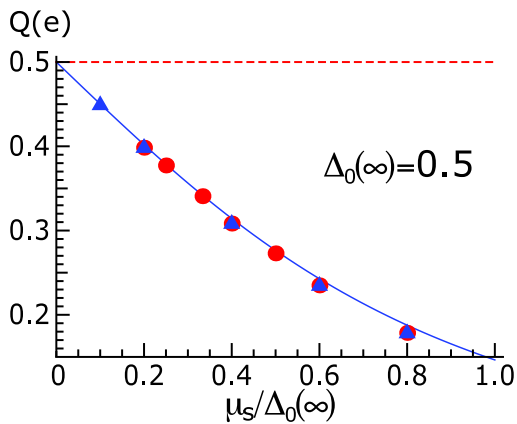}
\caption{
(Color online)
The induced fermionic charge of a quasiparticle,
a unit vortex in 
$\Delta(\boldsymbol{r})$ 
with or without attachment of an axial gauge half flux in 
$\boldsymbol{a}^{\ }_{5}(\boldsymbol{r})$,
as a function of the ratio 
$\mu^{\ }_{\mathrm{s}}/\Delta^{\ }_{0}(\infty)$. 
This charge is obtained by matching the numerical
Berry phase picked up when a quasiparticle hops along the
closed boundary of an area that encloses a magnetic flux
in the uniform background of 
$\mu^{\ }_{\mathrm{s}}$
to the corresponding Aharonov-Bohm phase
along the lines outlined in Sec.%
~\ref{subsec: Dynamic calculation of the charge}
and Appendix%
~\ref{appsec: Numerical Berry phase in the single-particle approximation}.
Hopping takes place on 
the square lattice with the lattice spacing $\mathfrak{a}$
and a magnetic flux of $\pi$ in units of the flux quantum 
$\phi^{\ }_{0}=hc/e$
per elementary plaquette, there being
$56\times 56$ elementary plaquettes. The closed path used
to compute the Berry phase is approximately circular with the radius
$r=14.5$ in units of the lattice spacing. 
The following parameters were chosen: the hopping $t=1$,
the magnitude of $\Delta(\boldsymbol{r},t)$ on the boundary is 
$\Delta^{\ }_{0}(\infty)=0.5$ 
while the flux is $\phi= 0.001 \phi^{\ }_{0}$ through each 
elementary plaquette. 
The (red) filled circles are the induced charges of a 
dynamical unit vortex in 
$\Delta(\boldsymbol{r},t)$ 
to which is also attached a dynamical axial gauge half flux in 
$\boldsymbol{a}^{\ }_{5}(\boldsymbol{r},t)$
as a function of 
$\mu^{\ }_{\mathrm{s}}/\Delta^{\ }_{0}(\infty)$. 
The (red) dashed line is the analytical charge $Q=1/2$. 
The (blue) filled triangles are the induced charges of a 
dynamical unit vortex in 
$\Delta(\boldsymbol{r},t)$ 
without the axial gauge half flux in 
$\boldsymbol{a}^{\ }_{5}(\boldsymbol{r})$
as a function of the ratio $\mu^{\ }_{\mathrm{s}}/\Delta^{\ }_{0}(\infty)$.
The (blue) solid line is the induced charge
computed from Eq.~(\ref{eq: derivation of Q})
as a function of the ratio $\mu^{\ }_{\mathrm{s}}/\Delta^{\ }_{0}(\infty)$. 
        }
\label{fig:AB-phase-mu-dependence}
\end{figure}

The results we obtain for the Berry phase when we wind the 
unscreened vortices around a closed path are shown in
Fig.~\ref{fig:AB-phase-mu-dependence} 
for the case of the first set-up (uniform applied magnetic field). 
We fix the parameters $\Delta=0.5$, $r=14.5$
(in a $56\times 56$ lattice) and $\phi=0.001 \phi^{\ }_{0}$ per plaquette,
and plot the charge $Q$ versus the parameter $\mu^{\ }_\mathrm{s}/\Delta$. 
The blue dots and red dots are the numerical results for a vortex without 
the axial gauge half flux and with the axial
gauge half flux, respectively, 
while the corresponding theoretical predictions from 
Ref.~\onlinecite{Chamon08a} and~\onlinecite{Chamon08b} 
are plotted in blue and red solid line. Notice
that the analytical and numerical results agree quite well for the
case of vortices unscreened by axial gauge half fluxes.

As anticipated, the analytical and numerical results are not consistent
for the case of screened vortices. The reason is precisely what
we highlighted in the beginning of this section, i.e., that the lattice
model studied numerically in this section does not contain the
U(1)$\times$U(1) symmetry, i.e.,
the axial gauge field dynamics present in
the Dirac Hamiltonian%
~(\ref{eq: def quantum Hamiltonian}). 
The same issue applies to the problem of computing the exchange statistics
of pairs of screened vortices.
We cannot study the statistical angle of
screened vortices within the approach of this section. 
In Sec.~\ref{sec: Microscopic models}, we will present a microscopic model
that does have the U(1)$\times$U(1) gauge symmetry.
However, this model cannot be studied by simply computing Slater determinants
(see Appendix%
~\ref{appsec: Numerical Berry phase in the single-particle approximation}) 
as has been done so far in this section.

Before closing Sec.~\ref{subsec: Dynamic calculation of the charge}, 
let us mention that we have checked the
results summarized by Fig.~\ref{fig:AB-phase-mu-dependence} that we
obtained by applying a uniform magnetic field against
those obtained with a single flux tube as in
Fig.~\ref{fig:flux-tube}a or with two
flux tubes as in Fig.~\ref{fig:flux-tube}b.

\subsection{
Fractional statistics for unscreened vortices
           }
\label{subsec: Fractional statistics for unscreened vortices}

We now present the numerical value 
of the statistical angle $\Theta$ in units of $\pi$
acquired under the exchange of two unit unscreened vortices 
in the bond-density-wave (Kekul\'e for graphene) 
order parameter $\Delta$, 
which we shall call quasiparticles from now on.
We have computed numerically the Berry phase $\gamma$ in units of $\pi$ 
accumulated when a first dynamical quasiparticle moves along a trajectory
that winds once around a second static quasiparticle
as outlined in Appendix%
~\ref{appsec: Numerical Berry phase in the single-particle approximation}.
The statistical angle $\Theta$ 
acquired under the exchange between these two quasiparticles is then 
\begin{equation}
\Theta= \frac{\gamma}{2}.
\end{equation}
Here, as we do not impose dynamically the axial gauge symmetry 
at the microscopic level as presented in Appendix%
~\ref{appsec: Numerical Berry phase in the single-particle approximation}
and unlike in Sec.~\ref{sec: Microscopic models}, 
we only treat unscreened vortices.
We have verified that $\gamma$, 
when computed along the lines of Appendix%
~\ref{appsec: Numerical Berry phase in the single-particle approximation},
does not change when axial gauge half flux tubes are attached to the 
vortices. 

To compare the microscopic exchange statistics with the
one computed within field theory in Sec.%
~\ref{subsec; Unscreened vortices},
we restrict the numerical computation to the half-filled case.
However, we will also test
the prediction of Sec.~\ref{subsec:one-more-electron}  
by working with one spinless fermion more than 
(or less than) at half-filling.

We will always take $m\approx 0.5$ in Eq.~(\ref{eq: def m}). 
Moreover, to limit finite size effects,
we assume that $\eta\ll m$, 
i.e., we work well below the transition point $|\eta|=m$
when the breaking of TRS is weak.

\begin{figure}
\includegraphics[angle=0,scale=0.6]{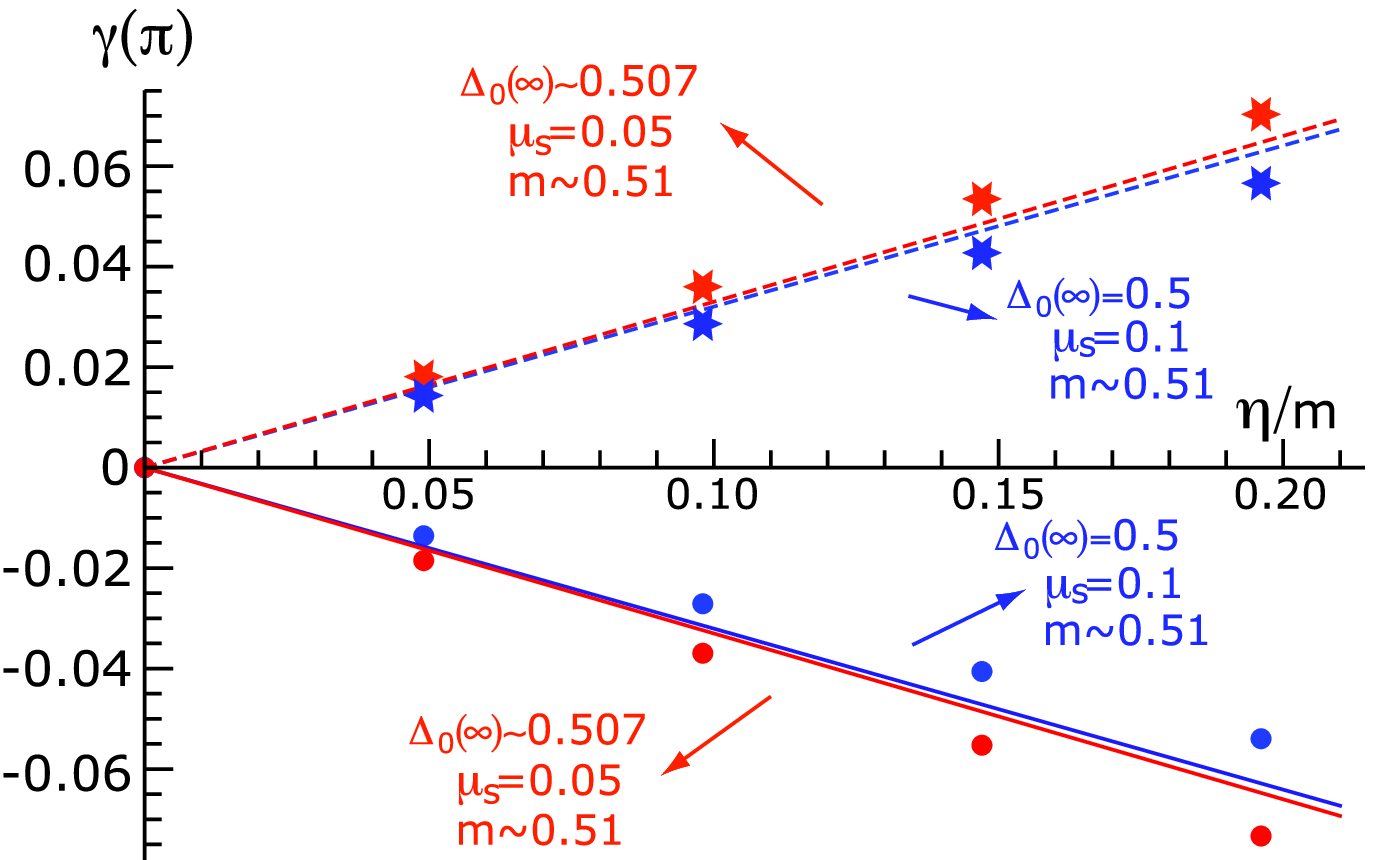}
\caption{
(Color online)
Berry phase in units of $\pi$
as a function of $\eta/m\ll1$ for fixed $m$
acquired during the exchange of two unscreened quasiparticles,
i.e., unit vortices in 
$\Delta(\boldsymbol{r},t)$ 
without the attachment of axial gauge half fluxes in 
$\boldsymbol{a}^{\ }_{5}(\boldsymbol{r},t)$.
Numerical computations 
along the lines outlined in Sec.%
~\ref{subsec: Dynamic calculation of the charge}
and
Appendix%
~\ref{appsec: Numerical Berry phase in the single-particle approximation}
were performed for
spinless fermions hopping on the square lattice
with lattice spacing $\mathfrak{a}$ and with a magnetic flux
of $\pi$ in units of the flux quantum $\phi^{\ }_{0}=hc/e$
threading each elementary plaquette
in the \textit{dynamic background} of a unit vortex in 
$\Delta(\boldsymbol{r},t)$
without the attachment of 
an axial gauge half flux in $\boldsymbol{a}^{\ }_{5}(\boldsymbol{r},t)$
and for a uniform value of $\eta$.
The square lattice is $72\times 72$ and the exchange path is
approximately circular with the radius $r=18.5$
in units of the lattice spacing. 
The following parameters were chosen: the hopping $t=1$,
$m=\sqrt{\Delta^{2}_{0}(\infty)+\mu^{2}_{\mathrm{s}}}\approx 0.51$ 
but with two different
value of $\mu^{\ }_{\mathrm{s}}=0.1$ and $0.025$.
Filled circles and solid lines represent the case when
the two quasiparticles carry the same unit vorticity.
Stars and dashed lines represent the case when 
the two quasiparticles carry the opposite unit vorticity.
Symbols are obtained numerically while the lines
are the predictions from Sec.~\ref{subsec; Unscreened vortices}.
        }
\label{fig: statistics a}
\end{figure}

The $\eta$ dependence of the Berry phase $\gamma$
with $m$ fixed is shown in Fig.~\ref{fig: statistics a} 
for different values of the uniform staggered chemical potentials
$\mu^{\ }_{\mathrm{s}}$. 
The magnitude of the Berry phase is seen to be independent
of whether the pair of quasiparticles have the same 
(filled circles) or opposite (star symbols) vorticities,
but it does depend on 
$\mu^{\ }_{\mathrm{s}}$,
i.e., on the induced fractional charge $Q$
given in Eq.~(\ref{eq: derivation of Q}). 
The $\eta$ dependence of $\gamma$ is linear, 
as predicted in Sec.~\ref{subsec; Unscreened vortices}, 
but with slopes deviating from the theoretical predictions,
i.e., Eq.~(\ref{eq:statistical-angle-eta-smaller-m}),
shown as the solid or dashed lines. 
The agreement between the Berry phase of the microscopic model
and Eq.~(\ref{eq:statistical-angle-eta-smaller-m})
is thus qualitatively but not quantitatively good. 

\begin{figure}
\includegraphics[angle=0,scale=0.55]{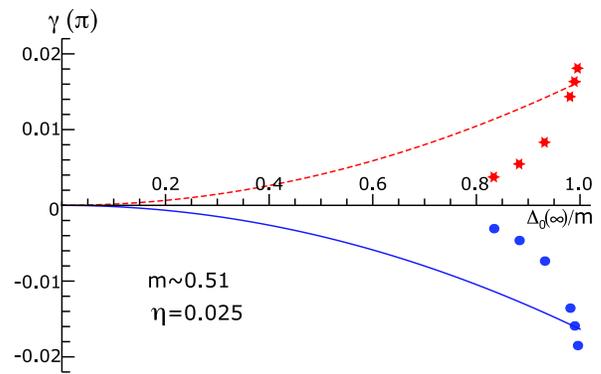}
\caption{
(Color online)
Berry phase in units of $\pi$
as a function of $\Delta^{\ }_{0}(\infty)/m$ for fixed 
$m$ and $\eta\ll m$ 
acquired during the exchange of two unscreened quasiparticles,
i.e., unit vortices in 
$\Delta(\boldsymbol{r},t)$ 
without the attachment of axial gauge half fluxes in 
$\boldsymbol{a}^{\ }_{5}(\boldsymbol{r},t)$.
Numerical computations 
along the lines outlined in Sec.%
~\ref{subsec: Dynamic calculation of the charge}
and
Appendix%
~\ref{appsec: Numerical Berry phase in the single-particle approximation}
were performed for
spinless fermions hopping on the square lattice
with lattice spacing $\mathfrak{a}$ and with a magnetic flux
of $\pi$ in units of the flux quantum $\phi^{\ }_{0}=hc/e$
threading each elementary plaquette
in the \textit{dynamic background} of a unit vortex in 
$\Delta(\boldsymbol{r},t)$
without the attachment of 
an axial gauge half flux in $\boldsymbol{a}^{\ }_{5}(\boldsymbol{r},t)$
and for a uniform value of $\eta$.
The square lattice is $72\times 72$ and the exchange path is
approximately circular with the radius $r=18.5$
in units of the lattice spacing. 
The following parameters were chosen: the hopping $t=1$,
$m=\sqrt{\Delta^{2}_{0}(\infty)+\mu^{2}_{\mathrm{s}}}=0.51$ 
and $\eta=0.025$.
Filled circles and solid lines represent the case when
the two quasiparticles carry the same unit vorticity.
Stars and dashed lines represent the case when 
the two quasiparticles carry the opposite unit vorticity.
Symbols are obtained numerically while the lines
are the predictions from Sec.~\ref{subsec; Unscreened vortices}.
        }
\label{fig: statistics b}
\end{figure}

The microscopic Berry phase $\gamma$ as a function of the ratio 
$\Delta/m$, which also parametrizes $Q(\Delta,\mu^{\ }_{\mathrm{s}})$, 
when $\eta=0.025$ is held fixed 
is shown in Fig.~\ref{fig: statistics b}
as filled circles when the quasiparticles carry the same vorticities
or as stars when the quasiparticles carry the opposite vorticities.
As expected, exchanging a pair of quasiparticles with equal unit
vorticities differs solely by a sign relative to
exchanging a pair of quasiparticles with opposite
unit vorticities. The lines (solid when the quasiparticles
have the same unit vorticity, dashed otherwise)
are given by
Eq.~(\ref{eq:statistical-angle-eta-smaller-m}). 
Evidently, the dependence on $Q\ll1/2$ of the
microscopic exchange statistics is not captured
by the field theory.

As discussed in Sec.~\ref{subsec:one-more-electron}, 
when adding (removing) one fermion to (from) half-filling, 
the Berry phase accumulated by a complete winding of 
quasiparticles of opposite unit vorticities
changes by $\pi$ for the case $\eta=0$. 
This extra phase is the response of the 
single-particle midgap states to varying 
the phase of $\Delta$ by $2\pi$. 
Numerically, this assertion is confirmed directly by computing 
the accumulated Berry phase and obtaining $\gamma=\pm \pi$ 
when filling or emptying one midgap state.

In summary, comparison of the microscopic Berry phase
accumulated by winding an unscreened quasiparticle around 
a static one with the field-theory computation of the
exchange statistic in Sec.~\ref{subsec; Unscreened vortices} 
shows that:
1)~The microscopic Berry phase $\gamma$ 
(and consequently the 
microscopic exchange statistical angle $\Theta=\gamma/2$) 
varies continuously as a function of $\eta$ and 
in a linear fashion for small $\eta$, 
in good agreement with the field-theory results.
2)~The slope $\gamma/\eta$ shows a monotonic dependence on the ratio 
$\Delta/m$, which is not in good quantitative agreement with the 
field-theory results. 
3)~ The magnitude $|\gamma|$ 
is independent of the relative sign of the quasiparticles vorticities.
This is expected for a vortex and its anti-vortex 
can annihilate. Consequently, winding a third vortex around 
a vortex anti-vortex pair must accumulate a vanishing Berry phase.
4)~Microscopic semion statistics $\Theta=\pm\pi/2$ 
is obtained when adding (removing) 
one fermion to (from) the half-filled system
in agreement with the prediction from the continuum theory.

\section{
Microscopic model
        }
\label{sec: Microscopic models}

We have seen in
Sec.~\ref{eq: Numerical calculation of the charge and Berry phase}
that
the fractional charge induced by an axial gauge half flux in
$\boldsymbol{a}^{\ }_{5}$ cannot be measured dynamically from the
Aharonov-Bohm phase inferred from the numerical computation of a Berry
phase.
This is so because the local axial gauge symmetry in the
continuum Hamiltonian~(\ref{eq: def quantum Hamiltonian}) is not
present in the lattice model used in
Sec.~\ref{eq: Numerical calculation of the charge and Berry phase}. Thus,
there is a dynamical contribution that is missing and that cannot be
captured by the simple models of one species of fermions hopping
either on the honeycomb or $\pi$-flux lattices used in 
Sec.~\ref{eq: Numerical calculation of the charge and Berry phase}.  For
the same reason, we could not obtain numerically the exchange
statistics in the case when the vortices are screened by the axial
gauge potential, since the exchange of the topological defects
necessarily acquires a dynamical contribution from
$\boldsymbol{a}^{\ }_{5}$.

We now construct
a lattice model sharing the \textit{same} local
U(1)$\times$U(1) symmetry and the \textit{same}
particle content as the dynamical theory
(\ref{eq: def Z}). The predictions for the
exchange statistics of screened vortices
done in Sec.~\ref{subsec: Screened vortices}
should be captured by this lattice model.
Unfortunately, we cannot verify this claim,
for the largest system sizes that we could 
treat numerically are of the order of the vortex core.

Consider a square lattice $\Lambda$ whose sites we denote with the
Latin letters $i,j,k$, and $l$.  We denote with
$\hat{1}\equiv\hat{\mathbf x},\hat{2}\equiv\hat {\mathbf y}$ the two
orthonormal vectors spanning the square lattice $\Lambda$ (and we will
index these two vectors as $\hat\mu=\hat{1},\hat{2}$, for
$\mu=1,2$). Links (or bonds) on the square lattice between
nearest-neighbor sites $i$ and $j$ are labeled by $\langle ij\rangle$
(or simply by $ij$ when used as an index to a field defined on the
links). We denote by $\Box^{\ }_{ijkl}$ the square plaquette with the 
corners $i$, $j$, $k$, and $l$.

We define four sets of operators. 
There are the bosonic operators $\hat{A}^{\ }_{ij}$ and 
$\hat{A}^{\ }_{5ij}$ living on the links of the square lattice $\Lambda$.
There are the bosonic operators $\hat{\phi}^{\ }_{i}$ 
and the fermionic operators $\hat{\psi}^{\ }_{i}$ living on the sites. 
The spinor-valued operator
$\hat{\psi}^{\ }_{i}$ has here four components on which the $4\times4$
matrices defined in Eqs.~(2.1d) and (2.1e) act.

These four sets of operators, together with their canonical conjugate
operators, satisfy the following relations:
\begin{subequations}
\label{eq:alg-AA5phipsi}
\begin{equation}
\begin{split}
&
\hat{A}^{\dag}_{kl}=
\hat{A}^{\ }_{kl}=
-\hat{A}^{\ }_{lk},
\quad
\hat{L}^{\dag}_{ij}=
\hat{L}^{\ }_{ij}=
-\hat{L}^{\ }_{ji},
\\
&
\left[
\hat{L}^{\ }_{ij},
\hat{A}^{\ }_{kl}
\right]=
-{i}\left(
\delta^{\ }_{ik}
\delta^{\ }_{jl}
-
\delta^{\ }_{il}
\delta^{\ }_{jk}
\right),
\end{split}
\end{equation}
\begin{equation}
\begin{split}
&
\hat{A}^{\dag}_{5kl}=
\hat{A}^{\ }_{5kl}=
-\hat{A}^{\ }_{5lk},
\quad
\hat{L}^{\dag}_{5ij}=
\hat{L}^{\ }_{5ij}=
-\hat{L}^{\ }_{5ji},
\\
&
\left[
\hat{L}^{\ }_{5ij},
\hat{A}^{\ }_{5kl}
\right]=
-{i}\left(
\delta^{\ }_{ik}
\delta^{\ }_{jl}
-
\delta^{\ }_{il}
\delta^{\ }_{jk}
\right),
\end{split}
\end{equation}
\begin{equation}
\begin{split}
&\hat{\phi}^{\dag}_{j}=
\hat{\phi}^{\ }_{j},
\quad
\hat{\Pi}^{\dag}_{i}=\hat{\Pi}^{\ }_{i},
\quad
\left[
\hat{\Pi}^{\ }_{i},
\hat{\phi}^{\ }_{j}
\right]=
-{i}\,
\delta^{\ }_{ij},
\end{split}
\end{equation}
and, finally,
\begin{equation}
\begin{split}
&
\left\{
\hat{\psi}^{\   }_{i},
\hat{\psi}^{\dag}_{j}
\right\}=
\openone^{\ }_{4}\;
\delta^{\ }_{ij},
\qquad
\left\{
\hat{\psi}^{\dag}_{i},
\hat{\psi}^{\dag}_{j}
\right\}=
\left\{
\hat{\psi}^{\   }_{j},
\hat{\psi}^{\   }_{i}
\right\}=0,
\label{eq: mathcal{F}psi}
\end{split}
\end{equation}
with the \textit{equal-time global constraint}
(half-filling constraint)
\begin{equation}
|\Lambda|^{-1}
\sum_{i\in\Lambda}
\hat{\psi}^{\dag}_{i}\hat{\psi}^{\   }_{i}=2.
\label{eq: half-filling constraint}
\end{equation}
\end{subequations}
(Since we are working with four flavors of fermions, half-filling
means average 2 particles per each site.)

We define the lattice model by
the quantum Hamiltonian 
\begin{subequations}
\label{eq: def lattice model with U(1xU(1) sym} 
\begin{equation}
\hat{H}:=
\hat{H}^{\ }_{g}
+
\hat{H}^{\ }_{g^{\ }_{5}}
+
\hat{H}^{\ }_{J}
+
\hat{H}^{\ }_{t}
+
\hat{H}^{\ }_{t'}
+
\hat{H}^{\ }_{m}.
\end{equation}
Here,
\begin{equation}
\hat{H}^{\ }_{g}:=
\frac{g^{2}}{2}
\sum_{\langle ij\rangle}
\hat{L}^{2}_{ij}
-
\frac{1}{g^{2}}
\sum_{\Box^{\ }_{ijkl}}
\mathrm{Re}\,
e^{
{i}
\left(
\hat{A}^{\ }_{ij}
+
\hat{A}^{\ }_{jk}
+
\hat{A}^{\ }_{kl}
+
\hat{A}^{\ }_{li}
\right)
  }
\end{equation}
describes a U(1) lattice gauge theory with
gauge coupling $g^{2}$, 
\begin{equation}
\hat{H}^{\ }_{g^{\ }_{5}}:=
\frac{g^{2}_{5}}{2}
\sum_{\langle ij\rangle}
\hat{L}^{2}_{5ij}
-
\frac{1}{g^{2}_{5}}
\sum_{\Box^{\ }_{ijkl}}
\mathrm{Re}\,
e^{
{i}
\left(
\hat{A}^{\ }_{5 ij}
+
\hat{A}^{\ }_{5 jk}
+
\hat{A}^{\ }_{5 kl}
+
\hat{A}^{\ }_{5 li}
\right)
  }
\end{equation}
describes another U(1) lattice gauge theory with
gauge coupling $g^{2}_{5}$, 
\begin{equation}
\hat{H}^{\ }_{J}:=
\frac{J^{2}}{2}
\sum_{i\in\Lambda}
\hat{\Pi}^{2}_{i}
-
\frac{1}{J^{2}}
\sum_{\langle ij\rangle}
\left(
e^{
+
{i}
\left(
\hat{\phi}^{\ }_{i}
-
\hat{\phi}^{\ }_{j}
\right)
+
2
{i}
\hat{A}^{\ }_{5 ij}
  }
+
\hbox{H.c.}
\right)
\end{equation}
describes a quantum rotor (XY) model with coupling
$J^{2}$, and
\begin{equation}
\begin{split}
\hat{H}^{\ }_{t}:=&\,
{i}
t
\sum_{i\in\Lambda}
\sum_{{\mu}=1,2}
\hat{\psi}^{\dag}_{i}
\alpha^{\ }_{\mu}
e^{
{i}
\hat{A}^{\ }_{i(i+\hat{\mu})}
+
{i}
\gamma^{\ }_{5}\;
\hat{A}^{\ }_{5i(i+\hat{\mu})}
  }
\;\hat{\psi}^{\ }_{(i+\hat{\mu})}
\\
&
+
\hbox{H.c.}
\end{split}
\end{equation}
describes the nearest-neighbor hopping with the real-valued amplitude
$t$ of 4 independent fermions per site. 
So far, there are 4 non-equivalent Dirac points at half-filling
which are located at
$\boldsymbol{k}=(0,0)$, $(0,\pi)$, $(\pi,0)$, and $(\pi,\pi)$. 
This is why we have added the term
\begin{equation}
\begin{split}
&
\hat{H}^{\ }_{t^{\prime}}:=
t^{\prime}
\sum_{i\in\Lambda}
\left[
\hat{\psi}^{\dag}_{i}\;
4
R\;
\hat{\psi}^{\ }_{i}
\vphantom{\sum_{i\in\Lambda}}
\right.
\\
&
\,
\left.
-
\sum_{{\mu}=1,2}
\left(
\hat{\psi}^{\dag}_{i}\;
R\;
e^{
{i}
\hat{A}^{\ }_{i(i+\hat{\mu})}
+
{i}
\gamma^{\ }_{5}
\hat{A}^{\ }_{5i(i+\hat{\mu})}
  }
\;
\hat{\psi}^{\ }_{(i+\hat{\mu})}
+
\hbox{H.c.}
\right)
\right]
\end{split}
\label{eq: Wilson t' mass}
\end{equation}
that opens a gap of order $t'$ at the points
$\boldsymbol{k}=(0,\pi),(\pi,0),(\pi,\pi)$, 
thus leaving $\boldsymbol{k}=(0,0)$ as the sole Dirac point.
This scheme is precisely Wilson's
procedure used to overcome the doubling problem in lattice gauge
theories.~\cite{Wilson77} 
An important comment is in order, however. One reason why 
this prescription is not fully satisfying
in lattice gauge theories is that any mismatch between the
first and second terms of Eq.~(\ref{eq: Wilson t' mass}) leads to
a gap at $\boldsymbol{k}=(0,0)$ as well, i.e., 
fine-tuning is needed to achieve the correct particle content.
Here, this is fine because we are interested in systems where 
there is such a gap. Notice in that regard that the
gap at $\boldsymbol{k}=(0,0)$ that arises from a small mismatch between
these two terms (a small fraction of $t'$) is much smaller than the
one at the edges of the Brillouin zone (order $t'$). Indeed, such a
term due to a mismatch is actually part of the final term that we
consider in the Hamiltonian, namely
\begin{equation}
\hat{H}^{\ }_{m}:=
\sum_{i\in\Lambda}
\hat{\psi}^{\dag}_{i}
\left(
\mu^{\ }_{\mathrm{s}}
R
+
\Delta^{\ }_{\mathrm{k}}
\beta
e^{
{i}
\gamma^{\ }_{5}
\hat{\phi}^{\ }_{i}
  }
+
{i}
\eta
\alpha^{\ }_{1}
\alpha^{\ }_{2}
\right)
\hat{\psi}^{\ }_{i}.
\end{equation}
\end{subequations}
This contribution does indeed
open a gap at the remaining Dirac point at 
$\boldsymbol{k}=(0,0)$.

For any smooth and static boson background,
the continuum limit of
Hamiltonian~(\ref{eq: def lattice model with U(1xU(1) sym})
upon linearization of the fermion spectrum 
at the two non-equivalent
Dirac points at half-filling is given by
Eq.~(2.1), as is also the case with the fermion spectrum
of graphene restricted to spinless fermions hopping with
sufficiently smooth modulations of the hopping amplitudes.

Contrary to graphene for spinless fermions,
Hamiltonian~(\ref{eq: def lattice model with U(1xU(1) sym})
is invariant under the local
U(1)$\times$U(1)
gauge transformation
\begin{subequations}
\label{eq: U(1)xU(1) local gauge symmetry}
\begin{equation}
\begin{split}
&
\hat{L}^{\ }_{ij}
\to
\hat{L}^{\ }_{ij},
\qquad
\hat{A}^{\ }_{ij}
\to
\hat{A}^{\ }_{ij}
-
\left(
\chi^{\ }_{i}
-
\chi^{\ }_{j}
\right),
\\
&
\hat{L}^{\ }_{5ij}
\to
\hat{L}^{\ }_{5ij},
\qquad
\hat{A}^{\ }_{5 ij}
\to
\hat{A}^{\ }_{5 ij}
-
\left(
\xi^{\ }_{i}
-
\xi^{\ }_{j}
\right),
\\
&
\hat{\Pi}^{\ }_{i}
\to
\hat{\Pi}^{\ }_{i},
\qquad
\hat{\phi}^{\ }_{i}
\to
\hat{\phi}^{\ }_{i}+2\xi^{\ }_{i},
\\
&
\hat{\psi}^{\dag}_{i}
\to
\hat{\psi}^{\dag}_{i} 
e^{
+
{i}\chi^{\ }_{i}
+
{i}
\gamma^{\ }_{5}
\xi^{\ }_{i}
  },
\qquad
\hat{\psi}^{\ }_{j}
\to
\psi^{\ }_{j} 
e^{
-
{i}\chi^{\ }_{j}
-
{i}
\gamma^{\ }_{5}
\xi^{\ }_{j}
  },
\end{split}
\end{equation}
generated by 
\begin{equation}
\hat{H}\to
\hat{G}(\chi,\xi)\;\hat{H}\;\hat{G}^{-1}(\chi,\xi)
\end{equation}
with
\begin{equation}
\begin{split}
\hat{G}(\chi,\xi):=&\,
\prod\limits_{i\in\Lambda}
\exp
\left[
{i}
\left(
\hat{\psi}^{\dag}_{i}
\hat{\psi}^{\   }_{i}
+
\sum\limits_{{\mu}=1,2}
\hat{L}^{\ }_{i(i+\hat{\mu})}
\right)
\chi^{\ }_{i}
\right.
\\
&\,
\left.
+
{i}
\left(
\hat{\psi}^{\dag}_{i}
\gamma^{\ }_{5}
\hat{\psi}^{\   }_{i}
+
2
\hat{\Pi}^{\ }_{i}
+
\sum\limits_{{\mu}=1,2}
\hat{L}^{\ }_{5i(i+\hat{\mu})}
\right)
\xi^{\ }_{i}
\right].
\end{split}
\end{equation}
\end{subequations}
where $\chi^{\ }_{i}$ and $\xi^{\ }_{i}$
are arbitrary real-valued numbers.

The physical subspace is the set of gauge invariant states, i.e.,
states that are tensor products of states in the Fock space generated
by the algebra Eqs.~(\ref{eq:alg-AA5phipsi}),
\begin{subequations}
\label{eq: def gauge inv states}
\begin{equation}
\begin{split}
|\Psi\rangle\equiv&\,
|\Psi^{\ }_{A}\rangle
\otimes
|\Psi^{\ }_{A^{\ }_{5}}\rangle
\otimes
|\Psi^{\ }_{\phi}\rangle
\otimes
|\Psi^{\ }_{\psi}\rangle
\end{split}
\end{equation}
such that Gauss law holds globally,
\begin{equation}
\hat{G}^{-1}(\chi,\xi)\;
|\Psi\rangle=
|\Psi\rangle
\end{equation}
for all real-valued function
$\chi$ and $\xi$,
or, equivalently, locally
\begin{equation}
\begin{split}
&
0=
\left(
\hat{L}^{\ }_{i(i+\hat{1})}
-
\hat{L}^{\ }_{i(i-\hat{1})}
+
\hat{L}^{\ }_{i(i+\hat{2})}
-
\hat{L}^{\ }_{i(i-\hat{2})}
+
\hat{\psi}^{\dag}_{i}
\hat{\psi}^{\   }_{i}
\right)
|\Psi\rangle,
\\
&
0=
\left(
\hat{L}^{\ }_{5i(i+\hat{1})}
-
\hat{L}^{\ }_{5i(i-\hat{1})}
+
\hat{L}^{\ }_{5i(i+\hat{2})}
-
\hat{L}^{\ }_{5i(i-\hat{2})}
\right.
\\
&\hphantom{0=}
\left.
+
\hat{\psi}^{\dag}_{i}
\gamma^{\ }_{5}
\hat{\psi}^{\   }_{i}
+
2
\hat{\Pi}^{\ }_{i}
\right)
|\Psi\rangle,
\end{split}
\end{equation}
\end{subequations}
for any $i\in\Lambda$.

We denote by
$|\Psi^{\ }_{i,j}\rangle$
a gauge invariant state~(\ref{eq: def gauge inv states}) 
with two fractional
charges localized around sites $i$ and $j$, respectively.
The statistical phase $\Theta$
induced by the physical process by
which two fractional charges are exchanged is given by
the difference between two Berry phases,~\cite{Levin03}
\begin{equation}
\begin{split}
\Theta:=&\,
\frac{1}{2}
\mathrm{arg}\,
\prod_{i^{\ }_{\iota}\in\mathcal{P}}^{j\subset\mathcal{P}}
\left\langle
\Psi^{\ }_{i^{\ }_{\iota+1},j}
\right|
\hat{H}
\left|
\Psi^{\ }_{i^{\ }_{\iota},j}
\right\rangle
\\
&\,
-
\frac{1}{2}
\mathrm{arg}\,
\prod_{i^{\ }_{\iota}\in\mathcal{P}}^{j\subset\bar{\mathcal{P}}}
\left\langle
\Psi^{\ }_{i^{\ }_{\iota+1},j}
\right|
\hat{H}
\left|
\Psi^{\ }_{i^{\ }_{\iota},j}
\right\rangle.
\end{split}
\label{eq: U(1)xU(1) latice Berry phase}
\end{equation}
For both Berry phases, 
one fractional charge hops along
the closed path $\mathcal{P}=\{i^{\ }_{\iota}\}$,
while the other fractional charge is static.
For the former Berry phase, $j$ is located inside the area
bounded by $\mathcal{P}$, a choice that we denote by
$j\subset\mathcal{P}$.
For the latter Berry phase, $j$ is located outside the area
bounded by $\mathcal{P}$, a choice that we denote by
$j\subset\bar{\mathcal{P}}$.

The dimensionality of the gauge-invariant Hilbert space
scales with the dimensionality of the fermionic Hilbert space%
~(\ref{eq: mathcal{F}psi}),
which itself scales exponentially fast with the number of
sites. Given the half-filling constraint%
~(\ref{eq: half-filling constraint}), 
this limits the numerical evaluation of the right-hand side of%
~(\ref{eq: U(1)xU(1) latice Berry phase}) 
to lattices with linear dimensions of the order of 
the core size $1/m$ of the defects,
i.e., on distances much too short for the right-hand side
of Eq.~(\ref{eq: U(1)xU(1) latice Berry phase}) 
to be interpreted as the statistical angle of point-like
quasiparticles.

If we are willing to give up the local U(1)$\times$U(1)
gauge invariance%
~(\ref{eq: U(1)xU(1) local gauge symmetry}), 
i.e., the strongly correlated nature
of the problem, we can compute the contribution to the 
statistical phase arising from the fermion hopping.
Indeed, the problem then reduces to a single-particle one
for which the dimensionality of the relevant Hilbert spaces 
only scales linearly with the number of sites.
We stress that this contribution alone violates
the local U(1)$\times$U(1) gauge invariance.

\begin{widetext}
\begin{table*}
\caption{
\label{tab: 36 masses} 
The 36 mass matrices with particle-hole symmetry (PHS),
see Eq.~(\ref{eq: def PHS}), 
for the massless Dirac Hamiltonian $\mathcal{K}^{\ }_{0}$
from Eq.~(\ref{eq: def K0})
are of the form%
~(\ref{eq: 256 matrices})
and anticommute with $\mathcal{K}^{\ }_{0}$.
Each mass matrix can be assigned an order parameter for the
underlying microscopic model, here graphene 
or the square lattice with $\pi$-flux phase.
The latin subindex of the order parameter's name corresponds to the
preferred quantization axis in SU(2) spin space. The pair of
numeral subindices 02 and 32 are used to distinguish 
the two unit vectors spanning two-dimensional space.
Each mass matrix preserves or breaks
time-reversal symmetry (TRS), see Eq.~(\ref{eq: def TRS}),
spin-rotation symmetry (SRS), see Eq.~(\ref{eq: def SRS}),
and sublattice symmetry (SLS), see Eq.~(\ref{eq: def SLS}).
To any of the 36 mass matrices corresponds a ``partner'' mass matrix
obtained through the involutive transformation%
~(\ref{eq: def partner mass matrix}) denoted $C$.
        }
\begin{ruledtabular}
\begin{tabular}{llllllll}
Mass matrix  & Order parameter & TRS & SRS & SLS & Partner by $C$ & Order parameter by $C$& $C$ invariant\\
&&&&&&\\ 
$X^{\ }_{3010}$& {ReVBS} & {True} & {True} & {True} &$X^{\ }_{3010}$& {ReVBS}           &{True} \\
$X^{\ }_{0020}$& {ImVBS} & {True} & {True} & {True} &$X^{\ }_{0020}$& {ImVBS}           &{True} \\
$X^{\ }_{3033}$& {CDW}   & {True} & {True} & {False}&$X^{\ }_{3333}$& {N\'eel}$^{\ }_{z}$ &{False}\\
&&&&&&\\ 
$X^{\ }_{3003}$& {QHE}   & {False}& {True} & {False}&$X^{\ }_{3003}$& {QHE}             &{True} \\
&&&&&&\\ 
$X^{\ }_{3110}$& {ReVBS}$^{\ }_{x}$& {False}& {False}& {True} &$X^{\ }_{2132}$& {ImTSC}$^{\ }_{32z}$ &{False}\\
$X^{\ }_{0210}$& {ReVBS}$^{\ }_{y}$& {False}& {False}& {True} &$X^{\ }_{1132}$& {ReTSC}$^{\ }_{32z}$ &{False}\\
$X^{\ }_{3310}$& {ReVBS}$^{\ }_{z}$& {False}& {False}& {True} &$X^{\ }_{3310}$& {ReVBS}$^{\ }_{z}$   &{True} \\
&&&&&&\\ 
$X^{\ }_{0120}$& {ImVBS}$^{\ }_{x}$& {False}& {False}& {True} &$X^{\ }_{1102}$& {ReTSC}$^{\ }_{02z}$ &{False}\\
$X^{\ }_{3220}$& {ImVBS}$^{\ }_{y}$& {False}& {False}& {True} &$X^{\ }_{2102}$& {ImTSC}$^{\ }_{02z}$ &{False}\\
$X^{\ }_{0320}$& {ImVBS}$^{\ }_{z}$& {False}& {False}& {True} &$X^{\ }_{0320}$& {ImVBS}$^{\ }_{z}$   &{True} \\
&&&&&&\\ 
$X^{\ }_{3103}$& {QSHE}$^{\ }_{x}$& {True} & {False}& {False}&$X^{\ }_{2121}$& {ImTSC}$^{\ }_{z}$ &{False}\\
$X^{\ }_{0203}$& {QSHE}$^{\ }_{y}$& {True} & {False}& {False}&$X^{\ }_{1121}$& {ReTSC}$^{\ }_{z}$ &{False}\\
$X^{\ }_{3303}$& {QSHE}$^{\ }_{z}$& {True} & {False}& {False}&$X^{\ }_{3303}$& {QSHE}$^{\ }_{z} $ &{True} \\
&&&&&&\\ 
$X^{\ }_{3133}$& {N\'eel}$^{\ }_{x}$& {False}& {False}& {False}&$X^{\ }_{2211}$& {ReSSC} &{False}\\
$X^{\ }_{0233}$& {N\'eel}$^{\ }_{y}$& {False}& {False}& {False}&$X^{\ }_{1211}$& {ImSSC} &{False}\\
$X^{\ }_{3333}$& {N\'eel}$^{\ }_{z}$& {False}& {False}& {False}&$X^{\ }_{3033}$& {CDW}   &{False}\\
&&&&&&\\ 
$X^{\ }_{2211}$& {ReSSC} & {True} & {True} & {False}&$X^{\ }_{3133}$& {N\'eel}$^{\ }_{x}$ &{False}\\
$X^{\ }_{1211}$& {ImSSC} & {False}& {True} & {False}&$X^{\ }_{0233}$& {N\'eel}$^{\ }_{y}$ &{False}\\
&&&&&&\\ 
$X^{\ }_{1002}$& {ReTSC}$^{\ }_{02y}$& {True} & {False}& {True} &$X^{\ }_{1002}$& {ReTSC}$^{\ }_{02y}$ &{True} \\
$X^{\ }_{2002}$& {ImTSC}$^{\ }_{02y}$& {False}& {False}& {True} &$X^{\ }_{2302}$& {ImTSC}$^{\ }_{02x}$ &{False}\\
$X^{\ }_{1102}$& {ReTSC}$^{\ }_{02z}$& {False}& {False}& {True} &$X^{\ }_{0120}$& {ImVBS}$^{\ }_{x}$   &{False}\\
$X^{\ }_{2102}$& {ImTSC}$^{\ }_{02z}$& {True} & {False}& {True} &$X^{\ }_{3220}$& {ImVBS}$^{\ }_{y}$   &{False}\\
$X^{\ }_{1302}$& {ReTSC}$^{\ }_{02x}$& {False}& {False}& {True} &$X^{\ }_{1302}$& {ReTSC}$^{\ }_{02x}$ &{True} \\
$X^{\ }_{2302}$& {ImTSC}$^{\ }_{02x}$& {True} & {False}& {True} &$X^{\ }_{2002}$& {ImTSC}$^{\ }_{02y}$ &{False}\\ 
&&&&&&\\ 
$X^{\ }_{1032}$& {ReTSC}$^{\ }_{32y}$& {False}& {False}& {True} &$X^{\ }_{1332}$& {ReTSC}$^{\ }_{32x}$ &{False}\\
$X^{\ }_{2032}$& {ImTSC}$^{\ }_{32y}$& {True} & {False}& {True} &$X^{\ }_{2032}$& {ImTSC}$^{\ }_{32y}$ &{True} \\
$X^{\ }_{1132}$& {ReTSC}$^{\ }_{32z}$& {True} & {False}& {True} &$X^{\ }_{0210}$& {ReVBS}$^{\ }_{y}$   &{False}\\
$X^{\ }_{2132}$& {ImTSC}$^{\ }_{32z}$& {False}& {False}& {True} &$X^{\ }_{3110}$& {ReVBS}$^{\ }_{x}$   &{False}\\
$X^{\ }_{1332}$& {ReTSC}$^{\ }_{32x}$& {True} & {False}& {True} &$X^{\ }_{1032}$& {ReTSC}$^{\ }_{32y}$ &{False}\\
$X^{\ }_{2332}$& {ImTSC}$^{\ }_{32x}$& {False}& {False}& {True} &$X^{\ }_{2332}$& {ImTSC}$^{\ }_{32x}$ &{True} \\
&&&&&&\\ 
$X^{\ }_{1021}$& {ReTSC}$^{\ }_{y}$& {True} & {False}& {False}&$X^{\ }_{1321}$& {ReTSC}$^{\ }_{x}$     &{False}\\
$X^{\ }_{2021}$& {ImTSC}$^{\ }_{y}$& {False}& {False}& {False}&$X^{\ }_{2021}$& {ImTSC}$^{\ }_{y}$     &{True} \\
$X^{\ }_{1121}$& {ReTSC}$^{\ }_{z}$& {False}& {False}& {False}&$X^{\ }_{0203}$& {QSHE}$^{\ }_{y}$      &{False}\\
$X^{\ }_{2121}$& {ImTSC}$^{\ }_{z}$& {True} & {False}& {False}&$X^{\ }_{3103}$& {QSHE}$^{\ }_{x}$      &{False}\\
$X^{\ }_{1321}$& {ReTSC}$^{\ }_{x}$& {False}& {False}& {False}&$X^{\ }_{1021}$& {ReTSC}$^{\ }_{y}$     &{False}\\
$X^{\ }_{2321}$& {ImTSC}$^{\ }_{x}$& {True} & {False}& {False}&$X^{\ }_{2321}$& {ImTSC}$^{\ }_{x}$     &{True} 
\end{tabular}
\end{ruledtabular}
\end{table*}
\end{widetext}

\section{
More species of fermions -- classification of all masses in
        graphene and $\pi$-flux phase }
\label{sec: Reinstating the electron spin}

So far we have ignored the spin-1/2 quantum number of
electrons. If so, in the linear approximation%
~(2.1) of graphene restricted to spinless fermions say, 
$\hat{H}^{\ }_{\mathrm{scalar}}$
exhausts all possible symmetry-breaking 
instabilities with a local order parameters
compatible with charge conservation.
The local order parameter for a charge-density wave
that breaks the sublattice symmetry but preserves the
time-reversal symmetry is the real-valued
order parameter 
$\mu^{\ }_{\mathrm{s}}(\boldsymbol{r})$ 
(introduced by Semenoff for graphene in Ref.~\onlinecite{Semenoff84}).
The local order parameter for a bond-density wave instability
that preserves the sublattice and time-reversal 
symmetries is the complex-valued order parameter 
$\Delta(\boldsymbol{r})$ 
(the U(1) Kekul\'e order parameter
introduced by Hou {\it et al.}\ for graphene in Ref.~\onlinecite{Hou07}).
The local order parameter for a bond-density wave instability
that breaks the sublattice and time-reversal 
symmetries is the real-valued order parameter 
$\eta(\boldsymbol{r})$ 
(introduced by Haldane for graphene in Ref.~\onlinecite{Haldane88}).

If we reinstate spin-1/2 in the most naive way and consider two
independent copies of the model in Eq.~(\ref{eq: def quantum Hamiltonian}), 
then the results we found for spinless electrons are modified in a trivial way.
Defects bind equal values for the fractional charge for both
species, up and down spin, thereby doubling the total induced fermionic
charge (which is to be associated with a spin-singlet state). The same
happens to the exchange statistical angle. It is simply doubled
with respect to the results in
Sec.~\ref{sec: Fractional statistical angle}.

However, if spin is not a good quantum number, a larger number of
instabilities can occur and more masses or order parameters 
(other than 
$\mathrm{Re}\,\Delta$,
$\mathrm{Im}\,\Delta$,
$\mu^{\ }_{\mathrm{s}}$, 
and $\eta$)
need to be taken into account. 
Thus, one must consider more generic Dirac Hamiltonians
and study all their allowed masses. Topological defects in these
order parameters could bind states, whose (fractional) charge
and statistics would depend on the effective action (as function of all the
mass order parameters and the $a^{\ }_{\mu}$ and $a^{\ }_{5 \mu}$
fields) that is obtained upon integrating all the species of
fermions. This effective action  would be the extension of the one derived in
Sec.~\ref{sec: Derivative expansion} for the case of the four
order parameters
($\mathrm{Re}\,\Delta$,
$\mathrm{Im}\,\Delta$,
$\mu^{\ }_{\mathrm{s}}$, 
and $\eta$).

We do not fully carry this program in this paper. 
Nonetheless, we classify all these
masses according to the microscopic symmetries. 

This classification applies as well to the microscopic model
of Sec.~\ref{sec: Microscopic models}. There, we chose a specific way
to add Wilson masses [see Eq.~(\ref{eq: Wilson t' mass})]
to selectively get rid of all but 2 Dirac points
in order to recover in the long-wavelength limit Hamiltonian%
~(\ref{eq: def quantum Hamiltonian}). The set of all
(64) Wilson masses can also classified as we do below.

\begin{widetext}
\begin{table*}
\caption{
\label{tab: 56 5-tuplets} 
Enumeration of the 56 distinct 5-tuplets
of maximally pairwise anticommuting PHS
$X^{\ }_{\mu^{\ }_{1}\mu^{\ }_{2}\mu^{\ }_{3}\mu^{\ }_{4}}$.
The 56 5-tuplets are broken into 28 pairs related
by the operation of $C$ conjugation%
~(\ref{eq: def partner mass matrix}).
        }
\begin{ruledtabular}
\begin{tabular}{ll}
5-tuplet & Partner 5-tuplet by $C$ conjugation\\
&\\
$\left\{
\text{ReVBS},
\text{ImVBS},
\text{ReSSC},
\text{ImSSC},
\text{CDW}
\right\}$  
&
$\left\{
\text{ReVBS},
\text{ImVBS},
\text{N\'eel}^{\ }_x,
\text{N\'eel}^{\ }_y,
\text{N\'eel}^{\ }_z
\right\}$
\\&\\
$\left\{
\text{ImVBS},
\text{CDW},
\text{ReVBS}^{\ }_{x},
\text{ReVBS}^{\ }_{y},
\text{ReVBS}^{\ }_{z}
\right\}$ 
&
$\left\{
\text{ImVBS},
\text{N\'eel}^{\ }_z,
\text{ImTSC}^{\ }_{32z},
\text{ReTSC}^{\ }_{32z},
\text{ReVBS}^{\ }_{z}
\right\}$ 
\\
$\left\{
\text{ReVBS},
\text{CDW},
\text{ImVBS}^{\ }_{x},
\text{ImVBS}^{\ }_{y},
\text{ImVBS}^{\ }_{z}
\right\}$ 
&
$\left\{
\text{ReVBS},
\text{N\'eel}^{\ }_z,
\text{ReTSC}^{\ }_{02z},
\text{ImTSC}^{\ }_{02z},
\text{ImVBS}^{\ }_{z}
\right\}$ 
\\
$\left\{
\text{ReSSC},
\text{ImSSC},
\text{QSHE}^{\ }_x,
\text{QSHE}^{\ }_y,
\text{QSHE}^{\ }_z
\right\}$ 
&
$\left\{
\text{N\'eel}^{\ }_x,
\text{N\'eel}^{\ }_y,
\text{ImTSC}^{\ }_z,
\text{ReTSC}^{\ }_z,
\text{QSHE}^{\ }_z
\right\}$ 
\\&\\
$\left\{
\text{ReVBS},
\text{ReSSC},
\text{ReTSC}^{\ }_{02x},
\text{ImTSC}^{\ }_{02y},
\text{ReTSC}^{\ }_{02z}
\right\}$ 
&
$\left\{
\text{ReVBS},
\text{N\'eel}^{\ }_x,
\text{ReTSC}^{\ }_{02x},
\text{ImTSC}^{\ }_{02x},
\text{ImVBS}^{\ }_{x}
\right\}$ 
\\
$\left\{
\text{ReVBS},
\text{ImSSC},
\text{ImTSC}^{\ }_{02x},
\text{ReTSC}^{\ }_{02y},
\text{ImTSC}^{\ }_{02z}
\right\}$ 
&
$\left\{
\text{ReVBS},
\text{N\'eel}^{\ }_y,
\text{ImTSC}^{\ }_{02y},
\text{ReTSC}^{\ }_{02y},
\text{ImVBS}^{\ }_{y}
\right\}$ 
\\
$\left\{
\text{ImVBS},
\text{ImSSC},
\text{ReTSC}^{\ }_{32x},
\text{ImTSC}^{\ }_{32y},
\text{ReTSC}^{\ }_{32z}
\right\}$ 
&
$\left\{
\text{ImVBS},
\text{N\'eel}^{\ }_y,
\text{ReTSC}^{\ }_{32y},
\text{ImTSC}^{\ }_{32y},
\text{ReVBS}^{\ }_{y}
\right\}$ 
\\
$\left\{
\text{ImVBS},
\text{ReSSC},
\text{ImTSC}^{\ }_{32x},
\text{ReTSC}^{\ }_{32y},
\text{ImTSC}^{\ }_{32z}
\right\}$ 
&
$\left\{
\text{ImVBS},
\text{N\'eel}^{\ }_x,
\text{ImTSC}^{\ }_{32x},
\text{ReTSC}^{\ }_{32x},
\text{ReVBS}^{\ }_{x}
\right\}$
\\
$\left\{
\text{CDW},
\text{ImSSC},
\text{ImTSC}^{\ }_x,
\text{ReTSC}^{\ }_y,
\text{ImTSC}^{\ }_z
\right\}$
&
$\left\{
\text{N\'eel}^{\ }_z,
\text{N\'eel}^{\ }_y,
\text{ImTSC}^{\ }_x,
\text{ReTSC}^{\ }_x,
\text{QSHE}^{\ }_x
\right\}$  
\\
$\left\{
\text{CDW},
\text{ReSSC},
\text{ReTSC}^{\ }_x,
\text{ImTSC}^{\ }_y,
\text{ReTSC}^{\ }_z
\right\}$ 
&
$\left\{
\text{N\'eel}^{\ }_z,
\text{N\'eel}^{\ }_x,
\text{ReTSC}^{\ }_y,
\text{ImTSC}^{\ }_y,
\text{QSHE}^{\ }_y
\right\}$  
\\&\\
$\left\{
\text{ImVBS}^{\ }_{x},
\text{QSHE}^{\ }_y,
\text{ImVBS}^{\ }_{z},
\text{ReTSC}^{\ }_{32y},
\text{ImTSC}^{\ }_{32y}
\right\}$ 
&
$\left\{
\text{ReTSC}^{\ }_{02z},
\text{ReTSC}^{\ }_z,
\text{ImVBS}^{\ }_{z},
\text{ReTSC}^{\ }_{32x},
\text{ImTSC}^{\ }_{32y}
\right\}$ 
\\
$\left\{
\text{ImVBS}^{\ }_{x},
\text{QSHE}^{\ }_y,
\text{ReVBS}^{\ }_{x},
\text{N\'eel}^{\ }_x,
\text{QSHE}^{\ }_z
\right\}$ 
&
$\left\{
\text{ReTSC}^{\ }_{02z},
\text{ReTSC}^{\ }_z,
\text{ImTSC}^{\ }_{32z},
\text{ReSSC},
\text{QSHE}^{\ }_z
\right\}$ 
\\
$\left\{
\text{ImVBS}^{\ }_{x},
\text{ReTSC}^{\ }_{32y},
\text{ImTSC}^{\ }_{32z},
\text{ImTSC}^{\ }_{02x},
\text{ImTSC}^{\ }_x
\right\}$ 
&
$\left\{
\text{ReTSC}^{\ }_{02z},
\text{ReTSC}^{\ }_{32x},
\text{ReVBS}^{\ }_{x},
\text{ImTSC}^{\ }_{02y},
\text{ImTSC}^{\ }_x
\right\}$ 
\\
$\left\{
\text{ImVBS}^{\ }_{x},
\text{ReTSC}^{\ }_{32z},
\text{ReTSC}^{\ }_{02x},
\text{ReTSC}^{\ }_x,
\text{ImTSC}^{\ }_{32y}
\right\}$ 
&
$\left\{
\text{ReTSC}^{\ }_{02z},
\text{ReVBS}^{\ }_{y},
\text{ReTSC}^{\ }_{02x},
\text{ReTSC}^{\ }_y,
\text{ImTSC}^{\ }_{32y}
\right\}$ 
\\
$\left\{
\text{ImVBS}^{\ }_{x},
\text{ReTSC}^{\ }_{32z},
\text{ImTSC}^{\ }_{32z},
\text{ImVBS}^{\ }_{y},
\text{QSHE}^{\ }_z
\right\}$ 
&
$\left\{
\text{ReTSC}^{\ }_{02z},
\text{ReVBS}^{\ }_{y},
\text{ReVBS}^{\ }_{x},
\text{ImTSC}^{\ }_{02z},
\text{QSHE}^{\ }_z
\right\}$ 
\\
$\left\{
\text{ImVBS}^{\ }_{x},
\text{ReTSC}^{\ }_x,
\text{ImTSC}^{\ }_x,
\text{CDW},
\text{ReVBS}^{\ }_{x}
\right\}$ 
&
$\left\{
\text{ReTSC}^{\ }_{02z},
\text{ReTSC}^{\ }_y,
\text{ImTSC}^{\ }_x,
\text{N\'eel}^{\ }_z,
\text{ImTSC}^{\ }_{32z}
\right\}$ 
\\&\\
$\left\{
\text{QSHE}^{\ }_y,
\text{ImVBS}^{\ }_{z},
\text{QSHE}^{\ }_x,
\text{ReVBS}^{\ }_{z},
\text{N\'eel}^{\ }_z
\right\}$ 
&
$\left\{
\text{ReTSC}^{\ }_z,
\text{ImVBS}^{\ }_{z},
\text{ImTSC}^{\ }_z,
\text{ReVBS}^{\ }_{z},
\text{CDW}
\right\}$ \\
$\left\{
\text{QSHE}^{\ }_y,
\text{ReTSC}^{\ }_{02y},
\text{ReTSC}^{\ }_y,
\text{ImSSC},
\text{ImTSC}^{\ }_{32y}
\right\}$ 
&
$\left\{
\text{ReTSC}^{\ }_z,
\text{ReTSC}^{\ }_{02y},
\text{ReTSC}^{\ }_x,
\text{N\'eel}^{\ }_y,
\text{ImTSC}^{\ }_{32y}
\right\}$ 
\\
$\left\{
\text{QSHE}^{\ }_y,
\text{ReTSC}^{\ }_{02y},
\text{ImTSC}^{\ }_{02y},
\text{ReVBS}^{\ }_{x},
\text{ReVBS}^{\ }_{z}
\right\}$ 
&
$\left\{
\text{ReTSC}^{\ }_z,
\text{ReTSC}^{\ }_{02y},
\text{ImTSC}^{\ }_{02x},
\text{ImTSC}^{\ }_{32z},
\text{ReVBS}^{\ }_{z}
\right\}$ 
\\
$\left\{
\text{QSHE}^{\ }_y,
\text{ReTSC}^{\ }_{32y},
\text{ImTSC}^{\ }_{02y},
\text{ImTSC}^{\ }_y,
\text{ReSSC}\right\}$ 
&
$\left\{
\text{ReTSC}^{\ }_z,
\text{ReTSC}^{\ }_{32x},
\text{ImTSC}^{\ }_{02x},
\text{ImTSC}^{\ }_y,
\text{N\'eel}^{\ }_x
\right\}$ 
\\&\\
$\left\{
\text{ReVBS}^{\ }_{y},
\text{N\'eel}^{\ }_y,
\text{QSHE}^{\ }_x,
\text{ImVBS}^{\ }_{y},
\text{QSHE}^{\ }_z
\right\}$ 
&
$\left\{
\text{ReTSC}^{\ }_{32z},
\text{ImSSC},
\text{ImTSC}^{\ }_z,
\text{ImTSC}^{\ }_{02z},
\text{QSHE}^{\ }_z
\right\}$ 
\\
$\left\{
\text{ReVBS}^{\ }_{y},
\text{ReTSC}^{\ }_y,
\text{ImTSC}^{\ }_y,
\text{CDW},
\text{ImVBS}^{\ }_{y}
\right\}$ 
&
$\left\{
\text{ReTSC}^{\ }_{32z},
\text{ReTSC}^{\ }_x,
\text{ImTSC}^{\ }_y,
\text{N\'eel}^{\ }_z,
\text{ImTSC}^{\ }_{02z}
\right\}$ \\
$\left\{
\text{ReVBS}^{\ }_{y},
\text{ReTSC}^{\ }_{32y},
\text{ImTSC}^{\ }_y,
\text{ImTSC}^{\ }_{02z},
\text{ImTSC}^{\ }_{02x}
\right\}$ 
&
$\left\{
\text{ReTSC}^{\ }_{32z},
\text{ReTSC}^{\ }_{32x},
\text{ImTSC}^{\ }_y,
\text{ImVBS}^{\ }_{y},
\text{ImTSC}^{\ }_{02y}
\right\}$ 
\\
$\left\{
\text{ReVBS}^{\ }_{y},
\text{ReTSC}^{\ }_{02x},
\text{ImTSC}^{\ }_{02x},
\text{QSHE}^{\ }_x,
\text{ReVBS}^{\ }_{z}
\right\}$ 
&
$\left\{
\text{ReTSC}^{\ }_{32z},
\text{ReTSC}^{\ }_{02x},
\text{ImTSC}^{\ }_{02y},
\text{ImTSC}^{\ }_z,
\text{ReVBS}^{\ }_{z}
\right\}$ 
\\&\\
$\left\{
\text{N\'eel}^{\ }_y,
\text{ReTSC}^{\ }_{32y},
\text{ImTSC}^{\ }_{02y},
\text{ImTSC}^{\ }_z,
\text{ImTSC}^{\ }_x
\right\}$ 
&
$\left\{
\text{ImSSC},
\text{ReTSC}^{\ }_{32x},
\text{ImTSC}^{\ }_{02x},
\text{QSHE}^{\ }_x,
\text{ImTSC}^{\ }_x
\right\}$ 
\\
$\left\{
\text{ImVBS}^{\ }_{z},
\text{ReTSC}^{\ }_{32y},
\text{ImTSC}^{\ }_{02z},
\text{ImTSC}^{\ }_z,
\text{ImTSC}^{\ }_{32x}
\right\}$ 
&
$\left\{
\text{ImVBS}^{\ }_{z},
\text{ReTSC}^{\ }_{32x},
\text{ImVBS}^{\ }_{y},
\text{QSHE}^{\ }_x,
\text{ImTSC}^{\ }_{32x}
\right\}$ 
\\
$\left\{
\text{ReTSC}^{\ }_{02y},
\text{ReTSC}^{\ }_y,
\text{ImTSC}^{\ }_{32z},
\text{ImTSC}^{\ }_{32x},
\text{ImVBS}^{\ }_{y}
\right\}$ 
&
$\left\{
\text{ReTSC}^{\ }_{02y},
\text{ReTSC}^{\ }_x,
\text{ReVBS}^{\ }_{x},
\text{ImTSC}^{\ }_{32x},
\text{ImTSC}^{\ }_{02z}
\right\}$ 
\\
$\left\{
\text{ReTSC}^{\ }_y,
\text{ReTSC}^{\ }_{02x},
\text{ImTSC}^{\ }_z,
\text{ImTSC}^{\ }_{32x},
\text{N\'eel}^{\ }_x
\right\}$ 
&
$\left\{
\text{ReTSC}^{\ }_x,
\text{ReTSC}^{\ }_{02x},
\text{QSHE}^{\ }_x,
\text{ImTSC}^{\ }_{32x},
\text{ReSSC}
\right\}$
\end{tabular}
\end{ruledtabular}
\end{table*}

\end{widetext}

\subsection{
Classification of masses in graphene and $\pi$-flux phases
           }
\label{subsec: Classification of masses in graphene and pi-flux phases}

To describe all symmetry-breaking instabilities with a local order parameter
in graphene or the square lattice with $\pi$-flux phase,
we consider the Bogoliubov-de Gennes (BdG) Hamiltonian 
\begin{subequations}
\begin{equation}
\hat{H}^{\ }_{\mathrm{BdG}}
=
\frac{1}{2}
\int d^{2}\boldsymbol{r}\,
\hat{\Psi}^{\dag}
\mathcal{K}
\hat{\Psi}
\end{equation}
where $\hat{\Psi}$ is the 16-component Nambu spinor
\begin{equation}
\hat{\Psi}:=
\begin{pmatrix}
\hat{\psi}_{\uparrow}^{\ }, & 
\hat{\psi}_{\downarrow}^{\ }, & 
\hat{\psi}_{\uparrow}^{\dag}, & 
\hat{\psi}_{\downarrow}^{\dag}
\end{pmatrix}^{\mathrm{t}}
\end{equation}
and $\hat{\psi}^{\ }_{s=\uparrow,\downarrow}$
is a 4-component fermion annihilation operator that accounts for
the 2 valley and the 2 sublattice degrees of freedom. 
The kernel of the BdG Hamiltonian has the block structure
\begin{equation}
\mathcal{K}=
\begin{pmatrix}
\mathcal{H}^{\ }_{\mathrm{p}\mathrm{p}}
& 
\mathcal{H}^{\ }_{\mathrm{p}\mathrm{h}}
\\
\mathcal{H}^{\dag}_{\mathrm{p}\mathrm{h}} 
& 
-\mathcal{H}^{\mathrm{t}}_{\mathrm{p}\mathrm{p}} 
\end{pmatrix}
\label{eq: BdG block structure}
\end{equation}
\end{subequations}
where the
$8\times 8$ blocks
$\mathcal{H}^{\ }_{\mathrm{p}\mathrm{p}}$ 
and 
$\mathcal{H}^{\ }_{\mathrm{p}\mathrm{h}}$
act on the combined space of 
valley, sublattice, and spin degrees of freedom, 
and represent the normal and anomalous part of the BdG Hamiltonian,
respectively. These blocks satisfy
\begin{equation}
\begin{split}
&
\mathcal{H}^{\dag}_{\mathrm{p}\mathrm{p}}=
\mathcal{H}^{\   }_{\mathrm{p}\mathrm{p}}
\qquad
\mbox{(Hermiticity)},
\\
&
\mathcal{H}^{\mathrm{t}}_{\mathrm{p}\mathrm{h}}=
-\mathcal{H}^{\ }_{\mathrm{p}\mathrm{h}}
\qquad
\mbox{(Fermi statistics)}.
\end{split}
\label{eq: phs}
\end{equation}

To represent the single particle Hamiltonian $\mathcal{K}$, 
define the 256 16-dimensional Hermitian matrices
\begin{equation}
X^{\ }_{\mu^{\ }_{1}\mu^{\ }_{2}\mu^{\ }_{3}\mu^{\ }_{4}}:= 
{\rho}^{\ }_{\mu^{\ }_{1}}
\otimes
{s}^{\ }_{\mu^{\ }_{2}}
\otimes
\sigma^{\ }_{\mu^{\ }_{3}}
\otimes
\tau^{\ }_{\mu^{\ }_{4}}
\label{eq: 256 matrices}
\end{equation}
where
$\mu^{\ }_{1,2,3,4}=0,1,2,3$.
Here, we have introduced the four families 
$  {\rho}^{\ }_{\mu^{\ }_{1}}$,
$     {s}^{\ }_{\mu^{\ }_{2}}$,
${\sigma}^{\ }_{\mu^{\ }_{3}}$,
and
$  {\tau}^{\ }_{\mu^{\ }_{4}}$
of unit $2\times2$ and Pauli matrices that encode
the particle-hole (Nambu), spin-1/2, valley, and sublattice 
degrees of freedom of graphene
or the square lattice with $\pi$-flux phase, respectively.

The Dirac kinetic energy $\mathcal{K}^{\ }_{0}$
of graphene or the square lattice with $\pi$-flux phase
that accounts for the BdG block structure%
~(\ref{eq: BdG block structure})
is assigned the two 16$\times$16 Dirac matrices
\begin{subequations}
\begin{equation}
\alpha^{\ }_{1}\equiv
X^{\ }_{0031},
\qquad  
\alpha^{\ }_{2}\equiv
X^{\ }_{3032},
\end{equation}
and is given by
\begin{equation}
\mathcal{K}^{\ }_{0}:=
\boldsymbol{\alpha}\cdot(-{i}\boldsymbol{\partial}).
\label{eq: def K0}
\end{equation}
Similarly, by introducing the 16$\times$16 Hermitian matrices
\begin{equation}
\beta\equiv
X^{\ }_{3010},
\qquad
R\equiv
X^{\ }_{3033},
\qquad
\gamma^{\ }_{5}\equiv
X^{\ }_{3030},
\end{equation}
the counterpart to 
$\hat{H}$
in Eq.~(2.1) is given by 
\begin{equation}
\mathcal{K}:=
\mathcal{K}^{\ }_{0}
+
\mathcal{K}^{\ }_{\mathrm{gauge}}
+
\mathcal{K}^{\ }_{\mathrm{scalar}},
\label{eq: def K Dirac}
\end{equation}
where 
\begin{equation}
\begin{split}
&
\mathcal{K}^{\ }_{\mathrm{gauge}}:=
\boldsymbol{\alpha}\cdot
\left(
-
\boldsymbol{a}
-
\boldsymbol{a}^{\ }_{5}\gamma^{\ }_{5}
\right),
\\
&
\mathcal{K}^{\ }_{\mathrm{scalar}}:=
|\Delta|\beta e^{{i}\theta\gamma^{\ }_{5}}
+
\mu^{\ }_{\mathrm{s}} R
+
{i}
\eta
\alpha^{\ }_{1}
\alpha^{\ }_{2}.
\end{split}
\label{eq: 16 dimensional Dirac single-particle Hamiltonian}
\end{equation}
\end{subequations}

Given the Dirac kinetic term $\mathcal{K}^{\ }_{0}$, 
we treat 
$X^{\ }_{\mu^{\ }_{1}\mu^{\ }_{2}\mu^{\ }_{3}\mu^{\ }_{4}}$ 
as a perturbation,
\begin{equation}
\mathcal{K}^{\ }_{m} :=
\mathcal{K}^{\ }_{0} 
+ 
m 
X^{\ }_{\mu^{\ }_{1}\mu^{\ }_{2}\mu^{\ }_{3}\mu^{\ }_{4}}
\label{eq: def K Dirac kinectic+m}
\end{equation}
where $m\in \mathbb{R}$ is constant in space and time. 
If $X^{\ }_{\mu^{\ }_{1}\mu^{\ }_{2}\mu^{\ }_{3}\mu^{\ }_{4}}$
anticommutes with the Dirac kinetic energy
$\mathcal{K}^{\ }_{0}$,
then it opens a gap in the massless Dirac spectrum of 
$\mathcal{K}^{\ }_{0}$.
We shall call such a perturbation a mass in short.
Each mass can be thought of as being induced by 
a breaking of a microscopic symmetry (see below).

There are $64=4\times16$ mass matrices 
(i.e., $X^{\ }_{\mu^{\ }_{1}\mu^{\ }_{2}\mu^{\ }_{3}\mu^{\ }_{4}}$
that anticommutes with $\mathcal{K}^{\ }_{0}$).
Of these 64 mass matrices, only 36 satisfy the condition
\begin{equation}
X^{\ }_{1000}\,
X^{\mathrm{t}}_{\mu^{\ }_{1}\mu^{\ }_{2}\mu^{\ }_{3}\mu^{\ }_{4}}
\,
X^{\ }_{1000}
=
-
X^{\ }_{\mu^{\ }_{1}\mu^{\ }_{2}\mu^{\ }_{3}\mu^{\ }_{4}}
\label{eq: def PHS}
\end{equation}
for particle-hole symmetry (PHS)
and are thus compatible with the symmetry condition
$(\rho^{\ }_{1}\otimes s^{\ }_{0}\otimes \sigma^{\ }_{0}\otimes \tau^{\ }_{0}
  \hat{\Psi})^{\mathrm{t}}=\hat{\Psi}^{\dag}$
on the Nambu spinors 
[i.e., compatible with Eq.\ (\ref{eq: phs})].
All mass matrices with PHS are 
enumerated in Table \ref{tab: 36 masses}.

All 36 mass matrices from Table \ref{tab: 36 masses}
can be classified in terms of
the following (microscopic) 3 symmetry properties.
(i) A BdG Hamiltonian is time-reversal symmetry (TRS) when
\begin{equation}
X^{\ }_{0211}\,
\mathcal{K}^{*}\,
X^{\ }_{0211}=
\mathcal{K}.
\label{eq: def TRS}
\end{equation}
(ii) A BdG Hamiltonian has SU(2) spin rotation symmetry (SRS) when
\begin{equation}
\left[ X^{\ }_{3100}, \mathcal{K}
\right]
=
\left[ X^{\ }_{0200}, \mathcal{K}
\right]
=
\left[ X^{\ }_{3300}, \mathcal{K}
\right]
=0.
\label{eq: def SRS}
\end{equation}
(iii) A BdG Hamiltonian has
sublattice symmetry (SLS) when
\begin{equation}
X^{\ }_{0033}\,
\mathcal{K}\,
X^{\ }_{0033}=
-
\mathcal{K}.
\label{eq: def SLS}
\end{equation}

For any lattice regularization of the
BdG Hamiltonian~(\ref{eq: def K Dirac kinectic+m}) 
supporting two sublattices
$\Lambda^{\ }_{\mathrm{A}}$
and
$\Lambda^{\ }_{\mathrm{B}}$,
as is the case for graphene
or the square lattice with $\pi$-flux phase,
the microscopic order parameter corresponding
to a mass matrix satisfying the SLS~(\ref{eq: def SLS})
is a non-vanishing expectation value for
a fermion bilinear with the two lattice fermions
residing on the opposite ends
of a bond connecting a site belonging to sublattice
$\Lambda^{\ }_{\mathrm{A}}$
and another site belonging to sublattice
$\Lambda^{\ }_{\mathrm{B}}$.
We shall say that such a mass matrix is associated to a
valence-bond solid (VBS) order parameter in analogy to 
the terminology used for quantum dimer models. A
VBS order picks up a microscopic orientation that 
translates into a complex-valued order parameter in
the continuum limit. Hence, we shall distinguish between
the real (ReVBS) and imaginary (ImVBS) parts of the VBS.
Triplet superconductivity is also possible on bonds
connecting the two sublattices. The terminology 
TSC will then also be used. To distinguish
TSC with or without TRS we shall reserve
the prefixes Re and Im for real and imaginary parts.
This is a different convention for the use of the prefixes
Re and Im than for a VBS.

Any mass matrix that does not satisfy the SLS~(\ref{eq: def SLS})
corresponds to a microscopic order parameter for which the
fermion bilinear has the two lattice fermions sitting on the same
sublattice. Microscopic examples are 
charge-density waves (CDW),
spin-density waves (SDW) such as N\'eel ordering,
orbital currents leading to the quantum Hall effect (QHE),
spin-orbit couplings leading to the quantum spin Hall effect (QSHE),
singlet superconductivity (SSC), or triplet superconductivity (TSC).

When SU(2) spin symmetry is broken by the order parameter, we add
a subindex $x$, $y$, or $z$ that specifies the relevant
quantization axis to the name of the mass matrix.
Moreover, TSC with SLS must be distinguished by the 2 possible
bond orientations (the underlying two-dimensional lattice has 2 
independent vectors connecting nearest-neighbor sites).
These 2 orientations are specified by the Pauli matrices used
in the valley and sublattice subspaces, 
i.e., by the 2 pairs of numbers 02 and 32.
Symmetry properties of all 36 PHS masses are 
summarized in Table \ref{tab: 36 masses}.

The set of all 36 PHS masses 
in Table \ref{tab: 36 masses}
is invariant under an involutive transformation defined by
\begin{equation}
\begin{split}
&
\hat{\Psi}
\to 
C\hat{\Psi}, 
\\
&
C 
=
\rho^{\ }_{0} \otimes s^{\ }_{+} \otimes \sigma^{\ }_{0} \otimes \tau^{\ }_{0}
+
\rho^{\ }_{1} \otimes s^{\ }_{-} \otimes \sigma^{\ }_{2} \otimes \tau^{\ }_{2},
\end{split}
\end{equation}
and which we shall call $C$ conjugation to distinguish it
from the particle-hole transformation~(\ref{eq: def PHS}).
Here, $s^{\ }_{\pm} = (s^{\ }_{3}\pm s^{\ }_{0})/2$.
For graphene or the square lattice with $\pi$-flux phase,
this transformation corresponds to
\begin{equation}
\begin{split}
&
 \hat{a}^{\   }_{\boldsymbol{r}^{\ }_{A} \uparrow}
\to 
 \hat{a}^{\   }_{\boldsymbol{r}^{\ }_{A} \uparrow},
\qquad
 \hat{b}^{\   }_{\boldsymbol{r}^{\ }_{B} \uparrow}
\to
 \hat{b}^{\   }_{\boldsymbol{r}^{\ }_{B} \uparrow},
\\
&
 \hat{a}^{\   }_{\boldsymbol{r}^{\ }_{A} \downarrow}
\to 
\hat{a}^{\dag}_{\boldsymbol{r}^{\ }_{A} \downarrow},
\qquad
 \hat{b}^{\   }_{\boldsymbol{r}^{\ }_{B} \downarrow}
\to 
-\hat{b}^{\dag}_{\boldsymbol{r}^{\ }_{B} \downarrow},
\end{split}
\end{equation}
where 
$\hat{a}^{\dag}_{\boldsymbol{r}^{\ }_{A} s}$
and
$\hat{b}^{\dag}_{\boldsymbol{r}^{\ }_{B} s}$
creates an electron with spin $s=\uparrow,\downarrow$ 
on sublattice
$\Lambda^{\ }_{\mathrm{A}}$
and sublattice
$\Lambda^{\ }_{\mathrm{B}}$, 
respectively
(see Appendix 
\ref{appsec: Numerical Berry phase in the single-particle approximation}).
Under this transformation 
\begin{equation}
X^{\ }_{\mu^{\ }_{1}\mu^{\ }_{2}\mu^{\ }_{3}\mu^{\ }_{4}}
\to
C^{\dag}
X^{\ }_{\mu^{\ }_{1}\mu^{\ }_{2}\mu^{\ }_{3}\mu^{\ }_{4}}
C.
\label{eq: def partner mass matrix}
\end{equation}
Hence, it leaves the massless Dirac kernel $\mathcal{K}^{\ }_{0}$ invariant.

The organization of the mass matrices in Table \ref{tab: 36 masses}
can be understood as follows. 

First, we preserve both SRS and charge conservation, 
i.e., we start with the 4 order parameters we have 
already encountered in the spinless case with charge conservation.
There are two valence bond solids, 
ReVBS ($\mathrm{Re}\,\Delta$)
and 
ImVBS ($\mathrm{Im}\,\Delta$).
They have maximal symmetry and are
invariant under the operation of $C$ conjugation%
~(\ref{eq: def partner mass matrix}).
The CDW order parameter ($\mu^{\ }_{\mathrm{s}}$)
breaks the SLS. It is mapped into
the N\'eel spin-density wave with quantization axis $z$
under the operation of $C$ conjugation%
~(\ref{eq: def partner mass matrix}).
The QHE order parameter ($\eta$)
breaks both the SLS and TRS symmetries.
It is invariant under the operation of $C$ conjugation%
~(\ref{eq: def partner mass matrix}).

Second, we break SRS with or without either
TRS or SLS while always preserving charge conservation.
The breaking of SRS is achieved by choosing 
a preferred quantization axis, say $x$, $y$, or $z$
in SU(2) spin space. 
Breaking SRS while preserving SLS 
is achieved with spin-polarized valence-bond
ordering in 6=3$\times$2 different ways, 
which we abbreviate by
ReVBS$^{\ }_{x}$, 
ReVBS$^{\ }_{y}$, 
ReVBS$^{\ }_{z}$, 
ImVBS$^{\ }_{x}$,  
ImVBS$^{\ }_{y}$, 
and
ImVBS$^{\ }_{z}$
in Table \ref{tab: 36 masses}. 
In doing so TRS is always broken.
Breaking SRS and SLS while preserving TRS 
is achieved through any 
of the 3 order parameters for the spin quantum Hall effect
(QSHE) introduced by Kane and Mele in 
Ref.~\onlinecite{Kane05},
which we abbreviate by
QSHE$^{\ }_{x}$, 
QSHE$^{\ }_{y}$,
and 
QSHE$^{\ }_{z}$ 
in Table \ref{tab: 36 masses}. 
Breaking SRS, SLS, and TRS is achieved through
any one of 3 colinear magnetic order in the form of
N\'eel order,
which we abbreviate by
N\'eel$^{\ }_{x}$, 
N\'eel$^{\ }_{y}$,
and 
N\'eel$^{\ }_{z}$ 
in Table \ref{tab: 36 masses}.

This brings the number of order parameters that
conserve the electronic charge to 16=4+6+3+3. There are thus
20=2+6+6+6  remaining order parameters that 
do not conserve the electronic charge.

Third, superconducting order is achieved microscopically
by pairing two electrons sitting on 
different or identical sublattices.
In the former case, SLS is preserved.
In the latter case, SLS is broken.
Pairing of the 2 electronic spins takes place
either in a singlet or in a triplet channel.
Antisymmetry under exchange of the two electrons
making up a spin-singlet Cooper pair can only be achieved 
in an even angular momentum channel. On-site pairing
is of course associated to vanishing angular momentum
so that singlet superconductivity can only be realized
when SLS is broken. This only leaves 2 possible
singlet superconducting order parameters that 
are distinguished by whether they preserve or
break TRS. They are denoted ReSSC and ImSSC,
respectively.
(Real and imaginary parts thus take a different
meaning here as for ReVBS and ImVBS.) 

Fourth, a triplet superconducting order parameter,
which we abbreviate by TSC 
in Table \ref{tab: 36 masses},
is characterized by a 
vector $\boldsymbol{d}$ in SU(2) spin space.
This vector can point along any one of the three
quantization axis $x$, $y$, and $z$ in
SU(2) spin space. 
Moreover, it can either 
preserve or break TRS for which cases we use
the notations ReTSC and ImTSC, respectively,
in Table \ref{tab: 36 masses}.
(Real and imaginary parts thus take a different
meaning here as for ReVBS and ImVBS.) 
When SLS is preserved
by the superconducting order parameter, 
there are $12=2\times2\times3$ 
independent order parameters,
for a second factor of 2 besides the one for TRS
arises since there are 2 directed nearest-neighbor lattice-bonds 
connecting nearest-neighbor sites of the two-dimensional lattice. 
This is abbreviated in Table \ref{tab: 36 masses}
by using the index 
bond=02,32 in
ReTSC$^{\ }_{\mathrm{bond}x}$,
ReTSC$^{\ }_{\mathrm{bond}y}$,
ReTSC$^{\ }_{\mathrm{bond}z}$,
ImTSC$^{\ }_{\mathrm{bond}x}$,
ImTSC$^{\ }_{\mathrm{bond}y}$,
and
ImTSC$^{\ }_{\mathrm{bond}z}$.
Finally, when SLS is broken
by the superconducting order parameter,
there are
$6=2\times3$ independent order parameters
that we abbreviate by 
ReTSC$^{\ }_{x}$,
ReTSC$^{\ }_{y}$,
ReTSC$^{\ }_{z}$,
ImTSC$^{\ }_{x}$,
ImTSC$^{\ }_{y}$,
and
ImTSC$^{\ }_{z}$
in Table \ref{tab: 36 masses}.

There are $12=4\times3$ order parameters that
are invariant under
the operation of $C$ conjugation%
~(\ref{eq: def partner mass matrix}).
They can be arranged in 4 groups of 3 each.
Each group of 3 obeys the same algebra. 
The 4 groups of 3 are:
(i)
ReVBS,
ImVBS,
QHE;
(ii)
ReVBS$^{\ }_{z}$,
ImVBS$^{\ }_{z}$,
QSHE$^{\ }_{z}$;
(iii)
ReTSC$^{\ }_{02x}$,
ImTSC$^{\ }_{32x}$,
ImTSC$^{\ }_{y}$;
(iv)
ReTSC$^{\ }_{02y}$,
ImTSC$^{\ }_{32y}$,
and
ImTSC$^{\ }_{x}$.

The operation of $C$ conjugation%
~(\ref{eq: def partner mass matrix})
is a useful tool to identify the possibility of
exotic topological effects.

For example, we observe that the pair of SSC
order parameters ReSSC and ImSSC,
studied in 
Refs.~\onlinecite{Beenakker06}
and \onlinecite{GhaemiWilczek}
in the context of graphene,
are conjugate by $C$ to the N\'eel order parameters
$\mbox{N\'eel}^{\ }_{x}$
and
$\mbox{N\'eel}^{\ }_{y}$,
respectively.
Furthermore, Table  \ref{tab: 36 masses}
indicates that several triplets of masses that obeys the SU(2) algebra
are related by the operation of $C$ conjugation%
~(\ref{eq: def partner mass matrix}).
They are
\begin{equation}
\begin{split}
&
\left\{
\mathrm{ReVBS},
\mathrm{ReVBS},
\mathrm{CDW}
\right\}
\\
&\hphantom{AAAAA}
\overset{C}{\longleftrightarrow}
\left\{
\mathrm{ReVBS},
\mathrm{ImVBS},
\hbox{N\'eel}^{\ }_{z}
\right\},
\\
&
\left\{
\mathrm{ReVBS}^{\ }_{x},
\mathrm{ReVBS}^{\ }_{y},
\mathrm{ReVBS}^{\ }_{z}
\right\}
\\
&\hphantom{AAAA}
\overset{C}{\longleftrightarrow}
\left\{
\mathrm{ImTSC}^{\ }_{32z},
\mathrm{ReTSC}^{\ }_{32z},
\mathrm{ReVBS}^{\ }_{z}
\right\},
\\
&
\left\{
\mathrm{ImVBS}^{\ }_{x},
\mathrm{ImVBS}^{\ }_{y},
\mathrm{ImVBS}^{\ }_{z}
\right\}
\\
&\hphantom{AAAA}
\overset{C}{\longleftrightarrow}
\left\{
\mathrm{ReTSC}^{\ }_{02z},
\mathrm{ImTSC}^{\ }_{02z},
\mathrm{ImVBS}^{\ }_{z}
\right\},
\\
&
\left\{
\mathrm{QSHE}^{\ }_{x},
\mathrm{QSHE}^{\ }_{y},
\mathrm{QSHE}^{\ }_{z}
\right\}
\\
&\hphantom{AAAAAAA}
\overset{C}{\longleftrightarrow}
\left\{
\mathrm{ImTSC}^{\ }_{z},
\mathrm{ReTSC}^{\ }_{z},
\mathrm{QSHE}^{\ }_{z}
\right\}.
\end{split}
\end{equation}
Vortex-like defective textures in any of these mass doublets or 
meron-like defective textures in any of these mass triplets
display fractionalization of some suitably defined quantum numbers.

Finally, the topological property that a band insulator supporting the
QSHE carries an odd number of Kramers doublets on its edges
carries over to the $C$ conjugate TSC. More precisely,
the fact that the superconductors with the ImTSC$^{\ }_{z}$ 
and ReTSC$^{\ }_{z}$ order parameters are examples of
$\mathbb{Z}^{\ }_{2}$ topological triplet superconductors
according to Refs.~\onlinecite{Roy06},
\onlinecite{Schnyder08}, 
\onlinecite{Roy08}, 
\onlinecite{Qi09}, 
and
\onlinecite{Kitaev08}
is here a mere consequence of their $C$ conjugation with
the QSHE$^{\ }_{x}$ and QSHE$^{\ }_{y}$
order parameters, respectively.

\subsection{
Classification of 5-tuplets of masses 
in graphene and $\pi$-flux phases
           }
\label{subsec: Classification of 5-tuplets of masses in graphene and pi-flux phases}

Mass matrices that commute pairwise generate 
competing local order parameters. Conversely,
mass matrices that anticommute pairwise generate
compatible local order parameters.

All but one PHS masses anticommute with 16 out 
of the 36 PHS masses. The Haldane mass is unique in that it
commutes with all PHS masses.

There are 560 sets of three mutually anticommuting PHS masses.
These triplets are generalizations of 
the triplet of compatible masses 
$\Delta=\mathrm{Re}\,\Delta+{i}\mathrm{Im}\,\Delta$ 
and $\mu^{\ }_{\mathrm{s}}$.
Integration over the Dirac fermions
in the presence of any one of these mass triplets
of mass $m$ in competition with the Haldane mass $\eta$
induces an O(3) NLSM in (2+1)-dimensional space and time
with 
or 
without a Hopf term for 
$m>|\eta|$
and $|\eta|>m$, 
respectively,
as was derived in Ref.~\onlinecite{Chamon08a}.

There are 280 sets of four mutually anticommuting PHS masses
and the maximum number of pairwise anticommuting
PHS mass matrices is 5. 
Out of $\binom{36}{5}=376992$ possibilities, 
there are 56 distinct 5-tuplets of compatible PHS mass matrices.
They are enumerated in Tables \ref{tab: 56 5-tuplets}.
(If PHS is not imposed,
the maximum number of pairwise
anticommuting mass matrices 
in the 64 mass matrices
is 7.
There are 288
distinct 7-tuplets of compatible mass matrices.)

In the background of each of these 5-tuplet, 
integration over the Dirac fermions
yields an O(5) NLSM in (2+1)-dimensional space and time
augmented by a Wess-Zumino-Witten (WZW) term
as was derived in Refs.~\onlinecite{Tanaka05a} and
\onlinecite{Tanaka05b}.
Defects-driven continuous phase transition between
phases of matter unrelated by symmetries (i.e., Landau forbidden)
become possible whenever the quantum numbers of the defective
order parameters in a given 5-tuplet are dual
in the sense of BF Chern-Simons field theories.%
~\cite{Ghaemi09}
We illustrate this idea with the following examples.

\begin{figure}
\includegraphics[angle=0,scale=0.55]{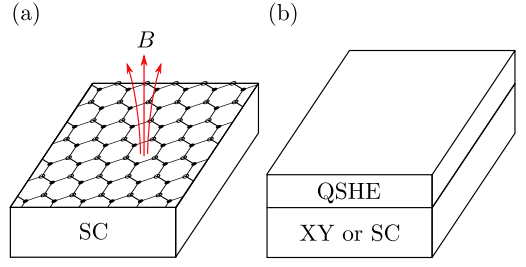}
\caption{
Two setups used to induce topological defects in
an order parameter that support fractional quantum numbers.
        }
\label{fig: setup}
\end{figure}

\subsubsection{
VBS-SSC-CDW 5-tuplet
              }
\label{subsubsec: VBS-SSC-CDW 5-tuplet}

The 5-tuplet 
\begin{equation}
\{ 
\mathrm{ReVBS},
\mathrm{ImVBS},
\mathrm{ReSSC},
\mathrm{ImSSC},
\mathrm{CDW}
\}
\label{eq: CDW-VBS-SSC 5-tuplet}
\end{equation}
embeds the triplet made of the CDW and the 2 VBS
order parameters into a 5-tuplet.%
~\cite{Ghaemi09}
Integration over the fermions yields an O(5) NLSM
augmented by a WZW term
for the corresponding 5-tuplets of bosonic fields
$n^{\ }_{1}$,
$n^{\ }_{2}$,
$n^{\ }_{3}$,
$n^{\ }_{4}$,
and 
$n^{\ }_{5}$
obeying the constraint that they add in quadrature to
unity. The O(5) symmetry can be broken, either spontaneously or explicitly,
down to the U(1)$\times$U(1) subgroup corresponding to holding
$\Delta^{2}_{\mathrm{CDW}}\equiv n^{2}_{5}$,
$\Delta^{2}_{\mathrm{BDW}}\equiv n^{2}_{1}+n^{2}_{2}$,
and 
$\Delta^{2}_{\mathrm{SSC}}\equiv n^{2}_{3}+n^{2}_{4}$
fixed
(except at the core of topological defects)
throughout space and time.
The corresponding Goldstone modes are the
phases 
$\theta^{\ }_{\mathrm{BDW}}$ 
and
$\theta^{\ }_{\mathrm{SSC}}$.
They become charge 2 Higgs fields if the U(1)$\times$U(1)
global symmetry they generate is gauged
through the introduction of the 
axial gauge fields
$a^{\mu}_{\mathrm{VBS}}$
and the electro-magnetic gauge fields 
$a^{\mu}_{\mathrm{SSC}}$,
respectively. Their dynamics is governed by
the Anderson-Higgs-Chern-Simons theory%
~(\ref{eq: final effective action})
with the identifications 
$\theta\to
 \theta^{\ }_{\mathrm{BDW}}$,
$a^{\mu}_{5}\to 
 a^{\mu}_{\mathrm{VBS}}$,
and
$a^{\mu}\to 
 a^{\mu}_{\mathrm{SSC}}
 -
 \partial^{\mu}\theta^{\ }_{\mathrm{SSC}}/2$.
The VBS phase is destroyed when the vortices
carried by the conserved topological current
$j^{\mathrm{vrt}\mu}_{\mathrm{VBS}}=
\epsilon^{\mu\nu\rho}
\partial^{\ }_{\nu}\partial^{\ }_{\rho}\theta^{\ }_{\mathrm{VBS}}/(2\pi)$
deconfine.
The SSC phase is destroyed when the vortices
carried by the conserved topological current
$j^{\mathrm{vrt}\mu}_{\mathrm{SSC}}=
\epsilon^{\mu\nu\rho}
\partial^{\ }_{\nu}\partial^{\ }_{\rho}\theta^{\ }_{\mathrm{SSC}}/(2\pi)$
deconfine. Because of the BF term in the
effective action, the quasiparticles supported
by $j^{\mathrm{vrt}\mu}_{\mathrm{VBS}}$ also carry 
a fraction of the gauge charge
of the gauge fields $a^{\mu}_{\mathrm{SSC}}$,
while the quasiparticles supported
by $j^{\mathrm{vrt}\mu}_{\mathrm{SSC}}$ also carry 
a fraction of the gauge charge
of the gauge fields $a^{\mu}_{\mathrm{VBS}}$.
Furthermore, both types of quasiparticles are bosons
(there is no TRS-breaking Haldane mass).
{}From these \textit{two} facts follows that
deconfinement of one type of quasiparticles implies
confinement of the second type of quasiparticles, i.e.,
a direct transition between the VBS and SSC phases.

An experimental setup to detect exotic quantum numbers
related to the 5-tuplet%
~(\ref{eq: CDW-VBS-SSC 5-tuplet})
is given in Fig.~\ref{fig: setup}(a).\cite{Ghaemi09}
We assume that graphene sits on top of a type-II s-wave SC substrate.
By the proximity effect, graphene develops a SSC order.
The SSC order can coexist with the CDW and VBS orders
in graphene according to Eq.~(\ref{eq: CDW-VBS-SSC 5-tuplet}). 
An applied magnetic field perpendicular to graphene
creates an Abrikosov lattice of vortices
in the substrate and, by the proximity effect, in graphene. 
The magnetic flux tubes threading graphene pin axial charges
according to Eq.~(\ref{eq: final effective action}).
(See also Refs.~\onlinecite{Seradjeh08} 
and \onlinecite{Ghaemi09}.)
Increasing the magnetic field so as to destroy SSC
deconfines the axial charges, i.e., stabilizes the
VBS. Conversely, destroying the VBS by the deconfinement
of VBS vortices also deconfines the electric charges, i.e.,
stabilizes the SSC.

\subsubsection{
VBS-N\'eel 5-tuplet
              }

The operation of $C$ conjugation%
~(\ref{eq: def partner mass matrix})
on the 5-tuplet~(\ref{eq: CDW-VBS-SSC 5-tuplet})
yields the 5-tuplet
\begin{equation}
\{ 
\mathrm{ReVBS},
\mathrm{ImVBS},
\hbox{N\'eel}^{\ }_{x},
\hbox{N\'eel}^{\ }_{y},
\hbox{N\'eel}^{\ }_{z}
\}.
\label{eq: Neel-VBS 5-tuplet}
\end{equation}
The triplet of N\'eel order parameters is here embedded
into a 5-tuplet by adding
the doublet of VBS order parameters.%
~\cite{Tanaka05a,Tanaka05b,Herbut07}
This 5-tuplet has been discussed in 
the context of deconfined quantum criticality of
two-dimensional $S=1/2$ quantum antiferromagnetic spin models.%
\cite{Senthil04a,Senthil04b,Senthil05,Ran06,
Sandvik07,
Melko07,
Kaul08}
The 5-tuplet~(\ref{eq: Neel-VBS 5-tuplet}) 
is the only 5-tuplet supporting the full SU(2) 
symmetry of the N\'eel vector. 
The symmetry analysis of 
Sec.~\ref{subsubsec: VBS-SSC-CDW 5-tuplet}
follows with the identifications
$\theta^{\ }_{\mathrm{VBS}}\to
 \theta^{\ }_{\mathrm{VBS}}$,
$\theta^{\ }_{\mathrm{SSC}}\to
 \theta^{\ }_{\hbox{\begin{scriptsize}N\'eel\end{scriptsize}}_{xy}}$,
$a^{\mu}_{\mathrm{VBS}}\to 
 a^{\mu}_{\mathrm{VBS}}$,
$a^{\mu}_{\mathrm{SSC}}\to 
 a^{\mu}_{\hbox{\begin{scriptsize}N\'eel\end{scriptsize}}_{xy}}$,
$j^{\mathrm{vrt}\mu}_{\mathrm{VBS}}\to
 j^{\mathrm{vrt}\mu}_{\mathrm{VBS}}$,
and
$j^{\mathrm{vrt}\mu}_{\mathrm{SSC}}\to
 j^{\mathrm{vrt}\mu}_{\hbox{\begin{scriptsize}N\'eel\end{scriptsize}}_{xy}}$.

\subsubsection{
SSC-QSHE 5-tuplet
              }

The 5-tuplet 
\begin{equation}
\{ 
\mathrm{ReSSC},
\mathrm{ImSSC},
\mathrm{QSHE}^{\ }_{x},
\mathrm{QSHE}^{\ }_{y},
\mathrm{QSHE}^{\ }_{z}
\}
\label{eq: SSC_QSHE 5-tuplet}
\end{equation}
embeds the triplet of QSHE order parameters into
a 5-tuplet by adding the two possible SSC order parameters.%
~\cite{Grover08,RanVishwanathLee08,QiZhang08}
The symmetry analysis of 
Sec.~\ref{subsubsec: VBS-SSC-CDW 5-tuplet}
follows with the identifications
$\theta^{\ }_{\mathrm{VBS}}\to
 \theta^{\ }_{\mathrm{SSC}}$,
$\theta^{\ }_{\mathrm{SSC}}\to
 \theta^{\ }_{\mathrm{QSHE}_{xy}}$,
$a^{\mu}_{\mathrm{VBS}}\to 
 a^{\mu}_{\mathrm{SSC}}$,
$a^{\mu}_{\mathrm{SSC}}\to 
 a^{\mu}_{\mathrm{QSHE}_{xy}}$,
$j^{\mathrm{vrt}\mu}_{\mathrm{VBS}}\to
 j^{\mathrm{vrt}\mu}_{\mathrm{SSC}}$,
and
$j^{\mathrm{vrt}\mu}_{\mathrm{SSC}}\to
 j^{\mathrm{vrt}\mu}_{\mathrm{QSHE}_{xy}}$.

An experimental setup to detect exotic quantum numbers
related to the 5-tuplet%
~(\ref{eq: SSC_QSHE 5-tuplet})
is given in Fig.~\ref{fig: setup}(b).
We bring in contact a (3D) bulk type-II SSC
with a material displaying the QSHE.
Instead of graphene for which the spin-orbit
coupling is very small, 
HgTe/(Hg,Cd)Te semiconductor quantum wells are suitable.%
~\cite{Bernevig-Taylor-Zhang06, Koenig07,Koenig08}
Any SSC vortex in the substrate induces by proximity effect
an ``$S^{\ }_{z}$ spin charge'' in the device supporting the QSHE,
while any $S^{\ }_{z}$ ``spin flux'' in the device supporting the QSHE
induces an electric charge.

\subsubsection{
XY-N\'eel-TSC-QSHE 5-tuplet 
              }

The operation of $C$ conjugation%
~(\ref{eq: def partner mass matrix})
on the 5-tuplet~(\ref{eq: SSC_QSHE 5-tuplet})
yields the 5-tuplet%
~\cite{RanVishwanathLee08,QiZhang08}
\begin{subequations}
\label{eq: Neel-TSC-QSHE 5-tuplet}
\begin{equation}
\{ 
\hbox{N\'eel}^{\ }_{x}, 
\hbox{N\'eel}^{\ }_{y}, 
\mathrm{ImTSC}^{\ }_{z},
\mathrm{ReTSC}^{\ }_{z},
\mathrm{QSHE}^{\ }_{z}
\}.
\end{equation}
By rotating SU(2) spin quantization axis,
i.e., by cyclic permutation of the indices
$x$, $y$, and $z$, 
we also get the 5-tuplets 
\begin{equation}
\{ 
\hbox{N\'eel}^{\ }_{y}, 
\hbox{N\'eel}^{\ }_{z}, 
\mathrm{ImTSC}^{\ }_{x},
\mathrm{ReTSC}^{\ }_{x},
\mathrm{QSHE}^{\ }_{x}
\}
\end{equation}
and 
\begin{equation}
\{ 
\hbox{N\'eel}^{\ }_{z}, 
\hbox{N\'eel}^{\ }_{x}, 
\mathrm{ImTSC}^{\ }_{y},
\mathrm{ReTSC}^{\ }_{y},
\mathrm{QSHE}^{\ }_{y}
\}.
\end{equation}
\end{subequations}
These 5-tuplets describe SLS- and SRS-breaking order parameters
consisting of an easy plane antiferromagnetic order parameter 
coexisting with the QSHE and TSC order parameters.
The symmetry analysis of 
Sec.~\ref{subsubsec: VBS-SSC-CDW 5-tuplet}
follows with the identifications
$\theta^{\ }_{\mathrm{VBS}}\to
 \theta^{\ }_{\hbox{\begin{scriptsize}N\'eel\end{scriptsize}}_{xy}}$,
$\theta^{\ }_{\mathrm{SSC}}\to
 \theta^{\ }_{\mathrm{TSC}_{z}}$,
$a^{\mu}_{\mathrm{VBS}}\to 
 a^{\mu}_{\hbox{\begin{scriptsize}N\'eel\end{scriptsize}}_{xy}}$,
$a^{\mu}_{\mathrm{SSC}}\to 
 a^{\mu}_{\mathrm{TSC}_{z}}$,
$j^{\mathrm{vrt}\mu}_{\mathrm{VBS}}\to
 j^{\mathrm{vrt}\mu}_{\hbox{\begin{scriptsize}N\'eel\end{scriptsize}}_{xy}}$,
and
$j^{\mathrm{vrt}\mu}_{\mathrm{SSC}}\to
 j^{\mathrm{vrt}\mu}_{\mathrm{TSC}_{z}}$,
say.

An experimental setup to detect exotic quantum numbers
related to the 5-tuplet%
~(\ref{eq: Neel-TSC-QSHE 5-tuplet})
is also given in Fig.~\ref{fig: setup}(b).
Any defect in the bulk $XY$ antiferromagnet,
i.e., a magnetic vortex, induces a localized midgap state
that carries a fraction of the electric charge
carried by the phase of the TSC
in the band insulator supporting the QSHE.
Any TSC vortex induces an ``$S^{\ }_{z}$ spin charge''
in the device supporting the QSHE. 
[A related fractional (electrical) charge is discussed 
at the helical edges of the QSHE.\cite{QiTaylorZhang07}]

\section{
Discussion
        }
\label{sec: Summary}

Motivated by the interplay between
charge-density ($\mu^{\ }_{\mathrm{s}}$),
bond-density ($\Delta=|\Delta|e^{-{i}\theta}$),
and integer quantum Hall ($\eta$)
instabilities
in graphene-like two-dimensional electronic systems,
we have computed the fractional charge and fractional statistics
of both screened and unscreened quasiparticles.

At the microscopic level,
screened quasiparticles are here the linear superpositions
of two bond-density waves ($\Delta$ and $\boldsymbol{a}^{\ }_{5}$), 
each of which carry a point defect.
Unscreened quasiparticles are defects in one type
($\Delta$) of bond-density wave.

In the long-wave-length and low-energy limit
and after integrating out the fermions,
the quantum dynamics of screened quasiparticles 
is controlled by the effective theory%
~(\ref{eq: final effective action})
of the Anderson-Higgs-Chern-Simons
type involving three fields. There are two U(1)
gauge fields and one phase field.

The first gauge field $a^{\ }_{\mu}$
is responsible for the
conservation of the total fermion number.
The second gauge field $a^{\ }_{5\mu}$
is responsible for the 
conservation of a relative fermion number, i.e.,
the difference in the fermion number located
at the two valleys of graphene say,
and is thus called an axial gauge field.
The phase field $\theta=-\mathrm{arg}\,\Delta$ 
originates microscopically
from the fact that bond distortions include 
atomic displacements away from the crystalline order 
that are parametrized by continuous 
angular degrees of freedom.

Screened quasiparticles are not yet explicitly
manifest in the field theory%
~(\ref{eq: final effective action}). 
They appear as point particles with 
the conserved topological current
$\bar{j}^{\mu}_{\mathrm{vrt}}=
(2\pi)^{-1}
\epsilon^{\mu\nu\lambda}
\partial^{\ }_{\nu}
\partial^{\ }_{\lambda}
\theta$
that carries no axial gauge charge,
once a duality transformation has been performed.
The Lagrangian dual to the Lagrangian%
~~(\ref{eq: final effective action})
can be presented as
a Chern-Simons theory for 4 gauge fields 
whose $K$ matrix\cite{Wen91}
is 4-dimensional
and couple through a 4-dimensional charge vector
to the vortex current.
Because the $K$ matrix
has a vanishing eigenvalue,\cite{Wen91} 
\textit{this dual theory is not a topological theory},
say such as a BF Chern-Simons theory.\cite{Blau91,Hansson04}
The vanishing eigenvalue of the $K$ matrix
signals the existence of low-energy excitations,
the screened quasiparticles. 
Their fractional charges $Q$
and statistical angle $\Theta$ 
can then be calculated and are presented in
the phase diagram of 
Fig.~\ref{fig:phase-diagram}.

When the U(1)$\times$U(1) local gauge symmetry holds, 
i.e., for screened quasiparticles that represent
vortices in the phase field $\theta$ whose axial
charges are dynamically screened by axial
gauge half fluxes in $a^{\ }_{5\mu}$, 
the fractional charge $Q$ and
the fractional statistical angle $\Theta$
in Fig.~\ref{fig:phase-diagram}
are complementary. One is non-vanishing if 
and only if the other vanishes. Moreover,
$Q$ and $\Theta$ are universal in the fully gaped
phases for which they are non-vanishing and
given by a rational number in some units.

When the U(1)$\times$U(1) local gauge symmetry is broken,
i.e., for unscreened quasiparticles that represent
vortices in the phase field $\theta$ 
without the attachment of axial gauge half fluxes,
the fractional statistical angle $\Theta$
is non-vanishing everywhere in
Fig.~\ref{fig:phase-diagram}
with a discontinuous jump at 
$m\equiv\sqrt{\mu^{2}_{\mathrm{s}}+|\Delta|^{2}}=|\eta|$
and a non-universal dependence on
the ratios $\mu^{\ }_{\mathrm{s}}/m$
and $\eta/m$. The fractional charge
$Q$ is only non-vanishing when $|\eta|<m$
where it is also non-universal.

Comparing the values of $\Theta$
in Fig.~\ref{fig:phase-diagram}
calculated from field theory with a numerical
evaluation of $\Theta$ for 
an underlying microscopic (lattice) model
is difficult for two reasons.

Defects in the phase $\theta$ have a characteristic
size of the order of $1/m$ for lattice models,
i.e., they bind a fermionic charge through midgap
states. The profile of defects in the axial gauge fields 
$a^{\ }_{5\mu}$
is power law,
i.e., they bind a fermionic charge through threshold
continuum states. Thus, the linear extend of 
any lattice model must be much
larger than $1/m$ for any reliable numerical calculation
of $\Theta$. On the one hand,
if we impose the U(1)$\times$U(1) local gauge invariance
at the lattice level, the system sizes accessible to
a numerical computation of $\Theta$ are, at best, of the order
$1/m$, i.e., too small for a comparison with field theory.
On the other hand, if the U(1)$\times$U(1) local gauge invariance
does not hold at the lattice level, say after performing
a mean-field approximation for which
the accessible system sizes are sufficient to measure
$Q$ with the help of a static probe such as the spectral asymmetry,
then the values of $Q$ and $\Theta$ are not universal anymore.
To put it differently, the values of $Q$ and $\Theta$ measured
dynamically depend sensitively on the dynamical rules used.
But these dynamical rules are model dependent when they are 
not fixed by imposing the local axial gauge symmetry.

The fractional charge $Q$
or the statistical angle $\Theta$ 
in the phase diagram of 
Fig.~\ref{fig:phase-diagram}
disagree with the results of Refs.%
~\onlinecite{Chamon08a}, 
\onlinecite{Seradjeh08}, 
and
\onlinecite{Milovanovic08}.

Although the charge assignment in
Ref.~\onlinecite{Chamon08a}
agrees with that 
in Fig.~\ref{fig:phase-diagram}
the statistical angle is ascribed the value
$\Theta=\mathrm{sgn}\,(\eta)\,\pi/4$
whenever $\eta\neq0$.%
~\cite{footnote: error in Chamon08a}
However, the statistical angle $\Theta$ 
is non-vanishing if and only if the
Hopf term is present in the 
O(3) non-linear-sigma model
derived in Ref.~\onlinecite{Chamon08a},
i.e., if and only if $|\eta|>m$,
in which case full agreement with
the charge and statistical angle assignments of
Fig.~\ref{fig:phase-diagram}
is recovered.

Seradjeh and Franz in Ref.~\onlinecite{Seradjeh08}
have computed the fractional charge $Q$
and fractional statistics $\Theta$ of dynamical
defects in $\Theta$
and $\boldsymbol{a}^{\ }_{5}$ for the field theory%
~(\ref{eq: def Z})
when $\mu^{\ }_{\mathrm{s}}=\eta=0$.
Their analysis has been repeated
by Milovanovic in Ref.~\onlinecite{Milovanovic08}.
They found the assignments $Q=\pm1/2$ and $\Theta=\pm\pi$.
Their semion statistics contradicts our result
$\Theta=0$ in Fig.~\ref{fig:phase-diagram}. 
This discrepancy can be traced to the fact that
Seradjeh and Franz used a singular chiral U(1) 
gauge transformation 
with the Pauli-Villars regularization
to derive an effective action
different than
Eq.~(\ref{eq: final effective action}).
As we show in Appendix%
~\ref{appsec: The pitfall of singular gauge transformations}
the effective action used by Seradjeh and Franz,
when suitably generalized to the case 
$\mu^{\ }_{\mathrm{s}}\neq\eta=0$,
fails to reproduce the fractional charge%
~(\ref{eq: derivation of Q})
of quasiparticles in the presence of a flux in $\boldsymbol{a}_5$ gauge field. 
Explicitly, it follows 
from Eq.\ 9 of their paper Ref.~\onlinecite{Seradjeh08}
that the fractional charge in the case when 
the mass vortex is accompanied by an axial half-flux,
enforcing the screening condition
$a^{\ }_{5\kappa} - \frac{1}{2}\partial^{\ }_{\kappa}\theta=0$,
is $Q=0$ !
However, the fractional charge $Q=1/2$ 
[see Eq.~(\ref{eq: derivation of Q})]
of screened quasiparticles is 
a result established 
from direct (static) numerical computation of $Q$ on
a suitable lattice regularization of the field theory
~(\ref{eq: def Z}).

The charge-density ($\mu^{\ }_{\mathrm{s}}$), 
bond-density ($\Delta$), 
and integer quantum Hall ($\eta$) instabilities
are the only instabilities compatible with 
the electron-number conservation
and SU(2) spin-rotation symmetry 
(these are, naturally, also the only four possible
instabilities for the spinless case). However, 
there can also be superconducting instabilities
or, if the electron spin is accounted for, magnetic instabilities. 
We have performed a systematic classification of all instabilities
for the 16 dimensional free Dirac Hamiltonian induced by
local order parameters that respect the
Bogoliubov-de-Gennes particle-hole symmetry.
We have found that the order parameter for the integer
quantum Hall effect (Haldane mass $\eta$) is unique, for it competes
with all other instabilities. We have also found that
the largest number of coexisting order parameters is 5
and enumerated all the corresponding 5-tuplets of masses.
Each of these 5-tuplet can be thought of as a generalization
of the 3-tuplet 
$(\mu^{\ }_{\mathrm{s}},\mathrm{Re}\,\Delta,\mathrm{Im}\,\Delta)$
that supports quasiparticles with fractional quantum numbers.
These 5-tuplets provide a rich playground for Landau-forbidden
continuous phase transitions.
Any U(1) order parameter in a 5-tuplet can be assigned a 
conserved charge and supports topological defects in the
form of vortices. A pair of U(1) order parameters in a 5-tuplet
is said to be dual if the vortices of one order parameter binds 
the charge of the other order parameter and vice versa. 
A continuous phase transition can then connect 
directly the two dual U(1) ordered phases through
a confining-deconfining transition of their vortices.

\section*{
Acknowledgments
         }

This work is supported in part by the DOE Grant DE-FG02-06ER46316 
(C-Y. H. and C. C.).
C.~M.\ acknowledges the kind hospitality of the Isaac Newton
Institute, Cambridge  and RIKEN.
We thank the Condensed Matter Theory Visitor's Program
at Boston University for support.
S.~R. thanks the Center for Condensed Matter Theory 
at University of California, Berkeley for its support.
S.~R. thanks P. Ghaemi, D.-H.\ Lee, and A. Vishwanath
for useful discussions.

\appendix

\section{
Calculations of the coefficients
$C^{(0)}_{11}$,
$C^{(1)}_{00}$,
$C^{(1)}_{33}$,
$C^{(1)}_{11}$,
and
$C^{(1)}_{03}$
        }
\label{appsec: coefficients}

Let $B^{\ }_{\mu}$ be a 4-dimensional
representation of an element of the Lie Algebra 
u(2) generated by the unit $4\times4$ matrix
and the $4\times4$ matrices
$\Sigma^{\ }_{\mathrm{a}}=
(\Sigma^{\ }_{1},
 \Sigma^{\ }_{2},
 \Sigma^{\ }_{3})$
whereby
\begin{equation}
[\Sigma^{\ }_{\mathrm{a}},
 \Sigma^{\ }_{\mathrm{b}}]=
{i}\epsilon^{\ }_{\mathrm{abc}}
\Sigma^{\ }_{\mathrm{c}},
\qquad
\left\{
\Sigma^{\ }_{\mathrm{a}},
 \Sigma^{\ }_{\mathrm{b}}
\right\}=
2\delta^{\ }_{\mathrm{ab}}.
\label{appeq: Sigma algebra}
\end{equation}
In this appendix,
we are going to integrate the 
Grassmann fields $\bar{\chi}$ and $\chi$
in the partition function
\begin{equation}
\begin{split}
&
Z:=
\int\mathcal{D}[\bar\chi,\chi]\,
e^{{i}S},
\\
&
S:=
\int dx^{0}dx^{1}dx^{2}\,\mathcal{L},
\\
&
\mathcal{L}:=
\bar{\chi}
\left(
{i}
\Slash{\partial}^{\ }_{\mu}
+
\Slash{B}
-
m\Sigma^{\ }_{3}
-
\eta
\right)
\chi.
\end{split}
\label{eq: def Z, S, L}
\end{equation}
Here, the Feynman slash notation
\begin{subequations}
\begin{equation}
\Slash{\partial}\equiv
\Gamma^{\mu}
\partial^{\ }_{\mu},
\qquad
\Slash{B}\equiv
\Gamma^{\mu}
B^{\ }_{\mu}
\end{equation}
is used when contracting 3-vectors with
the $4\times4$ matrices 
$\Gamma^{\ }_{\mu}=
(\Gamma^{\ }_{0},
 \Gamma^{\ }_{1},
 \Gamma^{\ }_{2})$
that realize a 4-dimensional representation of the 
algebra 
\begin{equation}
\{
\Gamma^{\ }_{\mu},
\Gamma^{\ }_{\nu}
\}=
2g^{\ }_{\mu\nu},
\qquad
g^{\ }_{\mu\nu}=
\mathrm{diag}(1,-1,-1),
\label{appeq: Gamma algebra}
\end{equation}
while they commute with 
$\Sigma^{\ }_{\mathrm{a}}=
(\Sigma^{\ }_{1},
 \Sigma^{\ }_{2},
 \Sigma^{\ }_{3})$,
\begin{equation}
\left[\Gamma^{\ }_{\mu},\Sigma^{\ }_{\mathrm{a}}\right]=0.
\label{eq: Gamma and Sigma commute}
\end{equation}
\end{subequations}

We shall work in momentum space. To this end,
we introduce the Fourier transforms
\begin{subequations}
\begin{equation}
\begin{split}
&
\bar{\chi}(x)=
\int_{k}
e^{-{i}k\cdot x}
\bar{\chi}(k),
\\
&
\chi(x)=
\int_{k}
e^{{i}k\cdot x}
\chi(k),
\\
&
B^{\ }_{\mu}(x)=
\int_{k}
e^{{i}k\cdot x}
B^{\ }_{\mu}(k),
\end{split}
\end{equation}
whereby the notations 
\begin{equation}
\begin{split}
&
k\cdot x\equiv
k^{\mu}x^{\ }_{\mu}=k^{\mu}g^{\ }_{\mu\nu}x^{\mu},
\\
&
k^{2}\equiv 
k^{\mu}k^{\ }_{\mu}=k^{\mu}g^{\ }_{\mu\nu}k^{\mu},
\\
&
\int_{k}\equiv
\int\frac{dk^{0}dk^{1}dk^{2}}{(2\pi)^{3}},
\end{split}
\end{equation}
\end{subequations}
will be used. The action and Lagrangian in 
Eq.~(\ref{eq: def Z, S, L})
are represented in momentum space by 
\begin{equation}
\begin{split}
&
S=
\int_{k^{\ }_{1},k^{\ }_{2}}
\mathcal{L},
\\
&
\mathcal{L}=
\bar{\chi}(k^{\ }_{1})
\left[
G^{-1}_{0}(k^{\ }_{1})
\delta(k^{\ }_{1}-k^{\ }_{2})
+
\Slash{B}(k^{\ }_{1}-k^{\ }_{2})
\right]
\chi(k^{\ }_{2}).
\end{split}
\end{equation}
The free propagator, here defined by
\begin{subequations}
\begin{equation}
\begin{split}
G^{\ }_{0}(k):=&\,
-
\frac{
1
     }
     {
\Slash{k}+\eta+m\Sigma^{\ }_{3}
     }
\\
=&\,
-
\frac{
\Slash{k}-\eta-m\Sigma^{\ }_{3}
     }
     {
k^{2}
-
\eta^{2}
-
m^{2}
-
2\eta m\Sigma^{\ }_{3}
     }
\\
=&\,
-
\frac{
\left(
\Slash{k}
-
\eta
-
m\Sigma^{\ }_{3}
\right)
\left(
k^{2}
-
\eta^{2}
-
m^{2}
+
2\eta m\Sigma^{\ }_{3}
\right)
     }
     {
\left(
k^{2}
-
\left(
\eta
-
m
\right)^{2}
\right)
\left(
k^{2}
-
\left(
\eta
+
m
\right)^{2}
\right)
     },
\end{split}
\end{equation}
can be decomposed into the sum of the unit $4\times4$
matrix weighted by the factor $P(k)$ and the
$4\times4$ matrix $\Sigma^{\ }_{3}$ 
weighted by the factor $Q(k)$;
\begin{equation}
\begin{split}
&
G^{\ }_{0}(k)=
P(k)
+
Q(k)\,\Sigma^{\ }_{3},
\\
&
P(k)=
-
\frac{
\Slash{k}
\left(
k^{2}
-
\eta^{2}
-
m^{2}
\right) 
-
\eta
\left(
k^{2}
-
\eta^{2}
+
m^{2}
\right) 
     }
     {
\left(
k^{2}
-
\left(
\eta
-
m
\right)^{2}
\right)
\left(
k^{2}
-
\left(
\eta
+
m
\right)^{2}
\right)
     },
\\
&
Q(k)=
-
\frac{
2\eta m
\Slash{k}
-
m
\left(
k^{2}
+
\eta^{2}
-
m^{2}
\right)
     }
     {
\left(
k^{2}
-
\left(
\eta
-
m
\right)^{2}
\right)
\left(
k^{2}
-
\left(
\eta
+
m
\right)^{2}
\right)
     }.
\end{split}
\label{appeq: def P Q}
\end{equation}
\end{subequations}

The induced effective action for the background
$\Slash{B}$ is defined by 
\begin{equation}
\begin{split}
&
\exp
\left(
{i}
S^{\ }_{\mathrm{eff}}[B]
\right)
\propto
\int
\mathcal{D}[\bar{\chi},\chi]\,
\exp({i}S[B]),
\\
&
S^{\ }_{\mathrm{eff}}[B]
:=
{i}
\sum_{n=1}^{\infty}
\frac{(-1)^n}{n}
\mathrm{Tr}\,(G^{\ }_{0}\Slash{B})^{n}
\equiv
{i}
\sum_{n=1}^{\infty}
S^{(n)}_{\mathrm{eff}}[B],
\end{split}
\label{eq: loop expansion}
\end{equation}
where it is understood that the Grassmann integration
is performed in a way that
preserves the local U(1)$\times$U(1) gauge symmetry%
~(\ref{eq: U(1) U(1) b}). 
The effective action in Eq.~(\ref{eq: L eff if m neq 0})
with the coefficients
from Table~\ref{tab:Cs}
follows by combining the
local U(1)$\times$U(1) gauge symmetry%
~(\ref{eq: U(1) U(1) b})
with the loop expansion%
~(\ref{eq: loop expansion}) 
up to the order $n=2$,
\begin{equation}
\begin{split}
&
S^{\ }_{\mathrm{eff}}[B]
\approx
{i}
S^{(1)}_{\mathrm{eff}}[B]
+
{i}
S^{(2)}_{\mathrm{eff}}[B]
+
\ldots,
\\
&
S^{(1)}_{\mathrm{eff}}[B]=
\int_{k}
\mathrm{tr}\,
\left[
G^{\ }_{0}(k)\Slash{B}(0)
\right],
\\
&
S^{(2)}_{\mathrm{eff}}[B]=
\frac{1}{2}
\int_{k,q}
\mathrm{tr}\,
\left[
G^{\ }_{0}(k)
\Slash{B}(q)
G^{\ }_{0}(k-q)
\Slash{B}(-q)
\right].
\end{split}
\end{equation}
One verifies by explicit calculation that
\begin{equation}
S^{(1)}_{\mathrm{eff}}[B]=0.
\end{equation}

To proceed with the evaluation of
$S^{(2)}_{\mathrm{eff}}[\Slash{B}]$, 
we note that the algebra
(\ref{appeq: Sigma algebra}),
(\ref{appeq: Gamma algebra}),
and (\ref{eq: Gamma and Sigma commute})
can always be realized with the choice
\begin{equation}
\Gamma^{\ }_{\mu}=
\gamma^{\ }_{\mu}
\otimes
\openone^{\ }_{\sigma},
\qquad
\Sigma^{\ }_{\mathrm{a}}=
\openone^{\ }_{\gamma}
\otimes
\sigma^{\ }_{\mathrm{a}},
\end{equation}
where 
$\gamma^{\ }_{\mu}=
(\gamma^{\ }_{0},
 \gamma^{\ }_{1},
 \gamma^{\ }_{2})$
and
$\sigma^{\ }_{\mathrm{a}}=
(\sigma^{\ }_{1},
 \sigma^{\ }_{2},
 \sigma^{\ }_{3})$
realize two-dimensional representations of the algebra
(\ref{appeq: Gamma algebra})
and
(\ref{appeq: Sigma algebra}),
respectively. With this choice, it is obvious that
a single trace over the $4\times4$ matrices spanned
by the unit $4\times4$ matrix,
$\Gamma^{\ }_{\mu}=
(\Gamma^{\ }_{0},
 \Gamma^{\ }_{1},
 \Gamma^{\ }_{2})$,
and
$\Sigma^{\ }_{\mathrm{a}}=
(\Sigma^{\ }_{1},
 \Sigma^{\ }_{2},
 \Sigma^{\ }_{3})$
factorizes into the product over two traces; 
one trace over the $2\times2$ matrices spanned
by the unit matrix 
$\openone^{\ }_{\gamma}$
and
$\gamma^{\ }_{\mu}=
(\gamma^{\ }_{0},
 \gamma^{\ }_{1},
 \gamma^{\ }_{2})$
and one trace 
over the $2\times2$ matrices spanned
by the unit matrix 
$\openone^{\ }_{\sigma}$
and
$\sigma^{\ }_{\mathrm{a}}=
(\sigma^{\ }_{1},
 \sigma^{\ }_{2},
 \sigma^{\ }_{3})
$.
It then follows that
\begin{subequations}
\begin{equation}
\begin{split}
S^{(2)}_{\mathrm{eff}}[B]=&\, 
\hphantom{+}
\frac{1}{2}
\int_{q}
\left[PP(q)\right]^{\nu\kappa}
\mathrm{tr}^{\ }_{\sigma}
\left[
B^{\ }_{\nu}(q)
B^{\ }_{\kappa}(-q)
\right]
\\
&\,
+
\frac{1}{2}
\int_{q}
\left[PQ(q)\right]^{\nu\kappa}
\mathrm{tr}^{\ }_{\sigma}
\left[
B^{\ }_{\nu}(q)
\sigma^{\ }_{3}
B^{\ }_{\kappa}(-q)
\right]
\\
&\,
+
\frac{1}{2}
\int_{q}
\left[QP(q)\right]^{\nu\kappa}
\mathrm{tr}^{\ }_{\sigma}
\left[
\sigma^{\ }_{3}
B^{\ }_{\nu}(q)
B^{\ }_{\kappa}(-q)
\right]
\\
&\,
+
\frac{1}{2}
\int_{q}
\left[QQ(q)\right]^{\nu\kappa}
\mathrm{tr}^{\ }_{\sigma}
\left[
\sigma^{\ }_{3}
B^{\ }_{\nu}(q)
\sigma^{\ }_{3}
B^{\ }_{\kappa}(-q)
\right]
\end{split}
\end{equation}
where $\left[HK(q)\right]^{\nu\kappa}$, $H$ and $K$ being $P$ or $Q$, is defined as
\begin{equation}
\left[HK(q)\right]^{\nu\kappa}:=
\int_{k}
\mathrm{tr}^{\ }_{\gamma}
\left[
H(k)
\gamma^{\nu}
K(k-q)
\gamma^{\kappa}
\right],
\label{appeq: def PP etc}
\end{equation}
and with the understanding that 
\begin{equation}
B^{\ }_{\mu}=
b^{\mathrm{a}}_{\mu}\sigma^{\mathrm{a}},
\qquad
\Slash{k}=
\gamma^{\mu} k^{\ }_{\mu}.
\end{equation}
\end{subequations}

If the integrals~(\ref{appeq: def PP etc})
are regularized so as to preserve
the Lorentz covariance, 
then they must be of the form
\begin{equation}
\left[HK(q)\right]^{\nu\kappa}=
g^{\nu\kappa}
\left[HK\right]^{(0)}
+
{i}
\epsilon^{\nu\rho\kappa}
q^{\ }_{\rho}
\left[HK\right]^{(1)}
+
\ldots,
\label{appeq: def PP small q etc}
\end{equation}
to linear order in $q$.
Furthermore, imposing a regularization
of the integrals~(\ref{appeq: def PP etc})
that preserves the local U(1)$\times$U(1) gauge symmetry%
~(\ref{eq: U(1) U(1) b})
demands that the coefficients 
\begin{equation}
\begin{split}
&
C^{(0)}_{00}=
C^{(0)}_{33}:=
\left[PP\right]^{(0)}
+
\left[QQ\right]^{(0)}=0,
\\
&
C^{(0)}_{03}:=
2
\left(
\left[PQ\right]^{(0)}
+
\left[QP\right]^{(0)}
\right)=0.
\end{split}
\end{equation}
This gives the effective Lagrangian
\begin{equation}
\begin{split}
\mathcal{L}^{(2)}_{\mathrm{eff}}=&\,
\hphantom{+}
C^{(0)}_{11}
\left(
b^{1\rho} 
b^{1}_{\rho}
+
b^{2\rho} 
b^{2}_{\rho}
\right)
\\
&\,
+
C^{(1)}_{00}
\epsilon^{\nu\rho\kappa} 
b^{0}_{\nu} 
\partial^{\ }_{\rho} 
b^{0}_{\kappa}
+
C^{(1)}_{33}
\epsilon^{\nu\rho\kappa}
b^{\mathrm{3}}_{\nu} 
\partial^{\ }_{\rho} 
b^{\mathrm{3}}_{\kappa}
\\
&\,
+
C^{(1)}_{11}
\epsilon^{\nu\rho\kappa}
\left(
b^{\mathrm{1}}_{\nu} 
\partial^{\ }_{\rho} 
b^{\mathrm{1}}_{\kappa}
+
b^{\mathrm{2}}_{\nu} 
\partial^{\ }_{\rho} 
b^{\mathrm{2}}_{\kappa}
+
\ldots
\right)
\\
&\,
+
C^{(1)}_{03}
\epsilon^{\nu\rho\kappa} 
b^{0}_{\nu}
\partial^{\ }_{\rho}
b^{3}_{\kappa}
\end{split}
\label{eq: L (2) eff}
\end{equation}
with the coefficients
\begin{equation}
\begin{split}
&
C^{(0)}_{11}:=
\left[PP\right]^{(0)}
-
\left[QQ\right]^{(0)},
\\
&
C^{(1)}_{00}=C^{(1)}_{33}:=
\left[PP\right]^{(1)}
+
\left[QQ\right]^{(1)},
\\
&
C^{(1)}_{11}:=
\left[PP\right]^{(1)}
-
\left[QQ\right]^{(1)},
\\
&
C^{(1)}_{03}:=
2
\left(
\left[PQ\right]^{(1)}
+
\left[QP\right]^{(1)}
\right).
\end{split}
\label{eq: C's terms PP ...}
\end{equation}

\begin{widetext}

\medskip
\medskip

The coefficients~(\ref{eq: C's terms PP ...})
are evaluated by performing a Wick rotation to
the Euclidean metric with the rules
\begin{equation}
\begin{split}
&
t\to
-{i}\tau,
\qquad
\boldsymbol{r}
\to
\boldsymbol{r},
\qquad
\gamma^{0}\to
\gamma^{0},
\qquad
\boldsymbol{\gamma}\to
{i}
\boldsymbol{\gamma},
\qquad
b^{\mathrm{a}}_{0}\to
{i}b^{\mathrm{a}}_{0},
\qquad
\boldsymbol{b}^{\mathrm{a}}\to
-\boldsymbol{b}^{\mathrm{a}},
\qquad
\mathrm{a}=1,2,3.
\end{split}
\end{equation}
Under these rules
\begin{equation}
g^{\mu\nu}\to 
-
\delta_{\ }^{\mu\nu},
\qquad
\epsilon^{\mu\nu\lambda}\to
{i}
\epsilon_{\ }^{\mu\nu\lambda},
\end{equation}
while the scalar functions
$P(k)$ and $Q(k)$ in the
propagator~(\ref{appeq: def P Q}) 
take the form
\begin{subequations}
\begin{equation}
\begin{split}
&
P(k)=
\left[
R(k)\Slash{k} 
+ 
S(k)
\right]
\otimes
\openone^{\ }_{\sigma},
\qquad
Q(k)=
\left[
T(k)
\Slash{k} 
+ 
U(k)
\right]
\otimes
\openone^{\ }_{\sigma},
\end{split}
\end{equation}
with $m^{\ }_{\pm}:=m\pm\eta$ and
\begin{equation}
\begin{split}
&
R(k)=
\frac{
{i}
\left(
k^{2}
+
\frac{
m^{2}_{+} 
+ 
m^{2}_{-}
     }
     {
2
     }
\right)
     }
     {
\left(
k^{2} 
+ 
m^{2}_{+} 
\right)
\left(
k^{2} 
+ 
m^{2}_{-} 
\right)
     },
\qquad
S(k)=
\frac{
-
\eta
\left(
k^{2}
- 
m_{+} 
m_{-} 
\right)
     }
     {
\left(
k^{2} 
+ 
m^{2}_{+} 
\right)
\left(
k^{2} 
+ 
m^{2}_{-} 
\right)
     },
\\
&
T(k)=
\frac{
-
2
\eta 
m 
{i}
     }
     {
\left(
k^{2} 
+ 
m^{2}_{+} 
\right)
\left(
k^{2} 
+ 
m^{2}_{-} 
\right)
      },
\qquad
U(k)=
\frac{
- 
m
\left(
k^{2}
+ 
m_{+} 
m_{-}
\right))
     }
     {
\left(
k^{2} 
+ 
m^{2}_{+} 
\right)
\left(
k^{2} 
+ 
m^{2}_{-} 
\right)}.
\end{split}
\end{equation}
\end{subequations}
Their small $q$ expansion are
\begin{subequations}
\begin{equation}
\begin{split}
&
R(k-q)=
R(k)
+
(k\cdot q)
R^{(1)}(k),
\qquad
S(k-q)=
S(k)
+
(k\cdot q)
S^{(1)}(k),
\\
&
T(k-q)=
T(k)
+
(k\cdot q)
T^{(1)}(k),
\qquad
U(k-q)=
U(k)
+
(k\cdot q)
U^{(1)}(k),
\end{split} 
\end{equation}
where 
\begin{equation}
\begin{split}
&
R^{(1)}(k)=
\frac{
2{i}
     }
     {
\left(
k^{2} 
+ 
m^{2}_{+}  
\right)^{2}
\left(
k^{2} 
+ 
m^{2}_{-} 
\right)^{2}}
\left(
k^{4} 
+ 
\left(
m^{2}_{+} 
+ 
m^{2}_{-}
\right) 
k^{2} 
-
m^{2}_{+} 
m^{2}_{-} 
+ 
\frac{ 
\left(
m^{2}_{+} 
+ 
m^{2}_{-}
\right)^{2}
     }
     {
2
     }
\right),
\\
&
S^{(1)}(k)=
\frac{
-2\eta 
     }
     {
\left(
k^{2} 
+ 
m^{2}_{+}  
\right)^{2}
\left(
k^{2} 
+ 
m^{2}_{-} 
\right)^{2}
     }
\left[
k^{4} 
- 
2 k^{2} 
m^{\ }_{+} 
m^{\ }_{-} 
-
m^{2}_{+} 
m^{2}_{-} 
-
m^{\ }_{+} 
m^{\ }_{-}  
\left(
m^{2}_{+} 
+ 
m^{2}_{-}
\right)
\right],
\\
&
T^{(1)}(k)=
\frac{
-
4
\eta 
m
{i} 
     }
     {
\left(
k^{2} 
+ 
m^{2}_{+} 
\right)^{2}
\left(
k^{2} 
+ 
m^{2}_{-} 
\right)^{2}
     }
\left[
2 k^{2} 
+ 
\left(
m^{2}_{+} 
+ 
m^{2}_{-}
\right)
\right],
\\
&
U^{(1)}(k)=
\frac{
-2m
     }
     {
\left(
k^{2} 
+ 
m^{2}_{+}  
\right)^{2}
\left(
k^{2} 
+ 
m^{2}_{-} 
\right)^{2}
     }
\left[
k^{4} 
+ 
2 k^{2} 
m^{\ }_{+} 
m^{\ }_{-} 
-
m^{2}_{+} 
m^{2}_{-} 
+
m^{\ }_{+} 
m^{\ }_{-}  
\left(
m^{2}_{+} 
+ 
m^{2}_{-}
\right)
\right].
\end{split}
\end{equation}
\end{subequations}

\end{widetext}

\noindent
At last, the coefficients%
~(\ref{eq: C's terms PP ...})
follow from
\begin{subequations}
\begin{equation}
\begin{split}
&
C^{(0)}_{11}=
-
\left(
\left[PP\right]^{(0)}
-
\left[QQ\right]^{(0)}
\right),
\\
&
C^{(1)}_{00}=C^{(1)}_{33}=
-i
\left(
\left[PP\right]^{(1)}
+
\left[QQ\right]^{(1)}
\right),
\\
&
C^{(1)}_{11}=
-
{i}
\left(
\left[PP\right]^{(1)}
-
\left[QQ\right]^{(1)}
\right),
\\
&
C^{(1)}_{03}=
-
2
{i}
\left(
\left[PQ\right]^{(1)}
+
\left[QP\right]^{(1)}
\right),
\end{split}
\label{eq: C's terms PP Euclidean}
\end{equation}
with
\begin{equation}
\begin{split}
&
\left[PP\right]^{(0)}=
\frac{4\pi}{(2\pi)^3}
\left(
-\frac{2}{3}
I^{\ }_{RR}
+
2 
\tilde{I}^{\ }_{SS}
\right),
\\
&
\left[PP\right]^{(1)}=
\frac{8\pi}{(2\pi)^3}
\left(
-
\frac{1}{3}
I^{\ }_{RS^{(1)}}
- 
\tilde{I}^{\ }_{SR}
+ 
\frac{1}{3}
I^{\ }_{S R^{(1)}}
\right),
\end{split}
\end{equation}

\begin{equation}
\begin{split}
&
\left[PQ\right]^{(0)}=
\frac{4\pi}{(2\pi)^3}
\left(
-\frac{2}{3}
I^{\ }_{RT}
+
2
\tilde{I}^{\ }_{SU}
\right),
\\
&
\left[PQ\right]^{(1)}=
\frac{8\pi}{(2\pi)^3}
\left(
-
\frac{1}{3}
I^{\ }_{R U^{(1)} }
-
\tilde{I}^{\ }_{ST}
+
\frac{1}{3}
I^{\ }_{ST^{(1)}}
\right),
\end{split}
\end{equation}

\begin{equation}
\begin{split}
&
\left[QP\right]^{(0)}=
\frac{4\pi}{(2\pi)^3}
\left(
-\frac{2}{3}
I^{\ }_{TR}
+
2
\tilde{I}^{\ }_{US}
\right),
\\
&
\left[QP\right]^{(1)}=
\frac{8\pi}{(2\pi)^3}
\left(
-
\frac{1}{3}
I^{\ }_{TS^{(1)}}
-
\tilde{I}^{\ }_{UR}
+
\frac{1}{3}
I^{\ }_{UR^{(1)}}
\right),
\end{split}
\end{equation}

\begin{equation}
\begin{split}
&
\left[QQ\right]^{(0)}=
\frac{4\pi}{(2\pi)^3}
\left(
-\frac{2}{3}
I^{\ }_{TT}
+
2 
\tilde{I}^{\ }_{UU}
\right),
\\
&
\left[QQ\right]^{(1)}=
\frac{8\pi}{(2\pi)^3}
\left(
-\frac{1}{3}
I^{\ }_{TU^{(1)}}
-
\tilde{I}^{\ }_{UT}
+
\frac{1}{3}
I^{\ }_{UT^{(1)}}
\right).
\end{split}
\end{equation}
Here, the integrals
\begin{equation}
\label{eq: integrals I...}
\begin{split}
&
I^{\ }_{XY}\equiv
\int\limits^{\infty}_{0}  dk
k^{4}
X(k)
Y(k),
\\
&
\tilde{I}^{\ }_{XY}\equiv
\int\limits^{\infty}_{0}  dk
k^{2}
X(k) 
Y(k),
\end{split}
\end{equation}
\end{subequations}
with 
$X(k)$ and $Y(k)$ 
denoting 
$R(k)$, 
$R^{(1)}(k)$, 
$S(k)$, 
$S^{(1)}(k)$, 
$T(k)$, 
$T^{(1)}(k)$, 
$U(k)$,
or $U^{(1)}(k)$
need to be regularized in a way
that preserves the local U(1)$\times$U(1) gauge symmetry%
~(\ref{eq: U(1) U(1) b})
and the Lorentz covariance.
The brute force method consisting in imposing the ultra-violet
cutoff $\Lambda$ in the integrals~(\ref{eq: integrals I...})
and ignoring all the terms linear in $\Lambda$
delivers the coefficients from Table~\ref{tab:Cs}.

\section{
Duality and statistics in the quantum $XY$ model with a Chern-Simons term
        }
\label{sec:duality}

The presentation of the effective
action~(\ref{eq: final effective action})
with the help of Table~\ref{tab:Cs}
is not optimal for the purpose of
extracting the statistical angle $\Theta$
acquired by the pairwise exchange
of unit vortices from
Sec.~\ref{sec: Fractional charge quantum number}.
Needed is a conserved vortex current that 
accounts for the local vortex density and 
the local vortex current generated by the physical
process involving the exchange of two vortices.
This vortex current can be non-vanishing
anywhere in the phase diagram in Fig.~\ref{fig:phase-diagram}.
Thus, an optimal presentation of
the effective action~(\ref{eq: final effective action})
should include this vortex current.
This can be achieved by taking advantage of the duality between
the quantum $XY$ model and compact quantum electrodynamics
in (2+1)-dimensional space and time,%
\cite{Banks77,Thomas78,Peskin78,Dasgupta81,Fisher89}
which we now briefly adapt for our purpose.

\subsection{
Duality
           }

We begin by defining the partition function for the quantum $XY$ model
in (2+1)-dimensional space and time with an additional Chern-Simons
term,
\begin{equation}
\begin{split}
&
Z^{CS}_{XY}:=
\int
\mathcal{D}[\theta]
e^{
{i}\int d^{3}x\,
\left(
\mathcal{L}^{\ }_{XY}[\theta] 
+
\mathcal{L}^{\ }_{CS}[\theta] 
\right)
  },
\\
&
\mathcal{L}^{\ }_{XY}[\theta]:=
\frac{K}{2}
\left(\partial^{\ }_{\mu}\theta\right)
\left(\partial^{\mu}\theta\right),
\\
&
\mathcal{L}^{\ }_{CS}[\theta]:=
\frac{\kappa}{4\pi}\,
\epsilon^{\mu\nu\lambda}
\left(\partial^{\ }_{\mu}\theta\right)
\partial^{\ }_{\nu}
\left(\partial^{\ }_{\lambda}\theta\right).
\end{split}
\label{eq: def Z quantum XY+CS 2+1}
\end{equation}
The Chern-Simons action 
$\mathcal{L}^{\ }_{CS}$ 
can be rewritten using an auxiliary vector gauge field
$d^{\ }_{\mu}$, 
\begin{subequations}
\begin{equation}
\begin{split}
&
Z^{CS}_{XY}:=
\int
\mathcal{D}[\theta]\;
\mathcal{D}[d_\mu]\;\;
e^{
{i}\int d^{3}x\,
\left(
\mathcal{L}^{\ }_{XY}[\theta] 
+
\mathcal{L}^{\ }_{d}[\theta,d_\mu] 
\right)
  }
\end{split}
\end{equation}
with
\begin{equation}
\begin{split}
\mathcal{L}^{\ }_{d}[\theta,d_\mu]:=&\,
\frac{\kappa}{4\pi}\,
\epsilon^{\mu\nu\lambda}
d^{\ }_{\mu}\partial^{\ }_{\nu}d^{\ }_{\lambda}
+
\frac{\kappa}{2\pi}\,
d^{\ }_{\mu}
\epsilon^{\mu\nu\lambda}
\partial^{\ }_{\nu}
\partial^{\ }_{\lambda}\theta.
\end{split}
\end{equation}
\end{subequations}
Observe that the Chern-Simons gauge field $d^{\ }_{\mu}$
couples to the current
\begin{equation}
\bar{j}^{\mu}_{\mathrm{vrt}}\equiv
\frac{1}{2\pi}
\epsilon^{\mu\nu\lambda}
\partial^{\ }_{\nu}
\partial^{\ }_{\lambda}
\theta.
\label{eq: def bar j vrt}
\end{equation}
This current is necessarily conserved
\begin{equation}
0=\partial^{\ }_{\mu}\,\bar{j}^{\mu}_{\mathrm{vrt}}.
\label{eq: topol current conserved}
\end{equation}

We wish to constrain all configurations $\theta$
appearing in the partition function
$Z^{CS}_{XY}$
by the condition~(\ref{eq: topol current conserved}),
i.e., we wish to restrict $\theta$ to any configuration such 
that it supports the conserved current 
$\bar{j}^{\mu}_{\mathrm{vrt}}$.
We call such configurations vortex configurations.

The condition of current conservation%
~(\ref{eq: topol current conserved})
can be enforced by the three Lagrange multipliers
$c^{\ }_{\mu}$ $\mu=0,1,2$. If so, the following
partition function restricted to vortex configurations 
follows,
\begin{subequations}
\label{eq: Z CS XY vrt}
\begin{equation}
\begin{split}
Z^{CS}_{XY\,\mathrm{vrt}}:=&\,
\int
\mathcal{D}[\theta]
\mathcal{D}[c^{\ }_{\mu}]
\mathcal{D}[d^{\ }_{\mu}]
e^{
{i}\int d^{3}x\,
\mathcal{L}^{CS}_{XY\,\mathrm{vrt}}
\left[
\theta,c^{\ }_{\mu},d^{\ }_{\mu}
\right] 
  }
\end{split}
\end{equation}
with
\begin{equation}
\begin{split}
\mathcal{L}^{CS}_{XY\,\mathrm{vrt}}
[\theta,c^{\ }_{\mu},d^{\ }_{\mu}]=&\,
\frac{K}{2}
\left(\partial^{\ }_{\mu}\theta\right)
\left(\partial^{\mu}\theta\right)
\\
&+
\frac{\kappa}{4\pi}\,
\epsilon^{\mu\nu\lambda}
d_\mu\partial_\nu d_\lambda
+
\kappa\,
d_\mu \;\bar{j}^{\mu}_{\mathrm{vrt}}
\\
&
-\frac{1}{2\pi}\;
{}^*\!f^\lambda\;
\partial^{\ }_{\lambda}
\theta
+c_\mu\;\bar{j}^{\mu}_{\mathrm{vrt}}
\;.
\end{split}
\end{equation}
\end{subequations}
Here, we have introduced the field
\begin{subequations}
\begin{equation}
{}^*\!f^\lambda\equiv
\epsilon^{\lambda\nu\mu}
\partial^{\ }_{\nu} c^{\ }_\mu,
\end{equation}
whose dual field is given by
\begin{equation}
f^{\ }_{\mu\nu}=
\epsilon_{\mu\nu\lambda}
\;{}^*\!f^\lambda
=
\partial^{\ }_{\mu} c^{\ }_{\nu}
-
\partial^{\ }_{\nu} c^{\ }_{\mu},
\end{equation}
\end{subequations}
and we dropped total derivatives 
after performing partial integrations.

The equation of motion for $\theta$ gives the condition
\begin{equation}
\partial^{\mu}
\theta=
\frac{1}{2\pi K}
{}^*\!f^\mu,
\end{equation}
from which we recover the inhomogeneous Maxwell equation
\begin{equation}
\bar{j}^{\mu}_{\mathrm{vrt}}=
\frac{1}{4\pi^2 K}
\partial^{\ }_{\nu}
f^{\mu\nu}.
\end{equation}

After integration over $\theta$, 
the partition function~(\ref{eq: Z CS XY vrt})
becomes
\begin{subequations}
\begin{equation}
\begin{split}
Z^{CS}_{XY\,\mathrm{vrt}}=&\,
\int
\mathcal{D}[c^{\ }_{\mu}]
\mathcal{D}[d^{\ }_{\mu}]
e^{
{i}\int d^{3}x\,
\mathcal{L}^{CS}_{\mathrm{vrt}}[c^{\ }_{\mu},d^{\ }_{\mu}] 
  }
\end{split}
\end{equation}
with
\begin{equation}
\begin{split}
\mathcal{L}^{CS}_{\mathrm{vrt}}[c^{\ }_{\mu},d^{\ }_{\mu}]=&\,
\frac{\kappa}{4\pi}\,
\epsilon^{\mu\nu\lambda}
d^{\ }_{\mu}\partial^{\ }_{\nu} d^{\ }_{\lambda}
+
\kappa
d^{\ }_{\mu}
\bar{j}^{\mu}_{\mathrm{vrt}}
\\
&\,
-
\frac{1}{16\pi^2 K}
f^{\mu\nu}\,f^{\ }_{\mu\nu}
+
c^{\ }_{\mu}
\bar{j}^{\mu}_{\mathrm{vrt}}.
\end{split}
\end{equation}
\end{subequations}
The dynamical gauge fields 
$c^{\ }_{\mu}$ and $d^{\ }_{\mu}$
that couple to the vortex
current $j^{\mu}_{\mathrm{vrt}}$ 
have a Maxwell for $c^{\ }_{\mu}$
and Chern-Simons for $d^{\ }_{\mu}$
kinetic energy. They endow the quantum theory with
an explicit U(1)$\times$U(1) local gauge symmetry.

\subsection{
Exchange statistics
           }

We turn our attention to the
computation of the exchange statistics of vortices 
with current $\bar{j}^{\mu}_{\mathrm{vrt}}$
interacting through the Chern-Simons action
\begin{equation}
\begin{split}
\mathcal{L}^{\ }_{\mathrm{eff}} [d^{\ }_{\mu}]:=&\,
\frac{\kappa}{4\pi}\,
\left(
\epsilon^{\mu\nu\lambda}
d^{\ }_{\mu}\partial^{\ }_{\nu}d^{\ }_{\lambda}
+
4\pi
d^{\ }_{\mu}\,
\bar{j}^{\mu}_{\mathrm{vrt}}
\right).
\end{split}
\end{equation}
The relationship between the current and the field that results from
the equations of motion is
\begin{equation}
\bar{j}^{\mu}_{\mathrm{vrt}}=
-
\frac{1}{2\pi}
\epsilon^{\mu\nu\lambda}
\partial^{\ }_{\nu} d^{\ }_{\lambda}.
\end{equation}
Hence, the vorticity 
\begin{equation}
n^{\ }_{\theta}=
\int d^{2}\boldsymbol{r}\,
\bar{j}^{0}_{\mathrm{vrt}}(\boldsymbol{r})
\end{equation}
supported by the vortex current
$\bar{j}^{\mu}_{\mathrm{vrt}}$
is related to the circulation from the gauge potential 
$d^{\ }_{\mu}$ 
through
\begin{equation}
n^{\ }_{\theta}=
-
\frac{1}{2\pi}
\int d^{2}\boldsymbol{r}\,
\left(
\partial^{\ }_{1} d^{\ }_{2}
-
\partial^{\ }_{2} d^{\ }_{1}
\right)
=
-
\frac{1}{2\pi}
\oint 
d\boldsymbol{l}
\cdot
\boldsymbol{d}.
\label{eq:n-d-rel}
\end{equation}

Consider now winding two vortices, with vorticities 
$n^{\ }_{1}$ and $n^{\ }_{2}$ around each other.
Without loss of generality,
suppose that we hold vortex 1 at the location 
$\boldsymbol{x}^{\ }_{1}$
fixed and move vortex 2 along any closed trajectory 
$\boldsymbol{x}^{\ }_{2}(t)$
that encircles once   
$\boldsymbol{x}^{\ }_{1}$. 
On the one hand, the field 
$d^{(1)}_{\mu}(\boldsymbol{x}^{\ }_{2})$ 
at the location of vortex 2 that is induced by vortex 1 must then satisfy,
according to Eq.~(\ref{eq:n-d-rel}),
\begin{equation}
\oint d\boldsymbol{x}^{\ }_{2}\, 
\cdot 
\boldsymbol{d}^{(1)}(\boldsymbol{x}^{\ }_{2})=
-2\pi n^{\ }_{1}.
\end{equation}
On the other hand,
the vector current resulting from moving vortex 2 around vortex 1 is
\begin{equation}
\bar{\boldsymbol{j}}^{(2)}_{\mathrm{vrt}}(t,\boldsymbol{x})=
n^{\ }_{2}
\frac{d\boldsymbol{x}^{\ }_{2}}{dt}\,
\delta
\left(
\boldsymbol{x}
-
\boldsymbol{x}^{\ }_{2}(t)
\right).
\end{equation}
Finally, the Berry phase acquired by winding 
vortex 2 around vortex 1 is
\begin{equation}
\begin{split}
2\Theta
=&\,
\frac{\kappa}{4\pi}
\left(
4\pi n^{\ }_{2}
\int dt\,
\boldsymbol{d}^{(1)}
\cdot
\frac{d\boldsymbol{x}^{\ }_{2}}{dt}
\right)
\\
=&\,
\frac{\kappa}{4\pi}
\left(
4\pi n^{\ }_{2}
\oint 
d\boldsymbol{x}^{\ }_{2}
\cdot
\boldsymbol{d}^{(1)}
\right)
\\
=&\,
\frac{\kappa}{4\pi}
\left(
-
8\pi^{2}
n^{\ }_{1}
n^{\ }_{2}
\right).
\end{split}
\end{equation}
We conclude that the statistical phase $\Theta$,
which is one-half of the Berry phase, is given by 
\begin{equation}
\begin{split}
\frac{\Theta}{\pi}
&=
-\kappa\;n_1 \;n_2
\;.
\end{split}
\end{equation}
In particular, for a positive unit vortex winding around a
negative unit vortex (anti-vortex), we find the statistical phase
\begin{equation}
\begin{split}
\frac{\Theta}{\pi}
&=
\kappa.
\end{split}
\end{equation}

\section{
Berry phase in the single-particle approximation
        }
\label{appsec: Numerical Berry phase in the single-particle approximation}

We are going to describe how Aharonov-Bohm phases 
$\gamma$ or, more generally, 
Berry phases $\Theta$ accumulated under the pairwise exchanges
of quasi-particles can be computed 
for non-interacting models of fermions defined on lattices.

We first discuss Berry phases for lattice models of non-interacting
fermions in all generality. 
We then specialize to the case of the $\pi$ flux phase
for which we define vortices, axial gauge half fluxes, etc.

\subsection{
Berry phase on the lattice
           }

Assume that we are given a lattice model, whose sites 
$\boldsymbol{r}$ and internal degrees of freedom 
are collectively denoted by the latin index 
$m$, that describes the
quantum dynamics of non-interacting fermions. In second
quantization, if the creation
$\hat{c}^{\dag}_{m}$
and annihilation
$\hat{c}^{\   }_{n}$
obey the usual fermion algebra
\begin{subequations}
\begin{equation}
\left\{
\hat{c}^{\   }_{m},
\hat{c}^{\dag}_{n}
\right\}=
\delta^{\ }_{m,n},
\qquad
\left\{
\hat{c}^{\dag}_{m},
\hat{c}^{\dag}_{n}
\right\}=
\left\{
\hat{c}^{\  }_{m},
\hat{c}^{\  }_{n}
\right\}=
0,
\end{equation}
then we take our non-interacting Hamiltonian to be
\begin{equation}
\hat{H}:=
-
\sum_{m,n}
t^{\ }_{mn}
\hat{c}^{\dag}_{m}
\hat{c}^{\   }_{n}
\end{equation}
where the matrix $t$ with the matrix elements $t^{\ }_{m n}$
is Hermitian,
\begin{equation}
t^{\ }_{mn}=t^{* }_{nm}.
\end{equation}
\end{subequations}

We shall call the matrix $t$ the background. Its 
uniform diagonal matrix elements (the chemical potential)
fixes the average number of fermions.
We shall assume that some choices for the matrix $t$
can be associated with point-like defects. These 
point-like defects can thus be labeled by their positions 
$\boldsymbol{r}^{\ }_{1}$,
$\boldsymbol{r}^{\ }_{2}$,
$\ldots$ on the lattice with their corresponding backgrounds
$t^{\ }_{\boldsymbol{r}^{\ }_{1},\boldsymbol{r}^{\ }_{2},\ldots}$.
For a given filling fraction, the many-body ground state
in the background 
$t^{\ }_{\boldsymbol{r}^{\ }_{1},\boldsymbol{r}^{\ }_{2},\ldots}$
of point-like defects is the Fermi sea
\begin{equation}
\left|
t^{\ }_{\boldsymbol{r}^{\ }_{1},\boldsymbol{r}^{\ }_{2},\ldots}
\right\rangle:=
{\prod_{m}}'\,
\hat{c}^{\dag}_{m}|0\rangle.
\end{equation}
Here, the prime over the product means that only the lowest
single-particle energy eigenstates are to be filled up to 
the given filling fraction out of the state $|0\rangle$
annihilated by the $\hat{c}^{\ }_{m}$.

Imagine that we move the $k$-th point-like defect along a closed path
$\mathcal{P}^{\ }_{k}$ of counterclockwise orientation while holding all
other point-like defects fixed.  We then discretize the path, 
thereby defining $N$ backgrounds 
$t^{(n)}_{\mathcal{P}^{\ }_{k}}$, $n=1,\dots,N$.

The gauge invariant phase 
$\gamma^{\ }_{\mathcal{P}^{\ }_{k}}$
is defined by
\begin{equation}
\gamma^{\ }_{\mathcal{P}^{\ }_{k}}
:=
-
\sum_{n=1}^{N}
\mathrm{arg}\,
\left\langle
\left.
t^{(n)}_{\mathcal{P}^{\ }_{k}}
\right|
t^{(n+1)}_{\mathcal{P}^{\ }_{k}}
\right\rangle.
\label{eq: def gamma l in P k}
\end{equation}
If we do this exercise for two cases, one when the path
$\mathcal{P}^{\ }_{k}$ encircles another defect $l$ and another one 
when the defect $l$ lies outside the path $\mathcal{P}^{\ }_{k}$, 
we obtain the statistical phase $\Theta^{\ }_{kl}$ acquired by the
counterclockwise exchange of point-like defects $k$ and $l$ from
\begin{equation}
\Theta^{\ }_{kl}:=
\frac{1}{2}
\left(
\gamma^{\ }_{\;l\;{\rm inside}\;\mathcal{P}^{\ }_{k}}
-
\gamma^{\ }_{\;l\;{\rm outside}\;\mathcal{P}^{\ }_{k}}
\right).
\label{eq: exchange stat phase}
\end{equation}
This phase does not depend on the presence of other
static point-defect inside the path 
$\mathcal{P}^{\ }_{k}$,
for their contributions to
$\gamma^{\ }_{\;l\;{\rm inside}\;\mathcal{P}^{\ }_{k}}$
cancel their contributions to
$\gamma^{\ }_{\;l\;{\rm outside}\;\mathcal{P}^{\ }_{k}}$.
 
The overlaps
\begin{equation}
\Gamma^{\ }_{n,n+1}:=
\left\langle
\left.
t^{(n)}_{\mathcal{P}^{\ }_{k}}
\right|
t^{(n+1)}_{\mathcal{P}^{\ }_{k}}
\right\rangle
\label{eq: overlap}
\end{equation}
from Eq.~(\ref{eq: def gamma l in P k})
can be presented as the determinants for the products 
between two matrices built out of the eigenvectors of
$t^{(n)}$ and $t^{(n+1)}$,
as we now show. For any background $t$, 
define the unitary transformation $U$ by
\begin{equation}
U\, t\, U^{\dag}=
\mathrm{diag}
\begin{pmatrix}
\varepsilon^{\ }_{m}
\end{pmatrix},
\end{equation}
i.e., $U$ is the matrix of eigenvectors with energies
$\varepsilon^{\ }_{m}$ of the single-particle
Hermitian matrix $t$.
For the two backgrounds entering
the overlap~(\ref{eq: overlap}),
these unitary transformations are
denoted by $U^{\ }_{n}$ and $U^{\ }_{n+1}$, respectively.
One then verifies that
\begin{equation}
\Gamma^{\ }_{n,n+1}=
\mathrm{det}
\left(
U^{\dag}_{n}
U^{\   }_{n+1}
\right).
\end{equation}
Evaluation of the phases~(\ref{eq: def gamma l in P k})
or~(\ref{eq: exchange stat phase})
requires $N$ diagonalizations and the
multiplication of $N$ determinants,
a computing exercise that scales as a power law in 
the number of sites in the lattice.

\subsection{
Lattice defects for the $\pi$ flux phase
           }

Consider a square Bravais lattice $\Lambda$
that is spanned by the orthogonal basis
of vectors 
$\boldsymbol{s}^{\ }_{1}$
and
$\boldsymbol{s}^{\ }_{2}$
of length  $\mathfrak{a}$, the lattice spacing.
We shall also define
$\boldsymbol{s}^{\ }_{3}\equiv-\boldsymbol{s}^{\ }_{1}$
and
$\boldsymbol{s}^{\ }_{4}\equiv-\boldsymbol{s}^{\ }_{2}$.
The square lattice is the union of two interpenetrating square lattices
$\Lambda^{\ }_{\mathrm{A}}$
and
$\Lambda^{\ }_{\mathrm{B}}$
with lattice spacing $\sqrt{2}\,\mathfrak{a}$.
Any site 
$\boldsymbol{r}^{\ }_{\mathrm{B}}\in\Lambda^{\ }_{\mathrm{B}}$
can be decomposed in a unique way
according to 
$\boldsymbol{r}^{\ }_{\mathrm{B}}=
 \boldsymbol{r}^{\ }_{\mathrm{A}}+\boldsymbol{s}^{\ }_{1}$
with $\boldsymbol{r}^{\ }_{\mathrm{A}}\in\Lambda^{\ }_{\mathrm{A}}$.

Because of the bipartite nature of the square lattice, we introduce
fermionic annihilation operators denoted 
$\hat{a}^{\ }_{\boldsymbol{r}^{\ }_{\mathrm{A}}}$ 
and 
$\hat{b}^{\ }_{\boldsymbol{r}^{\ }_{\mathrm{B}}}$ 
and their adjoints for any site 
$\boldsymbol{r}^{\ }_{\mathrm{A}}$
and 
$\boldsymbol{r}^{\ }_{\mathrm{B}}$
of the sublattice
$\Lambda^{\ }_{\mathrm{A}}$ 
and
$\Lambda^{\ }_{\mathrm{B}}$,
respectively. These operators obey the usual fermionic algebra
with the only non-vanishing anticommutators
\begin{eqnarray}
\left\{
\hat{a}^{\   }_{\boldsymbol{r}^{\ }_{\mathrm{A}}},
\hat{a}^{\dag}_{\boldsymbol{r}^{\ }_{\mathrm{A}}}
\right\}=1,
\qquad
\left\{
\hat{b}^{\   }_{\boldsymbol{r}^{\ }_{\mathrm{B}}},
\hat{b}^{\dag}_{\boldsymbol{r}^{\ }_{\mathrm{B}}}
\right\}=1.
\end{eqnarray}

The square lattice with a flux of $\pi$
per plaquette (the $\pi$ flux phase in short)
is the non-interacting tight-binding Hamiltonian
\begin{subequations}
\label{eq: def pi H}
\begin{equation}
\hat{H}^{\ }_{\pi}:=
-\sum_{\boldsymbol{r}\in\Lambda^{\ }_{\mathrm{A}}}
\sum_{j=1}^{4}
\left(
t^{(\pi)}_{\boldsymbol{r},\boldsymbol{r}+\boldsymbol{s}^{\ }_{j}}
\hat{b}^{\dag}_{\boldsymbol{r}+\boldsymbol{s}^{\ }_{j}} 
\hat{a}^{\ }_{\boldsymbol{r}}
+
\mathrm{H.c.}
\right)
\label{eq: def pi H a}
\end{equation}
with the (gauge dependent) choice of the tunneling amplitudes
\begin{equation}
\begin{split}
&
t^{(\pi)}_{\boldsymbol{r},\boldsymbol{r}+\boldsymbol{s}^{\ }_{1}}=
t^{(\pi)}_{\boldsymbol{r},\boldsymbol{r}+\boldsymbol{s}^{\ }_{3}}=
e^{{i}\pi/2} t=
{i}t, 
\\
&
t^{(\pi)}_{\boldsymbol{r},\boldsymbol{r}+\boldsymbol{s}^{\ }_{2}}=
t^{(\pi)}_{\boldsymbol{r},\boldsymbol{r}+\boldsymbol{s}^{\ }_{4}}=
t.
\end{split}
\label{eq: def pi H b}
\end{equation}
\end{subequations}

Time-reversal symmetry is the property that
$\hat{H}^{* }_{\pi}$
is locally gauge equivalent to
$\hat{H}^{\ }_{\pi}$.
Sublattice symmetry is the property that
$\hat{H}^{\ }_{\pi}\to-\hat{H}^{\ }_{\pi}$
under the local gauge transformation
\begin{equation}
\hat{a}^{\   }_{\boldsymbol{r}}\to
+\hat{a}^{\   }_{\boldsymbol{r}},
\qquad
\hat{b}^{\   }_{\boldsymbol{r}+\boldsymbol{s}^{\ }_{1}}\to
-
\hat{b}^{\   }_{\boldsymbol{r}+\boldsymbol{s}^{\ }_{1}}.
\label{eq: SLS gauge trsf}
\end{equation}

At half-filling, the Fermi surface collapses to two non-equivalent 
Fermi points due to the breaking of translation invariance,
for the unit cell is now the unit cell of the sublattice
$\Lambda^{\ }_{\mathrm{\mathrm{A}}}$ with two atoms per unit cell.
At half-filling, there are four non-equivalent ways to open a gap.

There is the charge-density wave instability through the perturbation
\begin{equation}
\hat{H}^{\ }_{\mu^{\ }_{\mathrm{s}}}:= t\,
\mu^{\ }_{\mathrm{s}}
\sum_{\boldsymbol{r}\in\Lambda^{\ }_{\mathrm{A}}}
\left(
\hat{a}^{\dag}_{\boldsymbol{r}} 
\hat{a}^{\   }_{\boldsymbol{r}}
-
\hat{b}^{\dag}_{\boldsymbol{r}+\boldsymbol{s}^{\ }_{1}} 
\hat{b}^{\   }_{\boldsymbol{r}+\boldsymbol{s}^{\ }_{1}}
\right)
\end{equation}
that breaks the sublattice symmetry 
$\hat{H}^{\ }_{\pi}\to-\hat{H}^{\ }_{\pi}$
of Hamiltonian~(\ref{eq: def pi H})
under the local gauge transformation%
~(\ref{eq: SLS gauge trsf})
but preserves time-reversal symmetry, 
$\hat{H}^{\ }_{\mu^{\ }_{\mathrm{s}}}=
 \hat{H}^{* }_{\mu^{\ }_{\mathrm{s}}}$.

There is the bond-density wave instability
through the perturbation
\begin{subequations}
\label{eq: BDW pi flux phase}
\begin{equation}
\hat{H}^{\ }_{\Delta}:=
-\sum_{\boldsymbol{r}\in\Lambda^{\ }_{\mathrm{A}}}
\sum_{j=1}^{4}
\left(
\delta
t^{\ }_{\boldsymbol{r},\boldsymbol{r}+\boldsymbol{s}^{\ }_{j}}\,
\hat{b}^{\dag}_{\boldsymbol{r}+\boldsymbol{s}^{\ }_{j}} 
\hat{a}^{\ }_{\boldsymbol{r}}
+
\mathrm{H.c.}
\right)
\label{eq: BDW pi flux phase a}
\end{equation}
with the tunneling amplitudes
\begin{equation}
\begin{split}
\delta
t^{\ }_{\boldsymbol{r},\boldsymbol{r}+\boldsymbol{s}^{\ }_{j}}:=&\,
\frac{t}{4}
\left[ 
{i}
\left(
\delta^{\ }_{j,1}
+ 
\delta^{\ }_{j,3}
\right)
+ 
\left(
\delta_{j,2}
+
\delta_{j,4} 
\right) 
\right] 
\\
&\,
\times
\left(
\Delta^{\ } 
e^{+{i}\frac{\pi}{2}j}
e^{
+
{i}
\boldsymbol{G}
\cdot
\boldsymbol{r}
  }
+
\mathrm{c.c}
\right),
\end{split}
\label{eq: BDW pi flux phase b}
\end{equation}
where the wave vector
\begin{equation}
\boldsymbol{G}=
\boldsymbol{K}^{\ }_{+}
-
\boldsymbol{K}^{\ }_{-}=
\frac{\pi}{\mathfrak{a}}\begin{pmatrix}1\\0\end{pmatrix}
\end{equation}
connects the two Fermi points
\begin{equation}
\boldsymbol{K}^{\ }_{\pm}:=
\frac{\pi}{2\mathfrak{a}}
\begin{pmatrix}\pm 1\\1\end{pmatrix}.
\end{equation}
\end{subequations}
[Here, 
$\boldsymbol{G}\cdot\boldsymbol{r}= 
 m^{\ }_{1}+m^{\ }_{2}$ 
if 
$\boldsymbol{r}= 
(m^{\ }_1 \boldsymbol{s}^{\ }_1
+ 
 m^{\ }_2 \boldsymbol{s}^{\ }_2)$
with $m^{\ }_{1}$ and $m^{\ }_{2}$ integers.]
It preserves the sublattice and time-reversal symmetries of
$\hat{H}^{\ }_{\pi}$.
Notice that 
$\delta t^{\ }_{\boldsymbol{r},\boldsymbol{r}+\boldsymbol{s}^{\ }_{j}}$ 
are purely imaginary (real) when $j=1,3$ ($j=2,4$).
When the complex-valued order parameter $\Delta$ 
is turned into a space-dependent order parameter 
$\Delta= \Delta^{\ }_0(\boldsymbol{r})\,e^{{i}f(\boldsymbol{r})}$ 
trough an amplitude $\Delta^{\ }_0(\boldsymbol{r})$
and phase $f(\boldsymbol{r})$ modulation, 
then Eq.~(\ref{eq: BDW pi flux phase b})
turns into
\begin{equation}
\begin{split}
\delta
t^{\ }_{\boldsymbol{r},\boldsymbol{r}+\boldsymbol{s}^{\ }_{j}}=&\,
\frac{t\, \Delta^{\ }_{0}(\boldsymbol{r})}{2}
\left[ 
{i}
\left(
\delta^{\ }_{j,1}
+ 
\delta^{\ }_{j,3}
\right)
+ 
\left(
\delta^{\ }_{j,2}
+
\delta^{\ }_{j,4} 
\right) 
\right]
\\
&\,\times
\cos
\left( 
f(\boldsymbol{r}) 
+ 
\frac{\pi}{2}\, j 
+ 
\boldsymbol{G} 
\cdot 
\boldsymbol{r}
\right).
\end{split}
\end{equation}
If the bond-density wave supports the unit vortex
$\Delta(\boldsymbol{r})=\Delta^{\ }_0(r)\,e^{\pm{i}\theta}$
at the origin of the lattice,
whereby we have introduced the polar coordinates
$\boldsymbol{r}\cdot\boldsymbol{s}^{\ }_{1}/\mathfrak{a}=r\cos\theta$
and
$\boldsymbol{r}\cdot\boldsymbol{s}^{\ }_{2}/\mathfrak{a}=r\sin\theta$, 
then
\begin{equation}
\begin{split}
\delta
t^{\ }_{\boldsymbol{r},\boldsymbol{r}+\boldsymbol{s}^{\ }_{j}}=&\,
\frac{t\, \Delta^{\ }_{0}(r)}{2}
\left[ 
{i}
\left(
\delta^{\ }_{j,1}
+ 
\delta^{\ }_{j,3}
\right)
+ 
\left(
\delta^{\ }_{j,2}
+
\delta^{\ }_{j,4} 
\right) 
\right]  
\\
&\,\times
\cos 
\left( 
\pm \theta 
+ 
\frac{\pi}{2}\, j 
+ 
\boldsymbol{G} 
\cdot 
\boldsymbol{r}
\right).
\end{split}
\end{equation}
The case of an arbitrary distribution of vortices
of integer charges $n^{(k)}$ at the sites $\boldsymbol{r}^{(k)}$
follows with the identifications
\begin{equation}
f(\boldsymbol{r})=
\sum_{k}
n^{(k)}
\arctan
\frac{
r^{\ }_{2}
-
r^{(k)}_{2}
     }
     {
r^{\ }_{1}
-
r^{(k)}_{1}
     }.
\end{equation}

There is the time-reversal and sublattice symmetry-breaking
bond-density wave
\begin{subequations}
\begin{equation}
\begin{split}
H^{\ }_{\eta}:=&\,
-
\sum_{\boldsymbol{r}\in\Lambda^{\ }_{\mathrm{A}}}
\sum_{j=\pm}
\left( 
t^{a}_{2,j}\,
\hat{a}^{\dag}_{\boldsymbol{r}+\boldsymbol{a}^{\ }_{j}} 
\hat{a}^{\   }_{\boldsymbol{r}}
+
\mathrm{H.c.}
\right) 
\\
&\,
-\sum_{\boldsymbol{r}\in\Lambda^{\ }_{\mathrm{B}}}
\sum_{j=\pm}
\left( 
t^{b}_{2,j}\,
\hat{b}^{\dag}_{\boldsymbol{r}+\boldsymbol{a}^{\ }_{j}} 
\hat{b}^{\   }_{\boldsymbol{r}}
+
\mathrm{H.c.}
\right)
\end{split}
\end{equation}
where 
$\boldsymbol{a}^{\ }_{\pm}=
 \boldsymbol{s}^{\ }_{1}\pm\boldsymbol{s}^{\ }_{2}$
and
\begin{equation}
t^{a}_{2,+}= 
t^{b}_{2,-}=
+\frac{\eta}{4}\,t , 
\qquad 
t^{a}_{2,-}= 
t^{b}_{2,+}= 
-\frac{\eta}{4}\,t.
\end{equation}
\end{subequations}

The lattice origin of the axial gauge field 
can also be identified with a staggered modulation of the
nearest-neighbor hopping through the perturbation
\begin{subequations}
\begin{equation}
\hat{H}^{\ }_{5}:=
-\sum_{\boldsymbol{r}\in\Lambda^{\ }_{\mathrm{A}}}
\sum_{j=1}^{4}
\left(
\delta
t^{(5)}_{\boldsymbol{r},\boldsymbol{r}+\boldsymbol{s}^{\ }_{j}}\,
\hat{b}^{\dag}_{\boldsymbol{r}+\boldsymbol{s}^{\ }_{j}} 
\hat{a}^{\   }_{\boldsymbol{r}}
+
\mathrm{H.c.}
\right)
\end{equation}
with 
\begin{equation}
\begin{split}
&
+\delta t^{(5)}_{\boldsymbol{r},\boldsymbol{r}+\boldsymbol{s}^{\ }_2}=
-\delta t^{(5)}_{\boldsymbol{r},\boldsymbol{r}+\boldsymbol{s}^{\ }_4}\equiv
A^{(5)}_{1}(\boldsymbol{r}) \, t,
\\
&
-\delta t^{(5)}_{\boldsymbol{r},\boldsymbol{r}+\boldsymbol{s}^{\ }_1}=
+\delta t^{(5)}_{\boldsymbol{r},\boldsymbol{r}+\boldsymbol{s}^{\ }_3}\equiv
{i}
A^{(5)}_{2}(\boldsymbol{r}) \, t .
\end{split}
\end{equation}
\end{subequations}
Motivated by the axial gauge flux
\begin{equation}
a^{\ }_{5i}(\boldsymbol{r})= 
- 
n\, 
a(r)\,
\epsilon^{\ }_{ij} 
\frac{r^j}{r^2},
\end{equation}
where $a(r)$ is any function that vanishes no slower than $r$ 
at the origin and saturates to $1/2$ at infinity
that screens a charge $n$ vortex in the continuum limit,
we identify the lattice axial gauge flux that screens a charge $n$ vortex
located at the origin with
\begin{subequations}
\label{eq:axial-vortex-hop}
\begin{equation}
\begin{split}
&
+\delta t^{(5)}_{\boldsymbol{r},\boldsymbol{r}+\boldsymbol{s}^{\ }_2}= 
-\delta t^{(5)}_{\boldsymbol{r},\boldsymbol{r}+\boldsymbol{s}^{\ }_4}=  
-\,t \frac{a(r)}{r}\sin\theta,
\\
&
+\delta t^{(5)}_{\boldsymbol{r},\boldsymbol{r}+\boldsymbol{s}^{\ }_1}=
-\delta t^{(5)}_{\boldsymbol{r},\boldsymbol{r}+\boldsymbol{s}^{\ }_3}= 
-{i} \,t\frac{a(r)}{r}\cos\theta,
\end{split}
\end{equation}
where we choose to regularize the vortex with
\begin{equation}
a(r)= 
\frac{1}{2}\tanh\frac{r}{\xi}.
\end{equation}
\end{subequations}
Here, $\xi$ is a characteristic length scale that determines the
core radius of the axial gauge flux. The function $a(r)$ regularizes
the singularity of $1/r$ at the origin. 
The case of a distribution of axial gauge fluxes located at
$\boldsymbol{r}^{(k)}$ follows with the substitutions
$n\to n^{(k)}$ for the integer vortex charges,
$\boldsymbol{r}\to\boldsymbol{r}-\boldsymbol{r}^{(k)}$
for the positions of the axial gauge fluxes, and a linear superposition
of the corresponding tunneling amplitudes.

Finally, a uniform magnetic flux or a magnetic flux localized to
one plaquette of the square lattice follows from the 
Peierls substitution
\begin{equation}
t^{(\pi)}_{\boldsymbol{r},\boldsymbol{r}+\boldsymbol{s}^{\ }_{j}}\to
e^{{i}\phi^{\ }_{\boldsymbol{r},\boldsymbol{r}+\boldsymbol{s}^{\ }_{j}}}
t^{(\pi)}_{\boldsymbol{r},\boldsymbol{r}+\boldsymbol{s}^{\ }_{j}}
\end{equation}
in Eq.~(\ref{eq: def pi H b})
with any suitable choice for the phases 
$\phi^{\ }_{\boldsymbol{r},\boldsymbol{r}+\boldsymbol{s}^{\ }_{j}}$.

\section{
Chiral singular gauge transformations
        }
\label{appsec: The pitfall of singular gauge transformations}

The effective theory~(\ref{eq: final effective action}) 
is one of the main results of this paper.
{}From it follows the charge and statistics
of quasiparticles. This effective theory was derived
by combining symmetry arguments to reach 
Eq.~(\ref{eq: L eff m not 0})
and a direct computation to fix the coefficients
that symmetry does not determine. Computation
of these coefficients can be achieved in many independent
ways. For example, the computation of the coefficient
$C^{(1)}_{03}$ is fixed by obtaining the charge
of quasiparticles. Hence, $C^{(1)}_{03}$ can be deduced
from Refs.~\onlinecite{Hou07},
\onlinecite{Chamon08a}, and \onlinecite{Chamon08b}
for some range of the parameters $\mu^{\ }_{\mathrm{s}}$ 
and $\eta$, 
or, 
more directly,
from numerics. The key step to derive
the effective theory~(\ref{eq: final effective action}) 
was the U(2) \textit{pure gauge} transformation%
~(\ref{eq: U(2) trsf on chi's})
[see also Eq.~(\ref{eq: parametrization U})].
In this Appendix, 
we compare these different ways of deriving effective actions
for computing charge and statistics.

We consider the field theory defined by the partition function
\begin{equation}
\begin{split}
&
Z[a^{\ }_{\mu},a^{\ }_{5\mu},\theta]:=
\int\mathcal{D}[\bar\psi,\psi]
\exp
\left(
{i}
\int d^{3}x\,
\mathcal{L}
\right),
\\
&
\mathcal{L}:=
\bar{\psi}
\left(
{i}\Slash{\partial}
- 
\Slash{a}
-
\Slash{a}^{\ }_{5}
\gamma^5 
-
|\Delta| 
e^{{i}\theta\gamma^5}
-
\mu^{\ }_{\mathrm{s}}R 
\right)
\psi.
\end{split}
\label{eqapp: def Z}
\end{equation}
We recognize Eq.~(\ref{eq: def Z})
whereby contraction with the $4\times4$ dimensional
gamma matrices is implied by the Feynman slash notation
and there is no TRS-breaking (Haldane) mass $\eta$.

Following Seradjeh and Franz in Ref.~\onlinecite{Seradjeh08},
we perform the family of chiral gauge transformations
\begin{equation}
\begin{split}
&
\bar{\psi}
=:
\bar{\chi}^{\ }_{\zeta}\,
e^{-{i}\theta\gamma^{\ }_{5}/2}
e^{+{i}(\zeta-1/2)\theta},
\\
&
\psi=:
\hphantom{\bar\chi\,}
e^{-{i}\theta\gamma^{\ }_{5}/2}\,
e^{-{i}(\zeta-1/2)\theta}\,
\chi^{\ }_{\zeta}.
\end{split}
\label{eqapp: def zeta family chiral trsf}
\end{equation}
The parameter $0\leq\zeta\leq1$ implements
a choice of ``partition'' in the terminology
of Ref.~\onlinecite{Seradjeh08}.
Each chiral transformation%
~(\ref{eqapp: def zeta family chiral trsf})
is singular if the phase $\theta$ supports vortices, 
otherwise it is a pure gauge transformation.
The (classical) transformation law of $\mathcal{L}$
in Eq.~(\ref{eqapp: def Z})
under the family of chiral transformations%
~(\ref{eqapp: def zeta family chiral trsf})
is
\begin{subequations}
\begin{equation}
\mathcal{L}\to
\mathcal{L}^{\ }_{\zeta}
\end{equation}
where
\begin{equation}
\mathcal{L}^{\ }_{\zeta}=
\bar{\psi}
\left(
{i}\Slash{\partial}
- 
\Slash{a}^{\ }_{\zeta}
-
\Slash{b}
\gamma^5 
-
|\Delta| 
-
\mu^{\ }_{\mathrm{s}}R 
\right)
\psi
\end{equation}
and
\begin{equation}
\Slash{a}^{\ }_{\zeta}=
\Slash{a}
-
\left(\zeta-\frac{1}{2}\right)
\Slash{\partial}\theta,
\qquad
\Slash{b}=
\Slash{a}^{\ }_{5}
-
\frac{1}{2}
\Slash{\partial}\theta.
\end{equation}
\end{subequations}
Observe that,
whenever $\theta$ supports vortices
and $\zeta\neq1/2$,
a physical magnetic flux has appeared
where there was none to begin with. Thus,
if we demand TRS, we must choose $\zeta=1/2$
and demand that $a^{\ }_{\mu}$ is pure gauge.

In the spirit of Ref.~\onlinecite{Seradjeh08}, for general $\zeta$
we \textit{define} the family of partition functions 
\begin{equation}
Z^{\ }_{\zeta}
[a^{\ }_{\zeta\mu},b^{\ }_{\mu}]:=
\int\mathcal{D}[\bar\chi^{\ }_{\zeta},\chi^{\ }_{\zeta}]
e^{
{i}
\int d^{3}x\,
\mathcal{L}^{\ }_{\zeta}
   }
\equiv
e^{
{i}
\int d^{3}x\,
\mathcal{L}^{\mathrm{eff}}_{\zeta}
  }
\end{equation}
and compute the effective action 
$\mathcal{L}^{\mathrm{eff}}_{\zeta}$
for the gauge fields 
$a^{\ }_{\zeta\mu}$ and $b^{\ }_{\mu}$
that follows from integrating the massive fermions 
$\bar\chi^{\ }_{\zeta}$ and $\chi^{\ }_{\zeta}$
to lowest order in a gradient expansion.
All ultra-violet divergences induced by the integration over
the fermions can be disposed of
with the help of the Pauli-Villars regularization scheme.
The effective action, expressed 
in terms of $a^{\ }_{\mu}$, $a^{\ }_{5\mu}$, and $\theta$, 
that follows to leading order in a gradient expansion, is 
\begin{equation}
\begin{split}
\mathcal{L}^{\mathrm{eff}}_{\zeta}=&\,
\frac{|\Delta|^2}{2\pi m} 
\left(
a^{\ }_{5\mu}-\frac{1}{2}\partial^{\ }_{\mu}\theta
\right)
\left(
a^{\mu}_{5}-\frac{1}{2}\partial^{\mu}\theta
\right)
\\
&\,
-
\frac{2Q}{2\pi}
\epsilon^{\nu\rho\kappa}
\left(
a^{\ }_{\nu}
-
\frac{2\zeta-1}{2}
\partial^{\ }_{\nu}\theta
\right)
\partial^{\ }_{\rho}
\left(
a^{\ }_{5\kappa}
-
\frac{1}{2}\partial^{\ }_{\kappa}\theta
\right),
\end{split}
\label{eqapp: Seradjeh+Franz eff L}
\end{equation}
where 
$Q=
{\rm sgn}\,\mu^{\ }_{\mathrm{s}}
\frac{1}{2}\left(1-\mu^{\ }_{\mathrm{s}}/m\right)$.
The effective action~(\ref{eqapp: Seradjeh+Franz eff L})
fails to capture the charge of screened quasiparticles. For example, 
the conserved induced fermionic current
\begin{equation}
j^{\mu}_{\zeta}:=
\frac{2Q}{2\pi}
\epsilon^{\mu\rho\kappa}
\partial^{\ }_{\rho}
\left(
a^{\ }_{5\kappa}
-
\frac{1}{2}\partial^{\ }_{\kappa}\theta
\right),
\end{equation}
which is independent of the parameter $\zeta$, does not reproduce the
induced fermionic charge~(\ref{eq: derivation of Q}) when the
axial gauge field screens the mass vortices. It follows from their
result that the charge bound to screened vortices 
(in which case 
$a^{\ }_{5\kappa}-\frac{1}{2}\partial^{\ }_{\kappa}\theta=0$) is $Q=0$ (!)
instead of $Q=1/2$ as found in Refs.%
~\onlinecite{Chamon08a},
\onlinecite{Chamon08b}, 
and in Sec.~\ref{sec: Fractional charge quantum number}.

Moreover, after proper dualization of 
the effective action~(\ref{eqapp: Seradjeh+Franz eff L})
[this dualization must include the Higgs mass,
i.e., the first line on the right-hand side of
Eq.~(\ref{eqapp: Seradjeh+Franz eff L}),
a fact that was ignored in Ref.~\onlinecite{Seradjeh08}]
it follows that the exchange statistics is $\zeta$
dependent. This is expected in view of the introduction
of magnetic fluxes whenever $\zeta\neq1/2$
contrary to the implicit assumption made in
Ref.~\onlinecite{Seradjeh08} when choosing $\zeta=0$.


\begin{thebibliography}{99}

\bibitem{Novoselov05} 
K. S. Novoselov \textit{et al.}, 
Nature (London) \textbf{438}, 197 (2005).

\bibitem{Zhang05}  
Y. Zhang \textit{et al.}, 
Nature (London) \textbf{438}, 201 (2005). 

\bibitem{Nair08}
R. R. Nair \textit{et al.}, 
Science \textbf{320}, 1308 (2008).

\bibitem{Lanzara07}
S. Y. Zhou, 
G.-H. Gweon, 
A. V. Fedorov, 
P. N. First, 
W. A. de Heer, 
D.-H. Lee, 
F. Guinea, 
A. H. Castro Neto, 
and
A. Lanzara,
Nature Mat.\ \textbf{6}, 770 (2007);
\textit{ibid} 916 (2007). 

\bibitem{Checkelsky09}
Joseph G. Checkelsky, Lu Li, and N. P. Ong,
Phys.\ Rev.\ B \textbf{79}, 115434 (2009).

\bibitem{Nomura09}
Kentaro Nomura, Shinsei Ryu, and Dung-Hai Lee, 
\texttt{arXiv:0906.0159}.

\bibitem{Chamon09}
C. Chamon, C.-Y. Hou, and C. Mudry,
unpublished.

\bibitem{Semenoff84}
G. W. Semenoff, 
Phys.\ Rev.\ Lett.\ \textbf{53}, 2449 (1984).

\bibitem{Haldane88} 
F. D. M. Haldane, 
Phys.\ Rev.\ Lett.\ \textbf{61}, 2015 (1988).

\bibitem{Hou07}
C.-Y. Hou, C. Chamon, and C. Mudry, 
Phys.\ Rev.\ Lett.\ \textbf{98}, 186809 (2007).

\bibitem{Callan85}
C. G. Callan and J. A. Harvey,
Nucl.\ Phys.\ B\textbf{250}, 427 (1985).


\bibitem{Jackiw76}
R. Jackiw and C. Rebbi, 
Phys.\ Rev.\ D \textbf{13}, 3398 (1976).

\bibitem{Su79} 
W. P. Su, J. R. Schrieffer, and A. J. Heeger, 
Phys.\ Rev.\ Lett.\textbf{42}, 1698 (1979), \textit{ibid}, 
Phys. Rev. B \textbf{22}, 2099 (1980).  

\bibitem{Jackiw81npb}
R. Jackiw and  J. R. Schrieffer, 
Nucl.\ Phys.\ B\textbf{190}, 253 (1981).

\bibitem{Chamon08a} 
C. Chamon, C.-Y. Hou, R. Jackiw, C. Mudry, 
S.-Y. Pi, and A. P. Schnyder, 
Phys.\ Rev.\ Lett.\ \textbf{100}, 110405 (2008).

\bibitem{Chamon08b}
C. Chamon, C.-Y. Hou, R. Jackiw, C. Mudry, S.-Y. Pi, 
and G. Semenoff,
Phys.~Rev.~B \textbf{77}, 235431 (2008).

\bibitem{Seradjeh08}
B. Seradjeh and M. Franz, 
Phys.~Rev.~Lett. \textbf{101}, 146401 (2008).

\bibitem{Milovanovic08}
M. V. Milovanovi\^c, 
Phys.~Rev.~B  \textbf{78}, 245424 (2008).

\bibitem{Jackiw07}
R. Jackiw and S.-Y. Pi, 
Phys.\ Rev.\ Lett.\ \textbf{98}, 266402 (2007).


\bibitem{footnote: covariant covention for gradient}
The gradient 
$\boldsymbol{\partial}\equiv\frac{\partial}{\partial\boldsymbol{r}}$
and the time derivative 
$\partial^{\ }_{t}\equiv\frac{\partial}{\partial t}$
form the covariant 3-vector
$\partial^{\ }_{\mu}=(\partial^{\ }_{t},\boldsymbol{\partial})$.

\bibitem{Itzykson80}
C. Itzykson and J.-B. Zuber,
\textit{Quantum field theory},
McGraw-Hill, New York (1980).

\bibitem{McClure56}
J. W. McClure, 
Phys.\ Rev.\ \textbf{104}, 666 (1956).

\bibitem{Morozov06}
S. V. Morozov \textit{et al.},
Phys.\ Rev.\ Lett.\ \textbf{97}, 016801 (2006).

\bibitem{Morpurgo06}
A. F. Morpurgo and F. Guinea,
Phys.\ Rev.\ Lett.\ \textbf{97}, 196804 (2006).

\bibitem{Gonzalez92}
J. Gonzalez, F. Guinea, and
M. A. H. Vozmediano, 
Phys.\ Rev.\ Lett.\ \textbf{69}, 172 (1992);
\textit{ibid}
Nucl.\ Phys.\ B \textbf{406}, 771 (1993).

\bibitem{Lammert00}
P. E. Lammert and V. H. Crespi, Phys. Rev. Lett. {\bf 85}, 5190 (2000).

\bibitem{Pachos08} 
J. K. Pachos, M. Stone, and K. Temme, 
Phys. Rev. Lett. \textbf{100}, 156806 (2008).

\bibitem{Affleck88} 
I. K. Affleck and J. B. Marston,
Phys.\ Rev.\ B \textbf{37}, 3774 (1988).

\bibitem{Wen89}  
X. G. Wen, F. Wilczeck, and A. Zee,
Phys.\ Rev.\ B \textbf{39}, 11413 (1989).

\bibitem{Fradkin91} 
E. Fradkin,
\textit{Field Theories of Condensed Matter Systems},
Addison-Wesley, Redwood City, CA (1991).

\bibitem{Mudry94}
C. Mudry and E. Fradkin,
Phys.\ Rev.\ B \textbf{49}, 5200 (1994);
\textit{ibid} \textbf{50}, 11409 (1994).

\bibitem{Ludwig94} 
A. W. W. Ludwig, M. P. A. Fisher, R. Shankar, and G. Grinstein, 
Phys.\ Rev.\ B \textbf{50}, 7526 (1994).

\bibitem{Hatsugai97} 
Y. Hatsugai, X.-G. Wen, and M. Kohmoto, 
Phys.\ Rev.\ B \textbf{56}, 1061 (1997).

\bibitem{Guruswamy00} 
S. Guruswamy, A. LeClair, and A. W. W. Ludwig, 
Nucl.\ Phys.\ B \textbf{583}, 475 (2000).

\bibitem{footnote H is real valued}
The microscopic tight-binding model can be represented
by a real-valued symmetric Hamiltonian.


\bibitem{Fujikawa04}
K. Fujikawa and H. Suzuki, 
``Path Integrals and Quantum Anomalies'', 
Oxford Univ. Press, Oxford (2004).

\bibitem{Jaroszewicz84}
T. Jaroszewicz, 
Phys.\ Lett.\ \textbf{146B}, 337 (1984).

\bibitem{Chen89}
Y.-H. Chen and F. Wilczek, 
Int.\ J. Mod.\ Phys.\ B \textbf{3}, 117 (1989).

\bibitem{Hlousek90} 
Z. Hlousek, D. Senechal, and S. H. Henry Tye, 
Phys.\ Rev.\ D \textbf{41}, 3773 (1990).

\bibitem{Yakovenko90}
V. M. Yakovenko,
Phys.\ Rev.\ Lett.\ \textbf{65}, 251 (1990).

\bibitem{Abanov00}
A. G. Abanov and P. B. Wiegmann,
Nucl.\ Phys.\ B \textbf{570}, 685 (2000).


\bibitem{Jaroszewicz86}
T. Jaroszewicz, 
Phys.~Rev.~D \textbf{34}, 3128 (1986).

\bibitem{Deser82}
S. Deser, R. Jackiw, and S. Templeton,
Ann.~Phys.~(N.Y.) \textbf{140}, 372 (1982).

\bibitem{Blau91}
M. Blau and G. Thompson, 
Ann.\ Phys.\ \textbf{205}, 130 (1991).

\bibitem{Hansson04}  
T. H. Hansson, Vadim Oganesyan, and S. L. Sondhi,
Ann.\ Phys.\ \textbf{313}, 497 (2004). 




\bibitem{Weeks08}
B. Seradjeh, C. Weeks, and M. Franz,
Phys.\ Rev.\ B \textbf{77}, 033104 (2008).


\bibitem{Wilson77} 
K. Wilson in 
\textit{New Phenomena in Subnuclear Physics}, 
Edited by A. Zichichi (Plenum, New York, 1977).

\bibitem{Levin03}
M. Levin, X.-G. Wen
Phys.\ Rev.\ B \textbf{67}, 245316 (2003). 


\bibitem{Kane05}
C. L. Kane and E. J. Mele,
Phys.\ Rev.\ Lett.\ \textbf{95}, 146802 (2005).

\bibitem{Beenakker06}
C. W. J. Beenakker,
Phys.\ Rev.\ Lett.\, \textbf{97}, 067007 (2006).

\bibitem{GhaemiWilczek}
Pouyan Ghaemi and Frank Wilczek,
\texttt{arXiv:0709.2626}.

\bibitem{Roy06}
Rahul Roy,
\texttt{arXiv:cond-mat/0608064}.

\bibitem{Schnyder08}
Andreas P. Schnyder,
Shinsei Ryu,
Akira Furusaki,
and
Andreas W. W. Ludwig,
Phys.\ Rev.\ B \textbf{78}, 195125 (2008).

\bibitem{Roy08}
Rahul Roy,
\texttt{arXiv:0803.2868}.

\bibitem{Qi09}
Xiao-Liang Qi, Taylor L. Hughes, Srinivas Raghu, and Shou-Cheng Zhang,
Phys.\ Rev.\ Lett.\ \textbf{102}, 187001 (2009).

\bibitem{Kitaev08}
A. Yu.\ Kitaev,
``Periodic table for topological insulators and superconductors'',
http://landau100.itp.ac.ru/Talks/kitaev.pdf.

\bibitem{Tanaka05a}
A. Tanaka and X. Hu, 
Phys.\ Rev.\ Lett.\ \textbf{95}, 036402 (2005).

\bibitem{Tanaka05b}
Akihiro Tanaka and Xiao Hu,
Phys.\ Rev.\ B \textbf{74}, 140407 (2006).

\bibitem{Ghaemi09}
Pouyan Ghaemi, Shinsei Ryu, and Dung-Hai Lee,
\texttt{arXiv:0903.1662}.


\bibitem{Herbut07}
I.\ Herbut,
Phys.\ Rev.\ Lett.\ \textbf{99}, 206404 (2007). 



\bibitem{Senthil04a}
T. Senthil, A. Vishwanath, L. Balents, S. Sachdev, 
and M. P. A. Fisher,
Science \textbf{303}, 1490 (2004).

\bibitem{Senthil04b}
T. Senthil, L. Balents, S. Sachdev, A. Vishwanath, 
and M. P. A. Fisher,
Phys.\ Rev.\ B \textbf{70}, 144407 (2004).

\bibitem{Senthil05}
T. Senthil and Matthew P. A. Fisher,
Phys.\ Rev.\ B \textbf{74}, 064405 (2006).

\bibitem{Ran06}
%
Ying Ran and Xiao-gang Wen,
Phys.\ Rev.\ Lett.\ \textbf{96}, 026802 (2006);
\textit{ibid} \texttt{arXiv:cond-mat/0609620}.

\bibitem{Sandvik07}

Anders W.\ Sandvik,
Phys.\ Rev.\ Lett.\ \textbf{98}, 227202 (2007).

\bibitem{Melko07}
Roger G.\ Melko, and Ribhu K.\ Kaul,
Phys.\ Rev.\ Lett.\ \textbf{100}, 017203 (2008).

\bibitem{Kaul08}
Ribhu K. Kaul, Roger G. Melko, Max A. Metlitski, and Subir Sachdev,
Phys.\ Rev.\ Lett.\ \textbf{101}, 187206 (2008).

\bibitem{Grover08}
Tarun Grover and T. Senthil
Phys.\ Rev.\ Lett.\ \textbf{100}, 156804 (2008).

\bibitem{RanVishwanathLee08}
Ying Ran, Ashvin Vishwanath, Dung-Hai Lee,
Phys.\ Rev.\ Lett.\ \textbf{101}, 086801 (2008).

\bibitem{QiZhang08} 
Xiao-Liang Qi and Shou-Cheng Zhang,
Phys.\ Rev.\ Lett.\ \textbf{101}, 086802 (2008).

\bibitem{Bernevig-Taylor-Zhang06}
B. Andrei Bernevig, Taylor L. Hughes, and Shou-Cheng Zhang,
Science \textbf{314}, 1757 (2006).

\bibitem{Koenig07}
Markus K\"onig,
Steffen Wiedmann, 
Christoph Bruene, 
Andreas Roth, 
Hartmut Buhmann, 
Laurens W. Molenkamp, 
Xiao-Liang Qi, 
and Shou-Cheng Zhang,
Science \textbf{318}, 766 (2007).

\bibitem{Koenig08}
Markus K\"onig,
Hartmut Buhmann, 
Laurens W. Molenkamp, 
Taylor L. Hughes, 
Chao-Xing Liu,
Xiao-Liang Qi, and 
Shou-Cheng Zhang,
J.\ Phys.\ Soc.\ Jpn.\ \textbf{77}, 031007 (2008).

\bibitem{QiTaylorZhang07}
Xiao-Liang Qi, 
Taylor L. Hughes, 
and 
Shou-Cheng Zhang, 
Nature Physics~\textbf{4}, 273  (2008). 


\bibitem{Wen91}
X. G. Wen and A. Zee,
Phys.\ Rev.\ B \textbf{44}, 274 (1991).

\bibitem{footnote: error in Chamon08a}
Equation (7) in Ref.~\onlinecite{Chamon08a}
and the heuristic argument that follows 
are both wrong.


\bibitem{Banks77}
T. Banks, R. J. Myerson, and J. Kogut, 
Nucl.\ Phys.\ \textbf{B129}, 493 (1977).

\bibitem{Thomas78} 
P. R. Thomas and M. Stone, 
Nucl.\ Phys.\ \textbf{B144}, 513 (1978).

\bibitem{Peskin78}  
M. Peskin, 
Ann.\ Phys.\ \textbf{113}, 122 (1978).

\bibitem{Dasgupta81}
C. Dasgupta and B. I. Halperin,
Phys.\ Rev.\ Lett.\ \textbf{47}, 1556 (1981).

\bibitem{Fisher89}
M. P. A. Fisher and D. H. Lee, 
Phys.\ Rev.\ B \textbf{39}, 2756 (1989). 


\end{thebibliography}
\end{document}